\documentclass[usenatbib,iop,numberedappendix]{aeb_emulateapj_2010}
\usepackage{amsmath}
\usepackage{amssymb}
\usepackage{epsfig}
\usepackage{xspace}
\usepackage{color}

\def\del#1{{}}

\defcitealias{2012ApJ...758...74B}{BBPS1}
\defcitealias{2012ApJ...758...75B}{BBPS2}
\defcitealias{BBPS3}{BBPS3}
\defcitealias{BBPS4}{BBPS4}
\defcitealias{BBPS5}{BBPS5}

\SetSymbolFont{symbols}{bold}{OMS}{cmsy}{b}{n}
\DeclareSymbolFont{bmisymbols}{OML}{cmm}{b}{it}
\DeclareMathSymbol{\balpha}{0}{bmisymbols}{"0B}
\DeclareMathSymbol{\bbeta}{0}{bmisymbols}{"0C}
\DeclareMathSymbol{\bgamma}{0}{bmisymbols}{"0D}
\DeclareMathSymbol{\bdelta}{0}{bmisymbols}{"0E}
\DeclareMathSymbol{\bepsilon}{0}{bmisymbols}{"0F}
\DeclareMathSymbol{\bzeta}{0}{bmisymbols}{"10}
\DeclareMathSymbol{\boldeta}{0}{bmisymbols}{"11}
\DeclareMathSymbol{\btheta}{0}{bmisymbols}{"12}
\DeclareMathSymbol{\biota}{0}{bmisymbols}{"13}
\DeclareMathSymbol{\bkappa}{0}{bmisymbols}{"14}
\DeclareMathSymbol{\blambda}{0}{bmisymbols}{"15}
\DeclareMathSymbol{\bmu}{0}{bmisymbols}{"16}
\DeclareMathSymbol{\bnu}{0}{bmisymbols}{"17}
\DeclareMathSymbol{\bxi}{0}{bmisymbols}{"18}
\DeclareMathSymbol{\bpi}{0}{bmisymbols}{"19}
\DeclareMathSymbol{\brho}{0}{bmisymbols}{"1A}
\DeclareMathSymbol{\bsigma}{0}{bmisymbols}{"1B}
\DeclareMathSymbol{\btau}{0}{bmisymbols}{"1C}
\DeclareMathSymbol{\bupsilon}{0}{bmisymbols}{"1D}
\DeclareMathSymbol{\bphi}{0}{bmisymbols}{"1E}
\DeclareMathSymbol{\bchi}{0}{bmisymbols}{"1F}
\DeclareMathSymbol{\bpsi}{0}{bmisymbols}{"20}
\DeclareMathSymbol{\bomega}{0}{bmisymbols}{"21}
\DeclareMathSymbol{\bvarepsilon}{0}{bmisymbols}{"22}
\DeclareMathSymbol{\bvartheta}{0}{bmisymbols}{"23}
\DeclareMathSymbol{\bvarpi}{0}{bmisymbols}{"24}
\DeclareMathSymbol{\bvarrho}{0}{bmisymbols}{"25}
\DeclareMathSymbol{\bvarsigma}{0}{bmisymbols}{"26}
\DeclareMathSymbol{\bvarphi}{0}{bmisymbols}{"27}

\newcommand{\mathbfit}[1]{\textbf{\textit{#1}}}

\newcommand{\bra}{\langle}
\newcommand{\ket}{\rangle}

\newcommand{\rmn}{\mathrm}
\newcommand{\dd}{\mathrm{d}}
\newcommand{\vecbf}{\mathbfit}

\newcommand{\bvel}{\bupsilon}

\newcommand{\KU}{K/U}
\newcommand{\CA}{c/a}

\slugcomment{Submitted to ApJ}

\shorttitle{Gas and Stellar Mass Fractions in Galaxy Clusters}
\shortauthors{Battaglia et al.}

\begin{document}

\title{On the Cluster Physics of Sunyaev-Zel'dovich and X-ray Surveys III: \\
  Measurement Biases and Cosmological Evolution of Gas and Stellar Mass
  Fractions}

\author{N. Battaglia\altaffilmark{1}, J. R. Bond\altaffilmark{2}, C. Pfrommer\altaffilmark{3,*}, J. L. Sievers\altaffilmark{4,5}}

\altaffiltext{1}{McWilliams Center for Cosmology, Carnegie Mellon University,
Department of Physics, 5000 Forbes Ave., Pittsburgh PA, USA, 15213}
\altaffiltext{2}{Canadian Institute for Theoretical Astrophysics, 60 St George, Toronto ON, Canada, M5S 3H8}
\altaffiltext{3}{Heidelberg Institute for Theoretical Studies, Schloss-Wolfsbrunnenweg 35, D-69118 Heidelberg, Germany}
\altaffiltext{4}{Joseph Henry Laboratories of Physics, Jadwin Hall, Princeton University, Princeton NJ, USA, 08544}
\altaffiltext{5}{Astrophysics and Cosmology Research Unit, University of Kwazulu-Natal, Westville, Durban 4000, South Africa}
\altaffiltext{*}{corresponding author, christoph.pfrommer@h-its.org}

\begin{abstract}
  Gas masses tightly correlate with the virial masses of galaxy clusters,
  allowing for a precise determination of cosmological parameters by means of
  large-scale X-ray surveys.  However, the gas mass fractions ($f_{\rmn{gas}}$)
  at the virial radius ($R_{200}$) derived from recent Suzaku observations of
  several groups are considerably larger than the cosmic mean, calling into
  question the accuracy of estimates of cosmological parameters.  Here, we use a
  large suite of cosmological hydrodynamical simulations to study measurement
  biases of $f_{\rmn{gas}}$. We employ different variants of simulated physics,
  including radiative gas physics, star formation, and thermal feedback by
  active galactic nuclei, which we show is able to arrest overcooling and to
  result in constant stellar mass fractions for redshifts $z<1$.  This implies
  that the stellar and dark matter masses increase at the same rate, which is
  realized if most of the stellar mass is already in place by the time it
  assembles in the cluster halo---in agreement with observations.  Computing the
  mass profiles in 48 angular cones, whose footprints partition the sphere, we
  find anisotropic gas and total mass distributions that imply an angular
  variance of $f_{\rmn{gas}}$ at the level of 30\%. This anisotropic
  distribution originates from the recent formation epoch of clusters and from
  the strong internal baryon-to-dark-matter density bias. In the most extreme
  cones, $f_{\rmn{gas}}$ can be biased high by a factor of two at $R_{200}$ in
  massive clusters ($M_{200}\sim 10^{15}\,\rmn{M}_\odot$), thereby providing a
  potential explanation for high $f_{\rmn{gas}}$ measurements by {\em
    Suzaku}. While projection lowers this factor, there are other measurement
  biases that may (partially) compensate. We find that at $R_{200}$,
  $f_{\rmn{gas}}$ is biased high by $20\%$ when assuming hydrostatic equilibrium
  masses, i.e., neglecting the kinetic pressure, and by another $\sim10-20\%$
  due to the presence of density clumping (depending on mass and dynamical
  state). At larger radii, both measurement biases increase dramatically. While
  the cluster sample variance of the true $f_{\rmn{gas}}$ decreases to a level
  of 5\% at $R_{200}$, the sample variance that includes both measurement biases
  remains fairly constant at the level of $10-20\%$ (depending on dynamical
  state). At the high-mass end ($M_{500} > 2\times 10^{14}\,\rmn{M}_\odot$), the
  true $f_{\rmn{gas}}$ within $R_{500}$ shows a constant redshift
  evolution. While this result is in principle encouraging for using gas masses
  to derive cosmological parameters, careful X-ray mocks are needed to control
  those various measurement biases.
\end{abstract}


\section{Introduction}

Galaxy clusters are the rarest and largest gravitationally-collapsed objects in
the universe, making them very sensitive tracers of the growth of structure
\citep[for a review, see][]{Voit2005}. Clusters exhibit a well-defined number
count that steeply falls as mass and redshift increase. The number density tail
is very sensitive to changes in cosmological parameters \citep[as demonstrated
by recent measurements, e.g., by][]{2008MNRAS.383..879A, 2009ApJ...692.1060V}
including the dynamical characteristics of dark energy, primordial
non-Gaussianity, and the theory of gravity. Hence, future large surveys of
clusters (in the optical, X-ray, and microwave/radio/sub-mm wavelengths) with
substantially increased statistics and resolution can potentially provide a gold
mine for fundamental cosmology \citep[e.g.,][]{2003PhRvD..68h3506B,
  2003PhRvD..67h1304H, 2003ApJ...585..603M, 2004ApJ...613...41M,
  2010PhRvD..82d1301K, 2012PhRvD..86d3516G}. However, this is only possible if
systematics associated with cluster mass calibrations and our incomplete
knowledge of the physics of the intracluster medium (ICM) can be understood and
controlled. As observations have been progressively refined and resolutions
improved, the clusters have been revealed to be too complex for simple
sphericalized analytical modelling, calling for detailed theoretical work with a
necessarily heavy computational component.

The virialized regions of clusters -- usually characterized by a sphere of
radius $R_{200}$ enclosing a mean density that is 200 times the critical density
of the Universe (and higher mean interior densities correspond to smaller radii)
-- are separated from the Hubble-flow. However, those regions maintain contact
with the surrounding filamentary cosmic web through ongoing accretion and
mergers as they evolve. Most of the cluster baryons are in the form of a hot
($kT\sim(1-10)\,$~keV) diffuse plasma, the ICM, while the remaining baryons are
confined to the cluster's numerous stars and galaxies. Since cold dark matter
(CDM) obeys the {\em collisionless} Boltzmann equation and since the dynamics of
the hot baryons additionally responds to pressure forces according to the ({\em
  collisional}) Euler equation (neglecting viscosity effects), we expect a
strong internal baryon-to-dark-matter (DM) density bias. We hope that when
averaged over cluster scales, the smoothed densities are nearly in the universal
Hubble-volume-smoothed proportion (or a constant, large fraction thereof). But
at which radius and cluster scale this exactly happens depends on baryonic
physics and needs to be carefully studied using a combination of hydrodynamical
cosmological simulations and ICM observations.

Using {\em Chandra} X-ray observations of a sample of relaxed clusters, earlier
work has successfully used the gas mass fraction in clusters to derive
cosmological parameters in agreement with the concordance $\Lambda$CDM model
\citep{2002MNRAS.334L..11A, 2004MNRAS.353..457A, 2008MNRAS.383..879A}. In these
analyses, masses were derived assuming spherical symmetry and hydrostatic
equilibrium while allowing for a modest constant non-thermal pressure
contribution. Similar conclusions were obtained through a joint X-ray and
Sunyaev-Zel'dovich (SZ) analysis \citep{2006ApJ...652..917L} as well as in a
{\em Chandra} sample of X-ray luminous clusters extending to a redshift of
$z\lesssim1.3$ \citep{2009A&A...501...61E}. Recently, a more sophisticated
analysis of an X-ray flux-selected sample of 238 clusters at $z\leq 0.5$, which
takes into account various selection effects, combines gas mass fractions, X-ray
luminosity, and temperature \citep{2010MNRAS.406.1759M,
  2010MNRAS.406.1773M}. These authors derive improved cosmological parameters
that are consistent with $\Lambda$CDM and have approximately equal constraining
power compared to other cosmological probes (at the time of publication) on the
DM density parameter, $\Omega_m$, the {\it rms} amplitude of the (linear)
density power spectrum on cluster-mass scales, $\sigma_8$, and the equation of
state of dark energy. These data were even used to obtain interesting
constraints on modified gravity and general relativity on cosmological scales
\citep{2009MNRAS.400..699R,2010MNRAS.406.1796R} as well as on neutrino
properties \citep{2010MNRAS.406.1805M}. However, in order to get unbiased
cosmological constraints, the gas mass fraction, $f_\rmn{gas} =
M_\rmn{gas}/M_\rmn{tot}$, either needs to be independent of both, redshift and
mass \citep{2005A&A...437...31S} or, if $f_\rmn{gas}$ varied with redshift
and/or mass, the associated biases and the scatter around them would have to be
accurately quantified.

\citet{2006ApJ...640..691V} studied gas and total mass profiles in a small
sample of {\em Chandra} clusters of high-quality data out to $R_{500}$. They
find that enclosed $f_\rmn{gas}$ profiles within $R_{2500}\simeq 0.4 R_{500}$
do not asymptotically approach the universal baryon fraction derived from
cosmic microwave background (CMB) observations (with a discrepancy of a
factor of 1.5-2). This discrepancy becomes smaller with increasing cluster
mass and radius, but is still noticeable at $R_{500}$ enforcing the need to
understand the behavior of $f_\rmn{gas}$ at even larger radii.

The recently launched {\em Suzaku} has enabled measurements of cluster X-ray
emission up to and even beyond $R_{200}$ due to its low-Earth orbit that ensures
a low and stable particle background, almost an order of magnitude lower than 
previous X-ray telescopes. These measurements are critical in
improving our understanding of the formation history of galaxy clusters, the
dynamical state of the ICM at greater radii without the need to extrapolate
measurements of the core regions of clusters, and to gain confidence in
assumptions used to derive cosmological parameters from cluster surveys.  There
is now a growing collection of {\em Suzaku} clusters, consisting of PKS0745-191
\citep{2009MNRAS.395..657G}, Abell 1795 \citep{2009PASJ...61.1117B}, Abell 2204
\citep{2009A&A...501..899R}, Abell 1413 \citep{2010PASJ...62..371H}, Abell 1689
\citep{2010ApJ...714..423K}, Abell 2142 \citep{2011PASJ...63S1019A}, Perseus
\citep{2011Sci...331.1576S}, a fossil group RX J1159+5531
\citep{2012ApJ...748...11H}, Abell 2029 \citep{2012MNRAS.422.3503W}, and Hydra A
\citep{2012PASJ...64...95S}.

While most of these {\em Suzaku} measurements observe an excess X-ray emission
for radii $r\gtrsim R_{500}$ over what is expected from cosmological
hydrodynamical simulations, there are two clusters, Abell 1795 and RX
J1159+5531, where the emission is compatible with the theoretical expectations
(within the error bars). The excess emission manifests itself in entropy
profiles which rise less steeply than the predictions of purely
gravitational hierarchical structure formation (and can even start to drop beyond
$R_{200}$). Alternatively, this implies increasing $f_\rmn{gas}$ profiles
reaching values above the cosmic mean and, in one case, even boosting the
universal baryon fraction by a factor of 1.6 at $R_{200}$ when accounting for a
stellar mass fraction of 12\% \citep{2011Sci...331.1576S}. In some cases, the
excess emission could be explained by the anisotropic nature of the cosmic
filaments connecting to clusters, in particular if there is an overlap of the
outskirts of two neighboring clusters (as it is possibly the case in Abell
2029). Possible physical explanations include substantial density clumping
and/or neglecting (respectively underestimating) non-thermal pressure in
deriving hydrostatic masses. Since the X-ray emissivity is proportional to the
square of the gas density, a clumpy ICM with dense regions will induce X-ray
surface brightness fluctuations \citep{2012MNRAS.421.1123C,2012MNRAS.421..726S}
that enhance the emission over the smooth component, biasing the entropy lower
and $f_\rmn{gas}$ higher in comparison to an extrapolation from the more
homogeneous inner regions. Observational or instrumental reasons for the
observed excess emission (which is typically a factor of 3-5 below the soft
extragalactic X-ray background) include fluctuations of unresolved point
sources, point-spread function leakage from masked point sources, or stray light
from the bright inner cluster core.

To better address possible systematics, a large sample of clusters with
well-defined selection criteria are currently being observed, exploiting the
strengths of three complementary X-ray observatories: {\em Suzaku} (low, stable
background), {\em XMM-Newton} (high effective area and sensitivity), and {\em
  Chandra} (good spatial resolution) \citep{2012AIPC.1427...13M}. Combining {\em
  Chandra} and {\em Suzaku} pointings in cluster outskirts increases the number
of detected point sources by a factor of 10 and, hence, reduces the
uncertainty on the X-ray surface brightness by a factor of $\sim3$. Early
results from this program on Abell 3378 show entropy and $f_\rmn{gas}$ profiles
that are in fact compatible with theoretical expectations. This result is
confirmed by a stacking analysis of {\em ROSAT} clusters that detects a
steepening of the density profiles beyond $\sim R_{500}$
\citep{2012A&A...541A..57E}, in agreement with simulation results that include
radiative physics and account for gas clumping \citep{2011ApJ...731L..10N}
-- apparently contradicting recent {\em Suzaku} results. 

Hydrodynamical simulations of the formation of clusters that include radiative
cooling, star formation and stellar feedback predict stellar mass fractions in
excess of what is observed, a decreasing gas mass fraction with redshift, and
overpredict the X-ray luminosity on group scales \citep{2005ApJ...625..588K,
  2006MNRAS.365.1021E, 2006MNRAS.373.1339R}. This problem is substantially
mitigated when including energetic feedback by active galactic nuclei (AGN)
\citep{Sijacki+2006,2007MNRAS.380..877S, Sijacki+2008, 2008ApJ...687L..53P,
  Booth+2009, 2010MNRAS.401.1670F, McCarthy+2010, Dubois+2010, Teyssier+2011,
  2011MNRAS.413..691Y,2013MNRAS.431.1487P}. Interestingly, including AGN
feedback brings the modeled SZ power spectrum into agreement with data while
purely radiative models overproduce the measured power on scales around 3 arcmin
($\ell \sim 3000$) \citep{2010ApJ...725...91B, 2012ApJ...758...75B,
  2010ApJ...725.1452S, 2011ApJ...727...94T}. Most notably, the SZ power spectrum
of models that only account for shock heating and neglect radiative physics and
hence the condensation of baryons into stars produce too much power on scales of
$\ell \sim 3000$ \citep{2010ApJ...725...91B}, in strong tension with the
measurements by SPT and ACT
\citep{2012ApJ...755...70R,2013arXiv1301.0824S}. While the gas mass fractions in
these models are higher than in the corresponding radiative models (with and
without AGN feedback), they are somewhat lower than the universal baryon
fractions. This demonstrates that for consistency reasons with the SZ power
spectrum, the {\em Suzaku}-inferred high values of $f_\rmn{gas}$ along some
radial arms either are not representative for the entire virial regions of those
clusters or appear to be biased high. Hence this emphasizes that a successful
model of the cluster physics not only needs to match stellar and gas mass
fractions as a function of radius and cluster mass, but also has to adhere to
constraints from SZ surveys, namely scaling relations and power spectrum
measurements.

This is the third in a series of papers addressing the cluster physics of SZ and
X-ray surveys. In the first two papers, we thoroughly scrutinized the influence
of feedback, non-thermal pressure and cluster shapes on $Y-M$ scaling relations
\citep[][hereafter BBPS1]{2012ApJ...758...74B} and the thermal SZ power spectrum
\citep[][hereafter BBPS2]{2012ApJ...758...75B}. The fourth paper will detail the
physics of density and pressure clumping due to infalling substructures,
accompanying the growth of clusters \citep[][hereafter BBPS4]{BBPS4}, and the
fifth will provide an information theoretic view of clusters and their
non-equilibrium entropies \citep[][hereafter BBPS5]{BBPS5}.

In this work, we are motivated by current and next generation surveys in the
X-rays ({\em eROSITA}) and those employing the Sunyaev-Zel'dovich effect such as
the {\em Atacama Cosmology Telescope} (ACT, \citealt{2007ApOpt..46.3444F}), the
{\em South Pole Telescope} (SPT, \citealt{2011PASP..123..568C}), and {\em
  Planck} to reassess critically the impact of structure formation and physics
on $f_\rmn{gas}$ measurements, in particular in the outskirts of galaxy
clusters. Using a large suite of cosmological simulations, we first quantify the
bias of $f_\rmn{gas}$ measurements due to non-equilibrium processes in the
cluster outskirts while allowing for cluster sample variance. Then, we provide
an extensive study of $f_\rmn{gas}$ in our various physical models to obtain a
solid theoretical foundation for cluster cosmology using gas mass fractions.  In
Section~\ref{sec:sims}, we introduce our simulations and modeled physics. In
Section~\ref{sec:biases}, we study how typical assumptions in measuring gas
fractions, e.g., adopting hydrostatic mass estimates and neglecting density
clumping, bias the measured values. In Section~\ref{sec:fgas}, we study how
$f_\rmn{gas}$ varies with radius, cluster mass, redshift, and simulated physics,
and compare our simulation values to observational data. We conclude in
Section~\ref{sec:conclusions}.

\section{Cosmological simulations and cluster data set}
\label{sec:sims}

We use smoothed particle hydrodynamic (SPH) simulations of large-scale periodic
boxes which provide us with large cluster samples to accurately characterize ICM
properties over large ranges of cluster masses and redshifts. We use a modified
version of the {\sc GADGET-2} \citep{2005MNRAS.364.1105S} code. Our sequence of
periodic boxes have lengths of $165$ and $330\,h^{-1}\,\rmn{Mpc}$ filled with
$N_\rmn{DM} = N_{\mathrm{gas}} = 256^3$ and $512^3$ dark matter (DM) and gas
particles, respectively. This choice maintains the same initial gas particle mass
$m_{\mathrm{gas}} = 3.2\times 10^9\, h^{-1}\,\mathrm{M}_{\odot}$, DM particle
mass $m_{\mathrm{DM}} = 1.54\times 10^{10}\, h^{-1}\,\mathrm{M}_{\odot}$, and a
minimum gravitational smoothing length $\varepsilon_s=20\, h^{-1}\,$kpc; our SPH
densities were computed with 32 neighbours. For our standard calculations, we
adopt a tilted $\Lambda$CDM cosmology, with total matter density (in units of
the critical) $\Omega_m$= $\Omega_{\mathrm{DM}}$ + $\Omega_b$ = 0.25, baryon
density $\Omega_b$ = 0.043, cosmological constant $\Omega_{\Lambda}$ = 0.75,
Hubble parameter $h$ = 0.72 in units of $100 \mbox{ km s}^{-1} \mbox{
  Mpc}^{-1}$, spectral index of the primordial power-spectrum $n_s$ = 0.96 and
$\sigma_8$ = 0.8.

We show results from simulations employing three different variants of the the
simulated physics: the first only accounts for gravitational {\em shock
  heating}; the second additionally accounts for radiative cooling, star
formation, supernova feedback, galactic winds, and cosmic rays physics and is
referred to as {\em radiative cooling} model; and the last accounts for
radiative physics (radiative cooling, star formation, supernova feedback) and
thermal feedback by AGNs and is referred to as {\em AGN feedback}
model. Radiative cooling and heating were computed assuming an optically thin
gas of a pure hydrogen and helium primordial composition in a time-dependent,
spatially uniform ultraviolet background. Star formation and supernovae feedback
were modelled using the hybrid multiphase model for the interstellar medium of
\citet{2003MNRAS.339..289S}. The CR population is modelled as a relativistic
population of protons described by an isotropic power-law distribution function
in momentum space with a spectral index of $\alpha=2.3$
\citep{2006MNRAS.367..113P,2007A&A...473...41E,2008A&A...481...33J}. With those
parameters, the CR pressure modifies the SZ effect at most at the percent level
and causes a reduction of the resulting integrated Compton-$y$ parameter
\citep{2007MNRAS.378..385P}. The AGN feedback prescription included in the
simulations \citep[for more details see][]{2010ApJ...725...91B} allows for lower
resolution and hence can be applied to large-scale structure simulations. It
couples the black hole accretion rate to the global star formation rate (SFR) of
the cluster, as suggested by \citet{2005ApJ...630..167T}. The thermal energy is
injected into the ICM such that it is proportional to the star formation within
a given central spherical region. Our AGN feedback model produces identical
results for $f_\rmn{gas}$ profiles to a more elaborate model
\citep{2007MNRAS.380..877S, Sijacki+2008} when run at comparable high resolution
\citep{2010ApJ...725...91B} and -- by extension -- is also able to reconcile the
X-ray luminosity-temperature scaling relation \citep{2008ApJ...687L..53P}.

We use a two-step algorithm to compute the virial mass of a cluster. First, we
find all clusters in a given snapshot using a friends-of-friends (FOF)
algorithm. Then, using a spherical overdensity method with the FOF values as
starting estimates, we iteratively calculate the center of mass, the virial
radius, $R_{\Delta}$, and mass, $M_{\Delta}$, contained within $R_{\Delta}$, and
compute the radially averaged profiles of a given quantity with radii scaled by
$R_{\Delta}$. We define $R_{\Delta}$ as the radius at which the mean interior
density equals $\Delta$ times the {\em critical density}, e.g., for $\Delta
=200$ or 500. All cluster profiles are computed as functions of $r/R_{200}$ and
are converted to other scaled radii (i.e., in units of $R_{500}$ or $R_{1000}$)
using the NFW density profile \citep{1997ApJ...490..493N}, the concentration
mass relation from \citet{2008MNRAS.390L..64D}, and the cosmological parameters
listed above. Our cluster sample (for each simulation type) consists of
$\simeq1300$ clusters with $M_{200} > 7\times 10^{13}\,\rmn{M}_\odot$ and
$\simeq800$ clusters with $M_{200} > 1\times 10^{14}\,\rmn{M}_\odot$ at $z=0$.


\section{Biases of gas mass fraction measurements}
\label{sec:biases}

We are interested in the enclosed gas mass fractions that we define as the ratio
of gas mass and total mass inside a given radius, i.e., $f_\rmn{gas}(<r) =
M_\rmn{gas}(<r) / M_\rmn{tot}(<r)$. All our values of $f_\rmn{gas}$ and the
stellar mass fraction, $f_\rmn{star}$, are normalized by the cosmic baryon
fraction $f_c = \Omega_b/\Omega_m$ that should be substituted by the current
best fit value from cosmological precision experiments and was assumed to be
$f_c =0.172$ in our simulations, which is on the high side in comparison to the
currently favored value of 0.155 \citep{2013arXiv1303.5076P}. Here, we study how
typical assumptions in measuring gas fractions, e.g., adopting hydrostatic mass
estimates and neglecting density clumping, bias the measured values. We
additionally address how the sample variance of clusters and the
three-dimensional (3D) angular orientation variance within clusters impacts
inferred values for $f_\rmn{gas}$. We also address how line-of-side projection
effects this angular orientation bias of the gas fraction, i.e., how an
$f_\rmn{gas}$ measurement from a mosaic X-ray observation along a radial arm
varies with angular orientation depending on whether that arm intersects a
(projected) cosmic filament or not.

\begin{figure*}
  \begin{minipage}[t]{0.5\hsize}
    \centering{\small All clusters, $z=0$:}
  \end{minipage}
  \begin{minipage}[t]{0.5\hsize}
    \centering{\small Relaxed clusters, $z=0$:}
  \end{minipage}
  \resizebox{0.5\hsize}{!}{\includegraphics{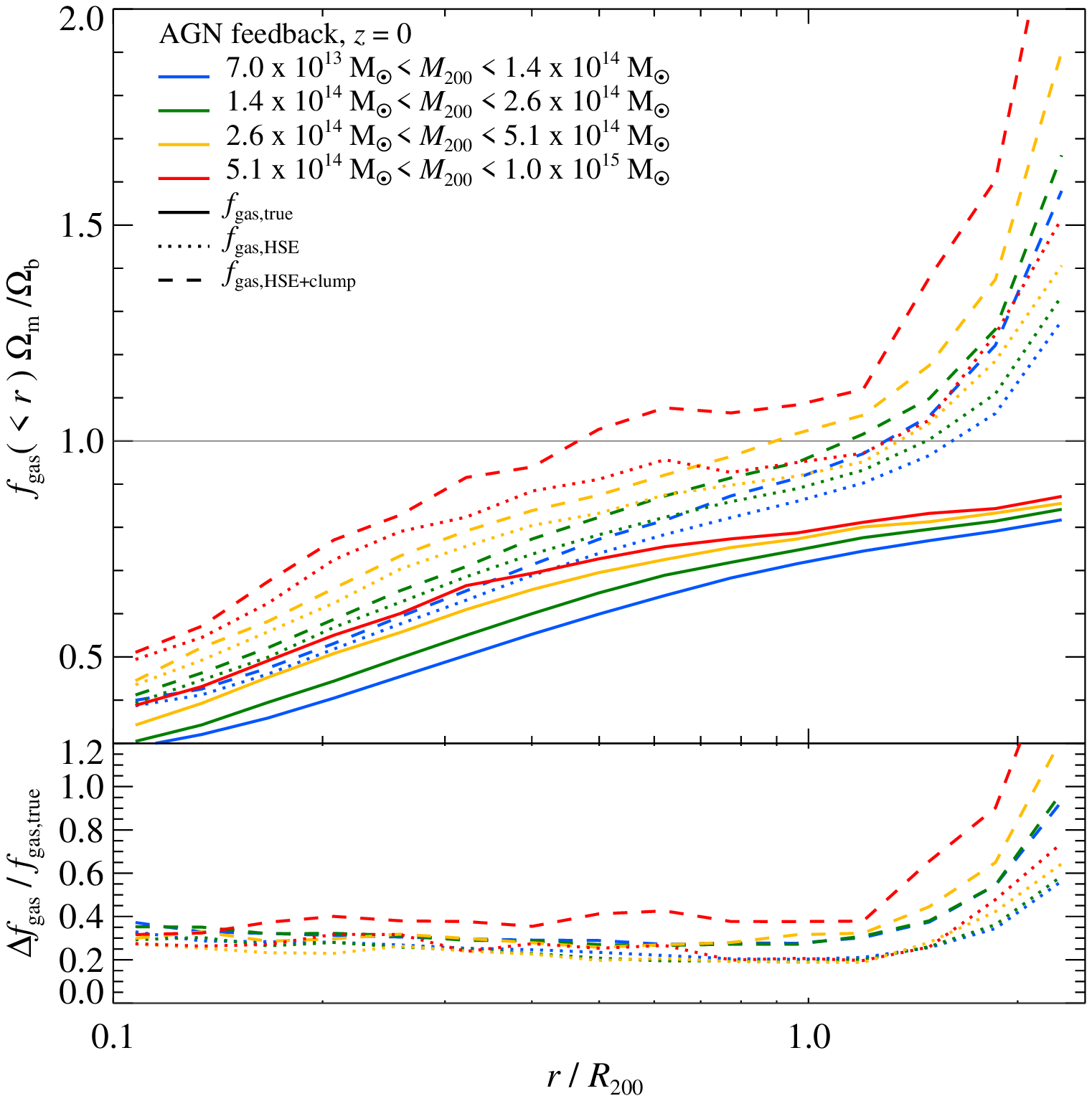}}%
  \resizebox{0.5\hsize}{!}{\includegraphics{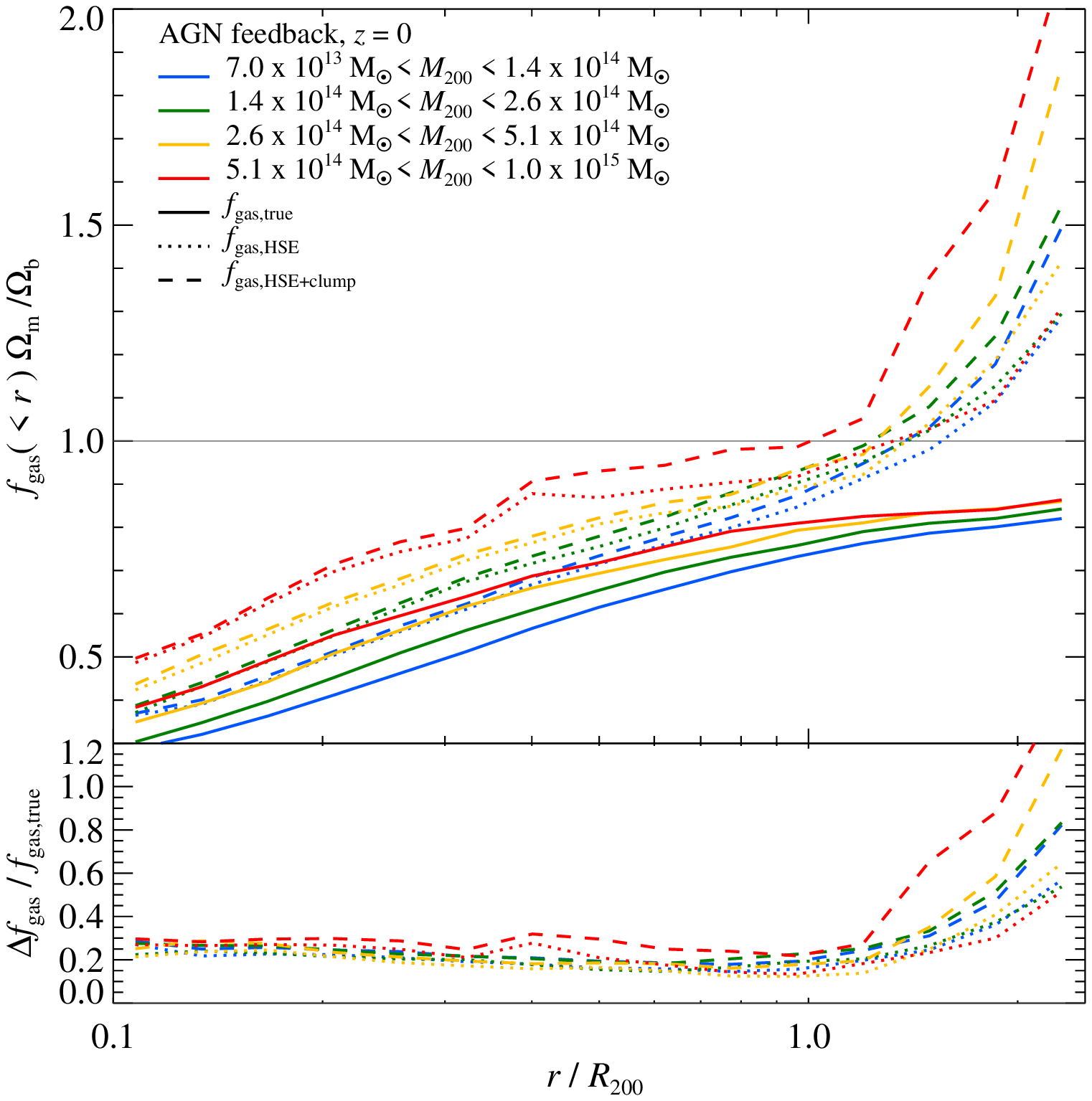}}\\
  \caption{The assumption of hydrostatic equilibrium and of a smoothly
    distributed ICM introduces a bias in the measurement of $f_{\rmn{gas}}$. We
    show the median radial profiles of the gas mass fraction, $f_{\rmn{gas}}
    (<r) \equiv M_{\rmn{gas}}(<r) / M_{\rmn{tot}} (<r)$, normalized by the
    universal baryon fraction for different assumptions in calculating
    $M_{\rmn{gas}}$ and $M_{\rmn{tot}}$ as a function of cluster mass.  We
    compare the true gas mass fraction, $f_{\rmn{gas,true}}$ (solid), a biased
    X-ray measurement of $f_{\rmn{gas}}$ due to the assumption of HSE, i.e.,
    without accounting for kinetic pressure support, $f_{\rmn{gas,HSE}}$
    (dotted), as well as a biased measurement of $f_{\rmn{gas}}$ due to the
    assumption of HSE and additionally neglecting density clumping,
    $f_{\rmn{gas,HSE+clump}}$ (dashed). The profiles in the left panel reflect
    the median of the distribution of all clusters and the right panel shows the
    median of a sub-sample of the third most relaxed clusters (with the lowest
    ratio of total kinetic-to-thermal energy, $\KU$). The bottom panels show the
    fractional difference from $f_{\rmn{gas,true}}$. There is a bias in
    $f_{\rmn{gas}}$ introduced by the assumption of hydrostatic equilibrium even
    for relaxed clusters. Here, the pressure derivative was smoothed with
    adjacent radial bins.}
\label{fig:fgas}
\end{figure*}

\begin{figure}
  \centering{\small All clusters, $z=1$:}
  \resizebox{\hsize}{!}{\includegraphics{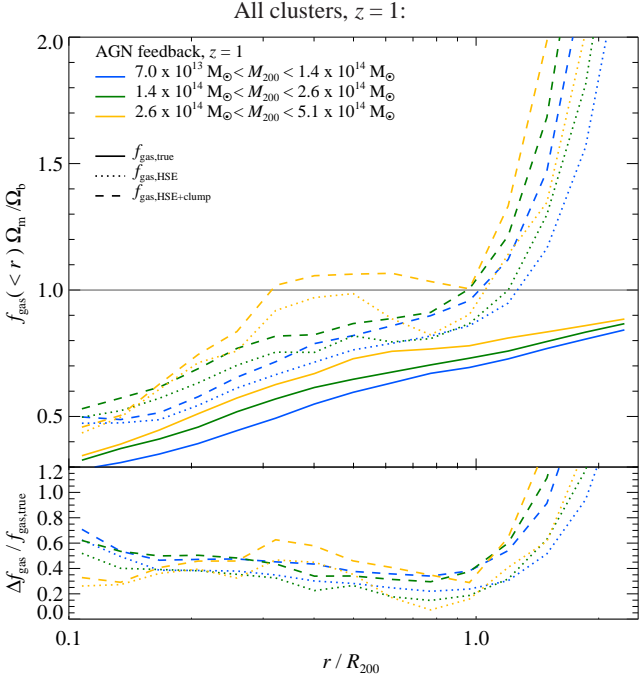}}\\
  \caption{Same as in Figure~\ref{fig:fgas}, but at $z=1$ for fixed cluster mass
    bins (at $z=1$ there is no cluster with $M_{200}>
    5\times10^{14}\rmn{M}_\odot$ in our simulations). While the intrinsic gas
    mass fraction shows no significant evolution with $z$, the assumption of
    hydrostatic equilibrium and of a smoothly distributed ICM introduces an
    increasing bias at $z=1$.}
\label{fig:fgas_z1}
\end{figure}

In Figure~\ref{fig:fgas}, we show the median enclosed gas mass fractions,
$f_\rmn{gas}(<r)$ as a function of radius for different cluster masses in our
AGN feedback model; on the left panel for our sample of all clusters, and on the
right panel for the 3rd most relaxed clusters. To define a {\em relaxed
  cluster}, we split the population of clusters into lower, middle, and upper
bands of the ratio of kinetic-to-thermal energy, $\KU$, within $R_{200}$, which
we use as a measure of dynamical state for the clusters (see Appendix
\ref{sec:KU_w}). We define the internal kinetic energy, $K$, and thermal energy,
$U$, of a cluster as
\begin{eqnarray}
\label{eq:KU}
K (<R_{200}) &\equiv& \sum_i \frac{3 m_{\rmn{gas},i}  P_{\mathrm{kin},i}}{2\rho_i}, \\
\label{eq:KU2}
U (<R_{200}) &\equiv& \sum_i \frac{3 m_{\rmn{gas},i}  P_{\mathrm{th},i}} {2\rho_i},
\end{eqnarray} 
where $m$ and $\rho$ are the gas mass and the SPH density, respectively for all
particles $i$ less than $R_{200}$, $P_{\mathrm{th}}=kT\rho/(\mu m_p)$ denotes
the thermal pressure (where $T$ and $\mu$ denote the temperature and mean
molecular weight, respectively), and the kinetic pressure is defined as
\begin{eqnarray}
P_{\mathrm{kin}} &=& \rho \, \delta \bvel^2 / 3, \\
\delta \bvel   &=& a\,\left(\bvel - \bar{\bvel}\right) +
a\,H(a)\left(\vecbf{x}-\bar{\vecbf{x}}\right) .
\end{eqnarray}
The code employs comoving peculiar velocities, which are translated into the
internal cluster velocities relative to the overall mean cluster velocity in the
Hubble flow by the relation given, where $H(a)$ is the Hubble function, $a$ is
the scale factor, $\bvel$ ($=\dd \vecbf{x}\,/\,\dd t$) is the peculiar velocity,
and $\vecbf{x}$ is the comoving position of each particle.  The
gas-particle-averaged cluster bulk flow within $R_{200}$ is $\bar{\bvel}$ and
the center of mass within $R_{200}$ is $\bar{\vecbf{x}}$.  Since the most
massive clusters are on average also the most disturbed systems, only two out of
25 clusters in our most massive cluster bin ($M_{200}>5.1\times
10^{14}\,\rmn{M}_\odot$) meet our criterion of being a relaxed cluster, i.e.,
are among the lowest third when selected by the ratio of kinetic-to-thermal
energy, $\KU$. Hence, in order to improve the statistics of our most massive bin
of relaxed clusters, we decided to also do a 3-split on $\KU$ in the largest
cluster mass bin.

As shown in Figure~\ref{fig:fgas}, $f_\rmn{gas}$ increases with radius and
approaches values within $2 R_{200}$ that range from 0.8 to 0.85 (in units of
$f_c$) for the mass interval from $M_{200}\simeq 10^{14}\,\rmn{M}_\odot$ to $5
\times 10^{14}\,\rmn{M}_\odot$. Smaller systems show smaller values of
$f_\rmn{gas}$ at every radius owing to the shallower potential wells in
comparison to bigger systems. In groups and small clusters, AGN feedback is able
to push the gas well beyond $R_{200}$ or halt its initial infall. In the process
of hierarchical structure formation, these gas-poor groups merge to form larger
clusters. Owing to their deeper potentials, those larger clusters provide a
greater gravitational pull on the gas that is sitting beyond $R_{200}$. As a
result, $f_\rmn{gas}$ increases at each radius in larger clusters. (We will show
later on in Section~\ref{sec:redshift} that the underlying assumption of a
constant $f_\rmn{gas}$ as a function of redshift at fixed cluster mass is
justified in our simulations.)

These simulated values of $f_\rmn{gas}/f_c<0.8$ for all cluster masses are
substantially smaller than Suzaku estimates of $f_\rmn{gas}/f_c\sim 1.4$ or even
1.6 within $R_{200}$ if one accounts for a stellar mass fraction of 12\%
\citep{2011Sci...331.1576S}. Two sources that could potentially bias the
measurement of gas fractions have been suggested, the assumption of hydrostatic
masses and neglecting density clumping due to infalling (sub-)structures
following the ongoing growth of clusters. Hence we will discuss each of these
effects in turn.

\subsection{Hydrostatic mass bias}

Even if clusters are in hydrostatic equilibrium (HSE), balancing the gravitational
force to the pressure gradient yields
\begin{equation}
\nabla P = \rho {\mathbfit g} \rightarrow  
-\rho GM(<r){\hat{\mathbfit{r}}}/ r^2 \ \mbox{for spherical symmetry}.
\label{eq:HSE}
\end{equation}
Assuming that all the pressure in Eq.~(\ref{eq:HSE}) is thermal ($P =
P_{\mathrm{th}}$) gives an incorrect estimate for the mass since the
non-thermal pressure, in particular the kinetic pressure has to be
included
\citep[e.g.,][]{1990ApJ...363..349E,2004MNRAS.351..237R,2006MNRAS.369.2013R,
  2009ApJ...705.1129L,2012NJPh...14e5018R,2012ApJ...751..121N}. For
clarity we define $M_{\rmn{HSE}}$ to be the mass derived using $P =
P_{\mathrm{th}}$ and $M_{\rmn{tot}} \equiv M_\rmn{DM} + M_\rmn{gas} +
M_\rmn{star}$. We define the bias owing to the assumption of HSE as
$M_{\rmn{HSE}}/M_{\rmn{tot}}$, which shows a weak increase in
radius. Note that this definition differs from other approaches in the
literature. Adopting $\tilde{M}_\rmn{tot} \propto \dd
(P_\rmn{th}+P_\rmn{kin}) /\dd r$ results in a radially decreasing
bias, presumably due to violation of spherical symmetry and
non-stationary, clumpy accretion that is not accounted for in the
derivation of $\tilde{M}_\rmn{tot}$ (see Appendix \ref{sec:lau}).

Comparing $M_{\rmn{HSE}}$ to the true mass inside a given radius,
$M_{\rmn{HSE}}$ on average underestimates $M_{\rmn{tot}}$ by $20-25$\% within
$R_{200}$ \citepalias{2012ApJ...758...74B} and biases $f_\rmn{gas}$ accordingly high
(dotted lines in Figure~\ref{fig:fgas}). This $M_{\rmn{HSE}}$ bias is almost
independent of cluster mass out to $R_{200}$ as can be seen by the fractional
difference of $f_\rmn{gas}$ (lower panels of Figure~\ref{fig:fgas}). However,
individual clusters can stray from this average deviation, since each cluster
has a unique dynamical state and formation history. This is suggested by a
scatter of $\sim5$\% between the $25^{\rmn{th}}$ and $75^{\rmn{th}}$ percentiles
of the complete distribution \citepalias{2012ApJ...758...74B}. For approaches that
assume a constant $M_{\rmn{HSE}}$ bias, the fractional difference of
$f_\rmn{gas,HSE}$ would be accordingly shifted downwards (by the same factor in
the lower panels of Figure~\ref{fig:fgas}).

When we select a sample of the third most relaxed clusters, the correction
factor will necessarily be smaller. We find that the bias on $f_\rmn{gas}$ for
this sample is reduced to $15-20$\% within $R_{200}$. In a small sample of
zoomed-cluster simulations that has been selected against major mergers the
hydrostatic mass correction was found to be of the order $M_{\rmn{HSE}}\sim
10-15$\% \citep{2006ApJ...650..128K}. However, in both cases, we see that
outside $R_{200}$, the hydrostatic bias rises steeply as
$P_{\rmn{kin}}/P_{\rmn{th}}$ approaches large values and eventually reaches
equipartition around $2R_{200}$ (at $z=0$) with the exact equipartition radius
depending on cluster mass.

\subsection{Clumping bias}

The emissivity of thermal bremsstrahlung scales with the square of the gas
density. Hence the X-ray-inferred gas density profile is biased high by the
presence of a clumped medium \citep{2011ApJ...731L..10N, 2013MNRAS.432.3030R}
and attains a bias of the square root of the density clumping factor, that we
define by
\begin{equation}
\tilde{\mathcal{C}}_{2,\rho} = \frac{\bra \rho^2 \ket}{\bra \rho \ket^2}.
\label{eq:C2}
\end{equation}
Averages are taken to be volume averages. We also define the SPH volume bias,
\begin{equation}
\tilde{\mathcal{C}}_{0} = \bra 1 \ket = \frac{1}{V}\,\sum_i \frac{M_i}{\rho_i}
= \frac{1}{V}\,\sum_i V_i,
\label{eq:C0}
\end{equation}
that we expect to be identical to unity everywhere.  While this is the case for
small radii, the SPH volume bias deviates progressively for larger radii. It
reaches a negative bias of 30\% at $3R_{200}$ ($z=0$), almost independent on the
simulated physics and cluster masses, but not of redshift since it doubles at
$z=1$ when radii are scaled to $R_{200}$ \citepalias{BBPS4}. To obtain unbiased
estimates, we decided to redefine the clumping factor, yielding
\begin{equation}
\mathcal{C}_{2,\rho} = \tilde{\mathcal{C}}_{2,\rho}\,\tilde{\mathcal{C}}_{0},
\label{eq:C_2,corr}
\end{equation}
where the multiplication with the volume bias factor, $\tilde{\mathcal{C}}_{0}$,
has been chosen to cancel the SPH volume in the weighting. Inside $R_{200}$, the
profile of $\sqrt{\mathcal{C}_{2,\rho}}$ increases moderately with radius but
remains $<1.2$ for $r<R_{200}$ for all clusters.  Outside $R_{200}$, it starts
to rise sharply and reaches values of 10 at $r\simeq3 R_{200}$ and $4 R_{200}$
for large clusters and groups, respectively, almost independent of the simulated
physics \citepalias{BBPS4}.  At each radius, there is a clear trend of an
increasing clumping factor for larger clusters, which have on average a higher
mass accretion rate, i.e., they are assembling at the current epoch and hence
are dynamically younger.

Cold, dense gas clumps that are ram-pressure stripped from galaxies and find
themselves in pressure equilibrium with the ambient hot ICM add an artificial
bias to the density clumping factor. This bias needs to be accounted for when
comparing to X-ray observations by employing, e.g., a temperature cut. Hence, in
our calculations of the density clumping factor, we only consider SPH particles
with $T>10^6\,\rmn{K}$ that can be observed by their X-ray emission.

To estimate the clumping bias of the enclosed gas mass fractions, we compute a
clumping-biased density profile for each cluster by means of
$\rho_{\rmn{clump}}(r) = \sqrt{\mathcal{C}_{2,\rho}(r)}\,\rho_{\rmn{true}}(r)$,
and integrate this to obtain a clumping-biased gas mass profile,
\begin{equation}
M_{\rmn{gas,clump}}(<r) = 4 \pi\int_0^r  \rho_{\rmn{clump}}(r')r'^2dr'.
\label{eq:M_clump}
\end{equation}

Here, we only asses the effect of clumping bias on $M_\rmn{gas}$. In principle,
density clumping should also impact the determination of $M_\rmn{tot}$. However,
there are different subtle effects that may partially compensate. Those include
larger Poisson scatter in density profiles implied by clumping, which then have
a noisier derivative, calling for some smoothing procedure which may bias the
inferred masses. Also, the X-ray emissivity is enhanced by the presence of
self-bound (dense) structures which increases the logarithmic derivative of the
density and hence $M_\rmn{tot}$. Conversely, these dense substructures have a
lower temperature. Fitting a single temperature spectrum to the data will then
bias the inferred temperature and hence $M_\rmn{tot}$ low. Simulated X-ray
observations are needed to carefully address these points that shall be subject
to future work.

Clumping bias increases inferred values for $f_{\rmn{gas}}$ -- similar to the
$M_{\rmn{HSE}}$ bias. In Figure~\ref{fig:fgas}, we show the effect of both
biases that we denote by $f_{\rmn{gas,HSE+clump}}$ (dashed lines). As a result
of both biases, $f_{\rmn{gas}}$ realizes its universal value at considerably
smaller radii in comparison to the true $f_{\rmn{gas}}$. In our sample of
relaxed clusters, the universal value is reached outside $R_{200}$ whereas in
our all-cluster sample, these high values are already obtained at $0.5 R_{200}$
for big clusters with $M_{200}> 5\times 10^{14}\,\rmn{M}_\odot$. Interestingly,
the biased values for $f_{\rmn{gas}}$ increase sharply outside the virial
radius, mainly driven by the clumping bias for large systems and with a
substantial contribution of $M_\rmn{HSE}$ bias for smaller
clusters. $f_{\rmn{gas}}$ is biased high by a factor of two at $2R_{200}$ for
large clusters and at $3R_{200}$ for groups (see Figure~\ref{fig:fgas}).

What is the relative contribution of these measurement biases as a function of
radius and cluster mass? In our all-cluster sample, clumping bias has a small
effect at the cluster center but is continuously increasing towards
$R_{200}$. At $R_{200}$, it approximately contributes half as much as the
$M_{\rmn{HSE}}$ bias to the overall bias with an absolute bias of $10-20$\%,
depending on cluster mass. In our relaxed cluster sample, the clumping bias
remains even more subdominant in comparison to the $M_{\rmn{HSE}}$
bias. However, in both cluster samples, the clumping bias increases dramatically
beyond $R_{200}$ with an increasing bias with cluster mass.

Clusters form as a result of hierarchical merging of smaller systems. Hence the
kinetic pressure support \citepalias{2012ApJ...758...74B} and the density
clumping (\citealt{2011ApJ...731L..10N}, \citetalias{BBPS4}) increase with $z$
at fixed mass and so do the associated measurement biases of
$f_{\rmn{gas}}$. This can be appreciated in Figure~\ref{fig:fgas_z1}, which
shows the radial profile of $f_{\rmn{gas}}$ at $z=1$.  While the intrinsic gas
mass fraction shows no significant evolution with $z$, the assumption of
hydrostatic equilibrium and of neglecting the density clumping introduces an
increasing bias at $z=1$ in comparison to the present epoch.

\begin{figure*}
  \begin{minipage}[t]{0.5\hsize}
    \centering{\small All clusters:}
  \end{minipage}
  \begin{minipage}[t]{0.5\hsize}
    \centering{\small Relaxed clusters:}
  \end{minipage}
  \resizebox{0.5\hsize}{!}{\includegraphics{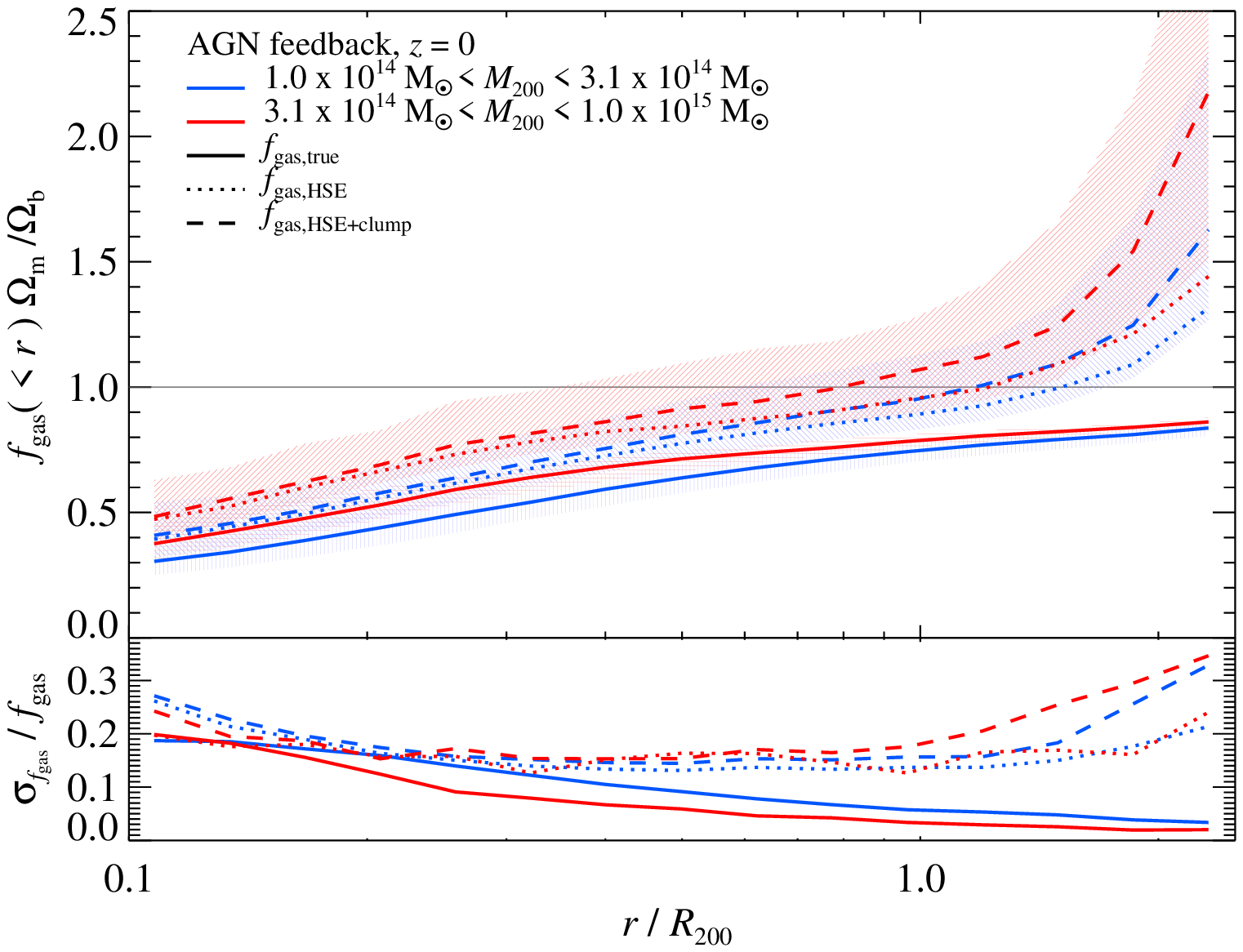}}%
  \resizebox{0.5\hsize}{!}{\includegraphics{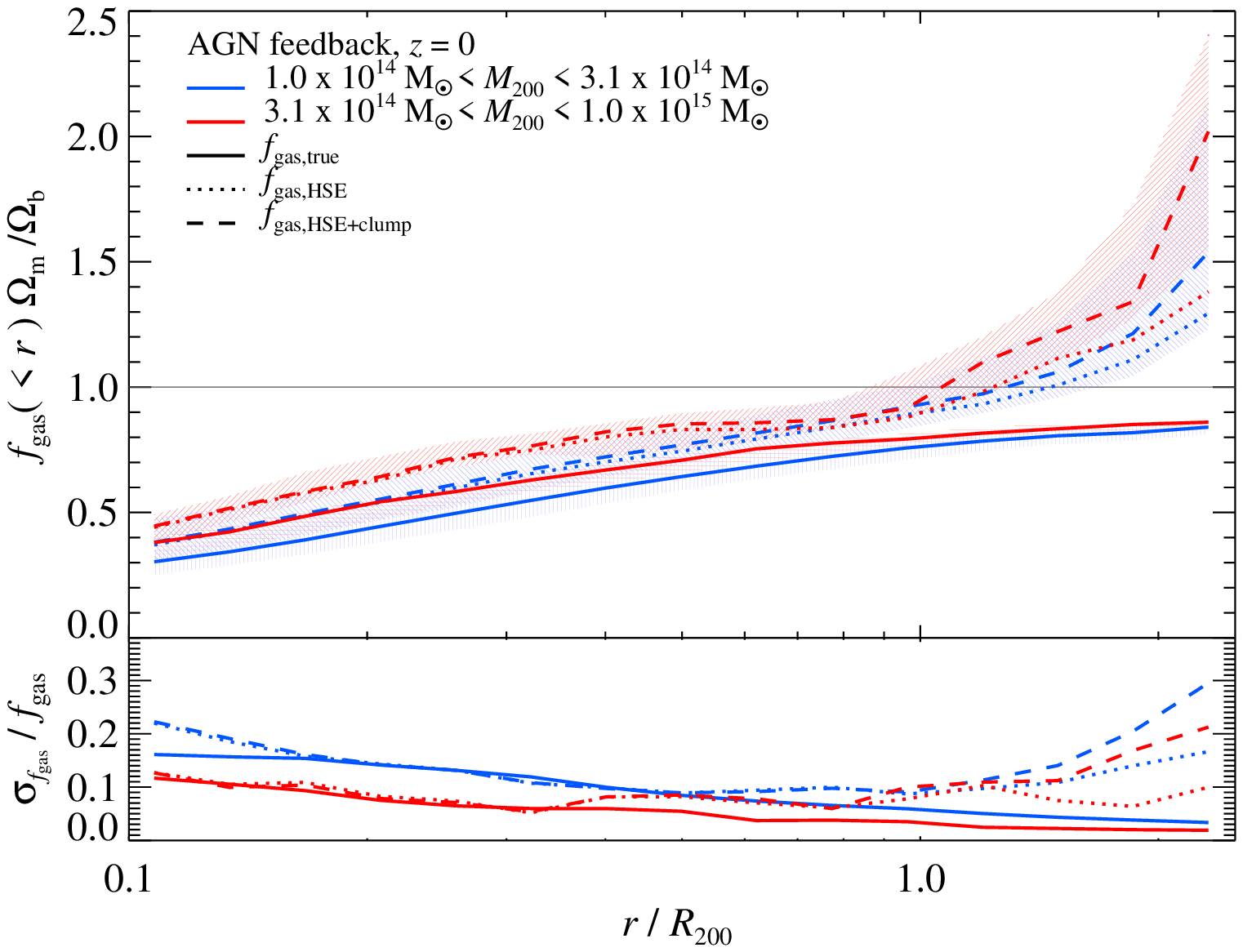}}\\
  \caption{Cluster sample variance of $f_{\rmn{gas}}$. Upper panels: we show the
    gas mass fraction $f_{\rmn{gas}}$ for different mass intervals (lines
    indicate median values, shaded regions are bounded by percentiles such that
    they contain $68.3\%$ of the probability, centered upon the median).  Lower
    panels: we show the relative cluster-to-cluster scatter of the gas mass
    fraction, $\sigma_{f_{\rmn{gas}}}$, that we define by the difference of the
    upper and lower ($1-\sigma$) percentiles normalized by twice the median of
    $f_{\rmn{gas}}$. Here, we compare radial profiles of the true gas mass
    fraction, $f_{\rmn{gas,true}}$ (solid), a biased X-ray measurement of
    $f_{\rmn{gas}}$ due to the assumption of HSE, i.e., without accounting for
    kinetic pressure support, $f_{\rmn{gas,HSE}}$ (dotted), as well as a biased
    measurement of $f_{\rmn{gas}}$ due to the assumption of HSE and additionally
    neglecting density clumping, $f_{\rmn{gas,HSE+clump}}$ (dashed). The
    profiles in the left panel include all clusters and the right panel shows
    the median of a sub-sample of the third most relaxed clusters (with the
    lowest ratio of total kinetic-to-thermal energy, $\KU$). While the scatter
    of $f_{\rmn{gas,true}}$ decreases for larger radii, the scatter of the
    biased $f_{\rmn{gas,HSE+clump}}$ remains fairly constant at a level of
    $\sigma_{f_\rmn{gas}}/f_\rmn{gas} \simeq 0.1 - 0.2$ (depending on the
    dynamical state and cluster mass) within $R_{200}$ and even starts to
    increase for larger radii due to the strong clumping term. (We do not show
    the shaded regions of $f_{\rmn{gas,HSE}}$ for visual clarity, but show this
    information in the lower panels.)}
\label{fig:fgas_scatter}
\end{figure*}

\begin{figure*}
\centering{
  \resizebox{0.325\hsize}{!}{\includegraphics{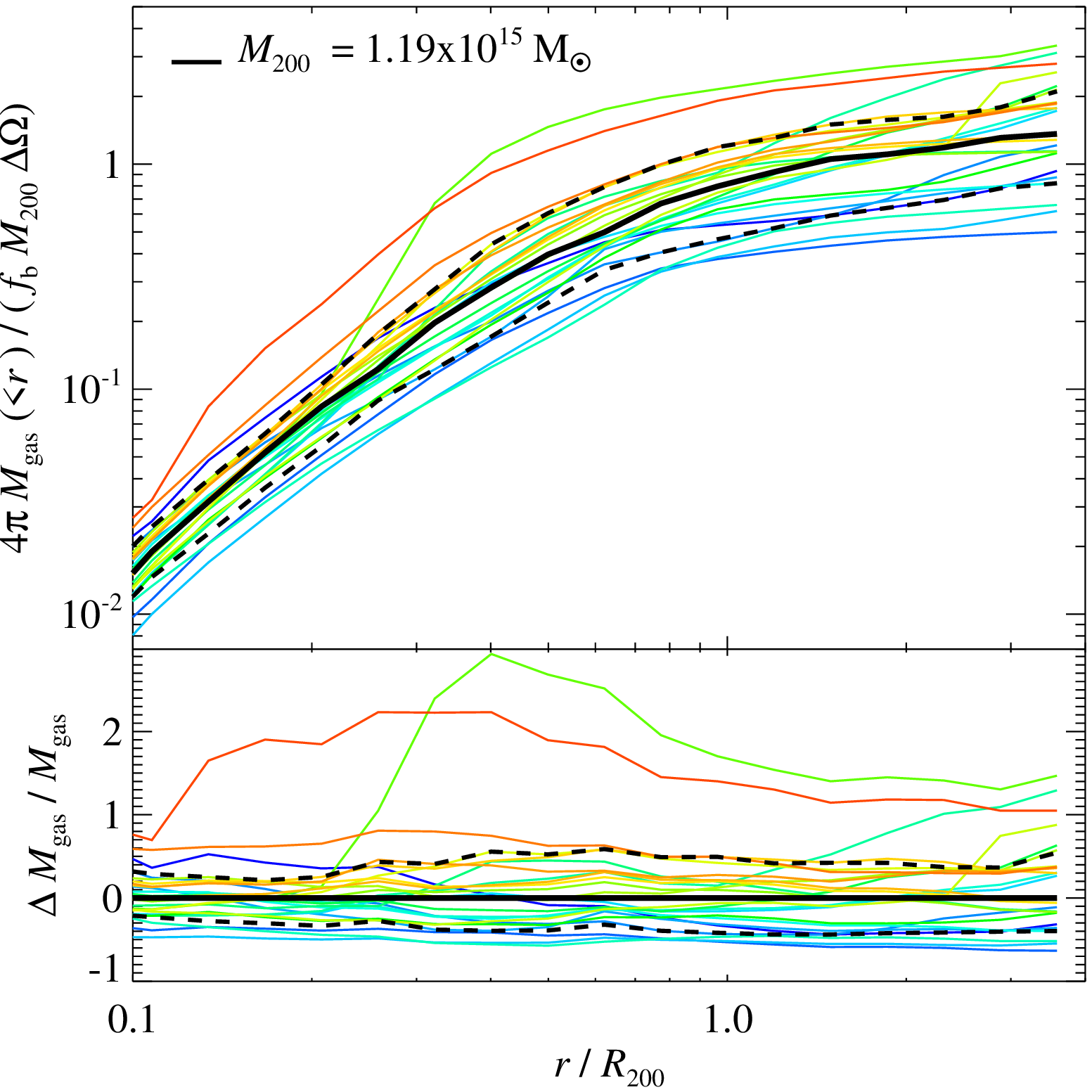}}\hfill%
  \resizebox{0.325\hsize}{!}{\includegraphics{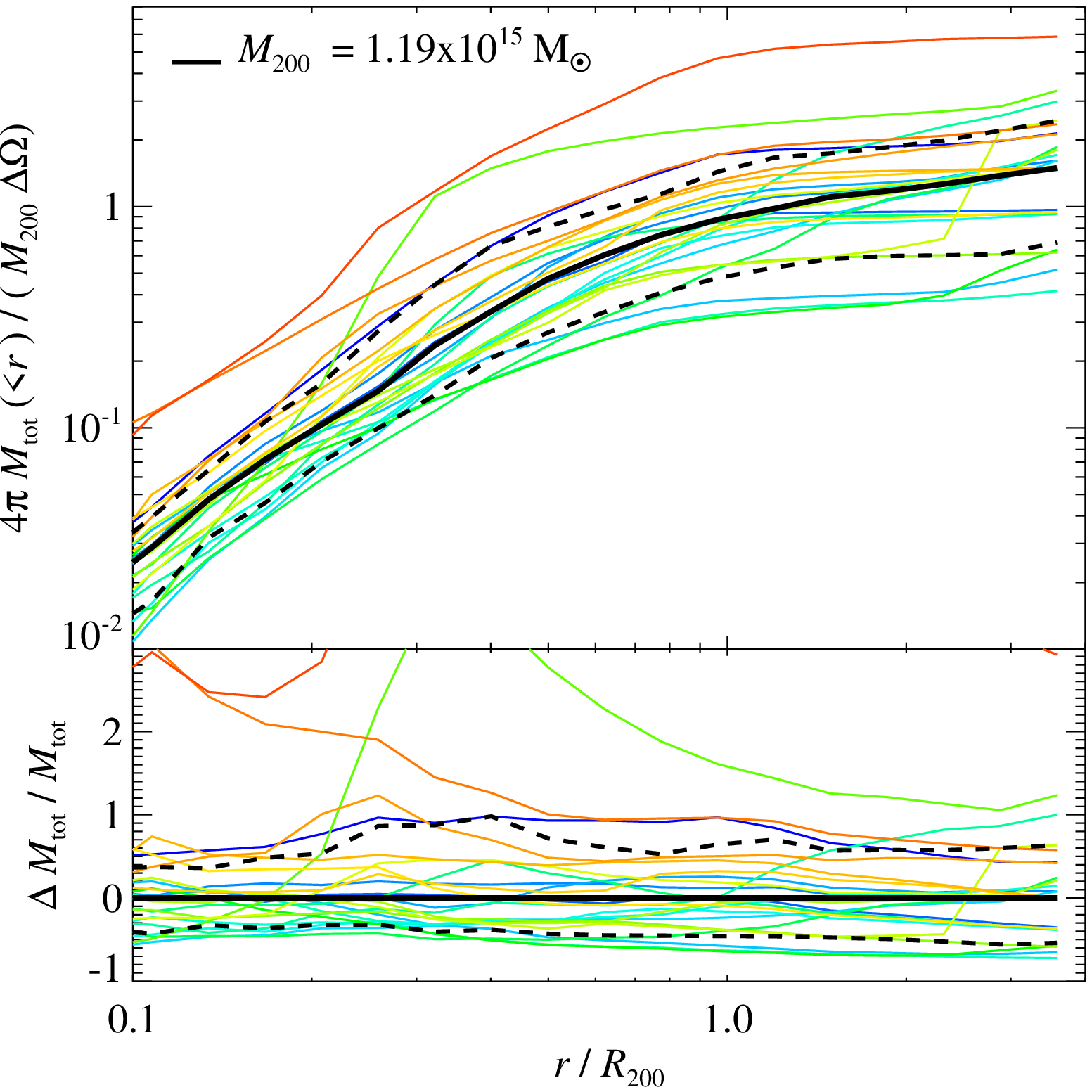}}\hfill%
  \resizebox{0.325\hsize}{!}{\includegraphics{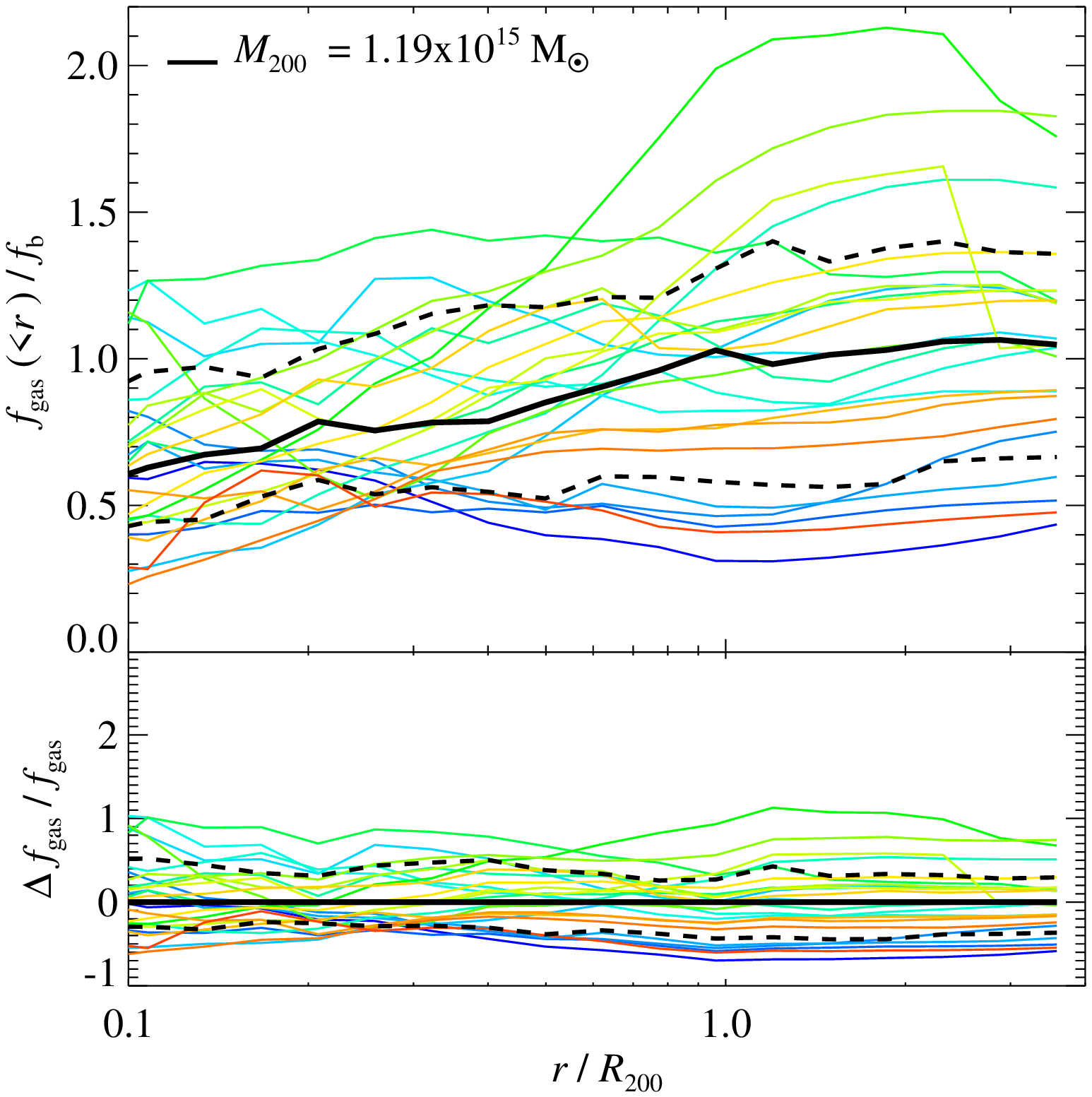}}}\\
\caption{Upper panels: we show the profiles of $M_{\rmn{gas}}$ (left),
  $M_{\rmn{tot}}$ (middle), and $f_{\rmn{gas}}$ (right) in 48 angular HEALPix
  cones (which subtend an equal area pixelization on the sphere), centered on
  one cluster of mass $M_{200} \approx 1.2\times 10^{15}\,\rmn{M}_\odot$. (Note
  that we only show every other profile for visual purposes.)  We overplot the
  median (solid black) as well as the upper and lower ($1-\sigma$) percentiles
  (dotted black). Lower panels: we show the variations of $M_{\rmn{gas}}$,
  $M_{\rmn{tot}}$, and $f_{\rmn{gas}}$ in the different cones relative to their
  corresponding median. Note the different scalings at the $y$-axis (logarithmic
  for the mass profiles, linear scale for $f_{\rmn{gas}}$). There is
  considerable scatter as well as the presence of a few enormous outliers in the
  $M_{\rmn{gas}}$ and $M_{\rmn{tot}}$ cones -- depending on the angular
  direction -- which is reduced for $f_{\rmn{gas}}$. This implies spatial
  correlations of the DM and gas distribution through self-bound substructures
  that are still holding on to (some) of their gas.}
\label{fig:ang large}
\end{figure*}

\begin{figure*}
\centering{
  \resizebox{0.325\hsize}{!}{\includegraphics{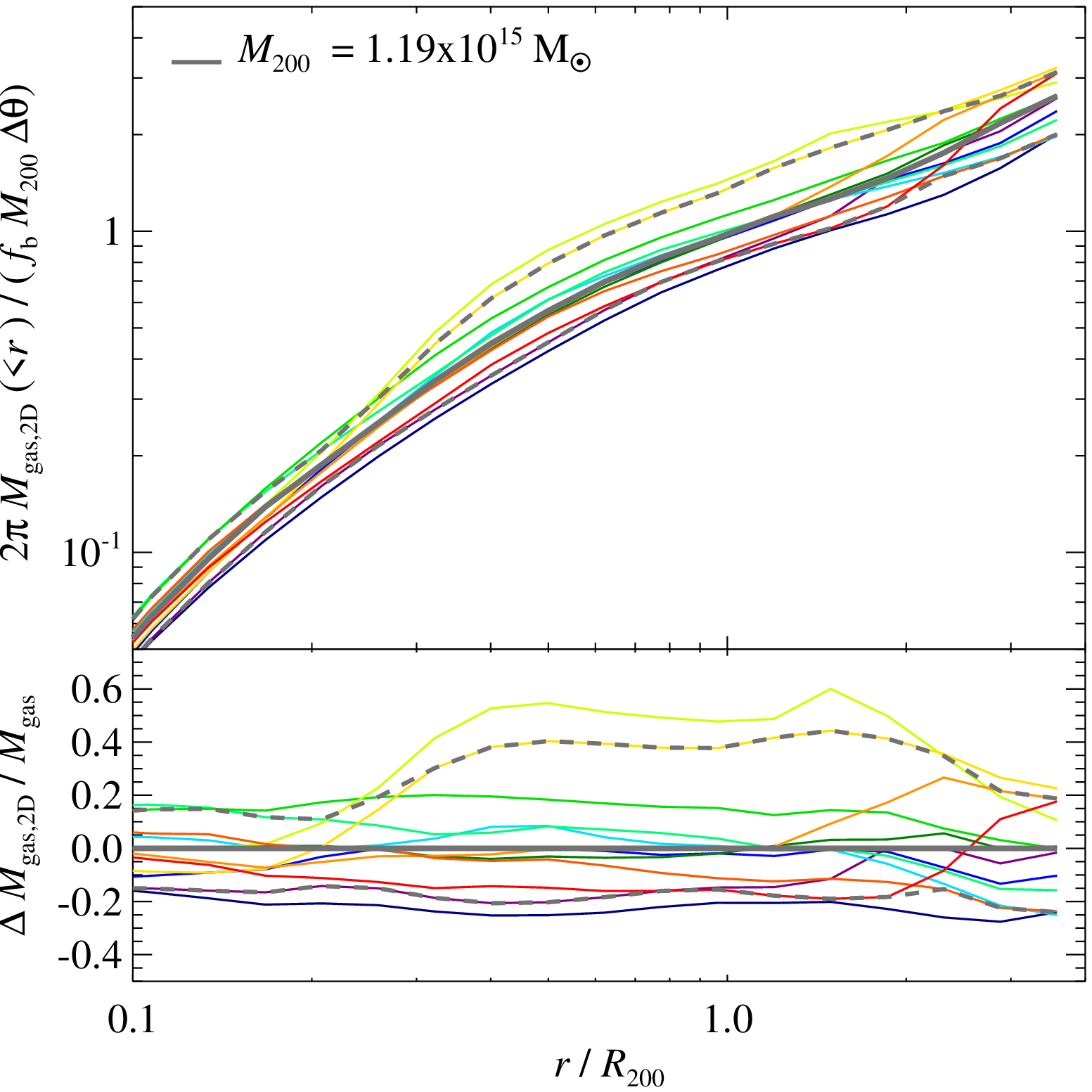}}\hfill%
  \resizebox{0.325\hsize}{!}{\includegraphics{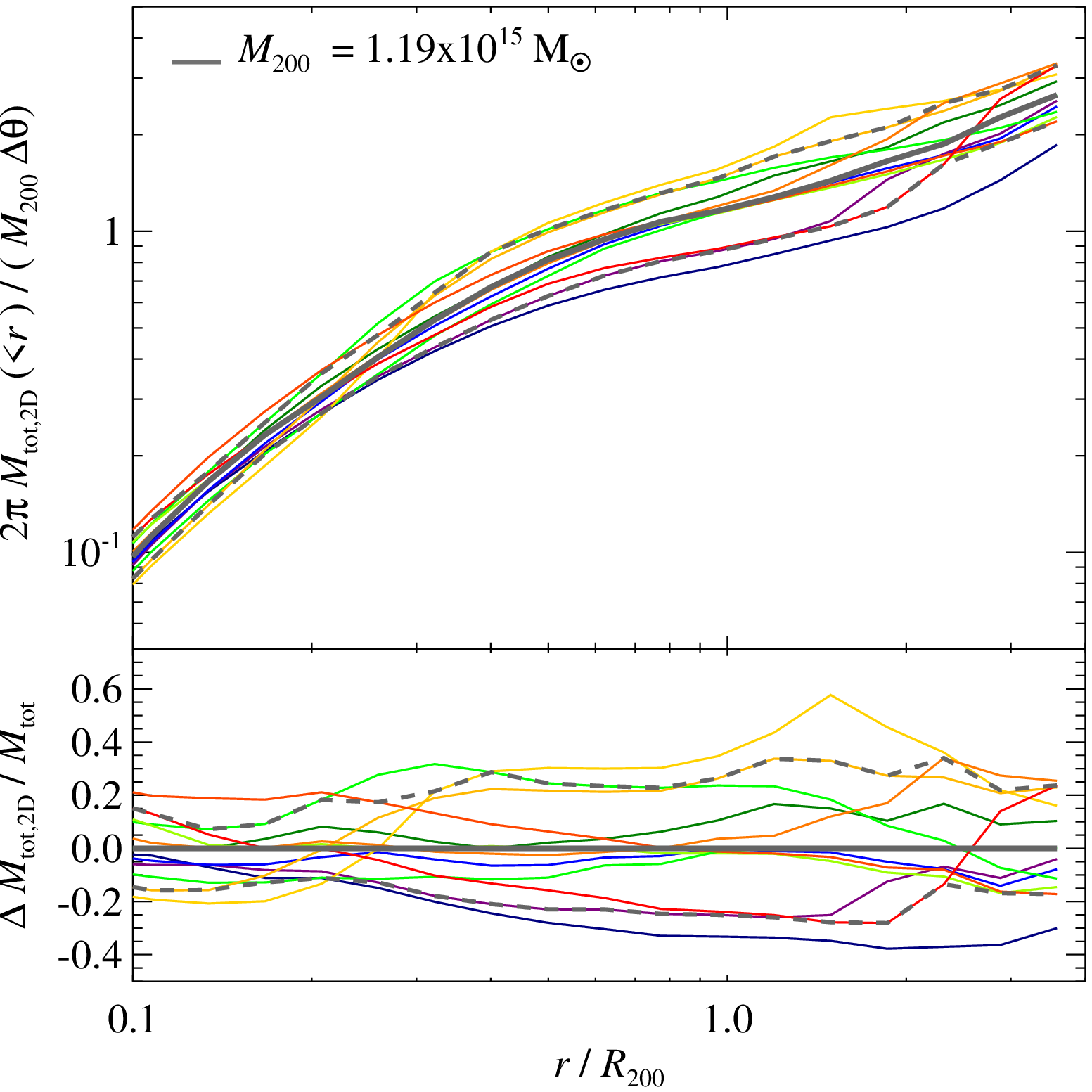}}\hfill%
  \resizebox{0.325\hsize}{!}{\includegraphics{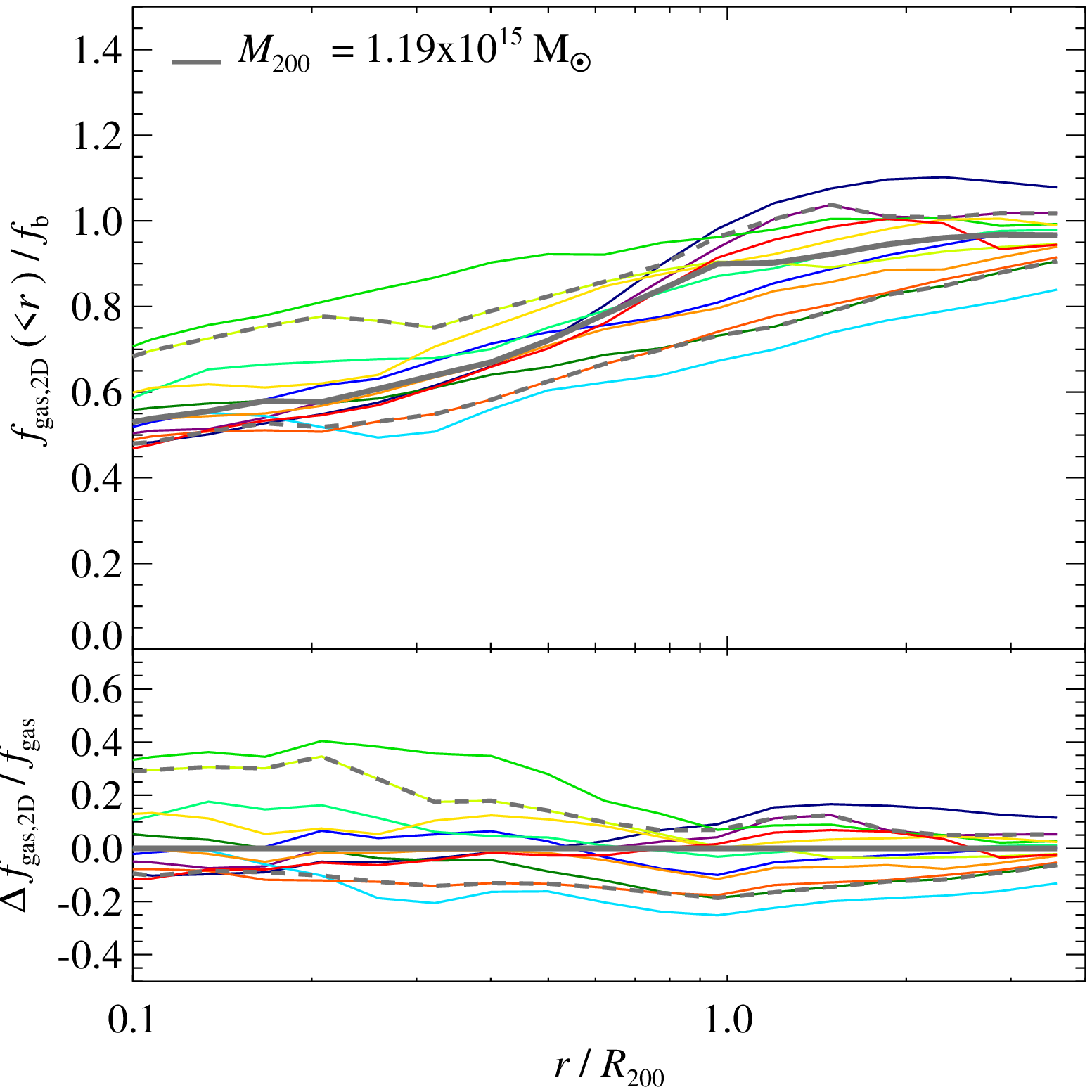}}}\\
\caption{Upper panels: we show the projected (cylindrical) profiles of
  $M_{\rmn{gas}}$ (left), $M_{\rmn{tot}}$ (middle), and $f_{\rmn{gas}}$ (right)
  in 12 equally large angular sectors, centered on one cluster of mass $M_{200}
  \approx 1.2\times 10^{15}\,\rmn{M}_\odot$. We overplot the median (solid
  black) as well as the upper and lower ($1-\sigma$) percentiles (dashed
  black). Lower panels: we show the variations of $M_{\rmn{gas}}$,
  $M_{\rmn{tot}}$, and $f_{\rmn{gas}}$ in the different sectors relative to
  their corresponding median. Note the different scalings at the $y$-axis
  (logarithmic for the mass profiles, linear scale for $f_{\rmn{gas}}$). In
  comparison to the 3D cones, there is less scatter and less dramatic outliers
  in the projected sector distribution of $M_{\rmn{gas}}$, $M_{\rmn{tot}}$, and
  $f_{\rmn{gas}}$.}
\label{fig:ang large2D}
\end{figure*}

\begin{figure*}
  \begin{minipage}[t]{0.325\hsize}
    \centering{\small Relaxed clusters, 3D:}
  \end{minipage}
  \begin{minipage}[t]{0.325\hsize}
    \centering{\small All clusters, 3D:}
  \end{minipage}
  \begin{minipage}[t]{0.325\hsize}
    \centering{\small All clusters, 2D:}
  \end{minipage}
  \centering{
    \resizebox{0.325\hsize}{!}{\includegraphics{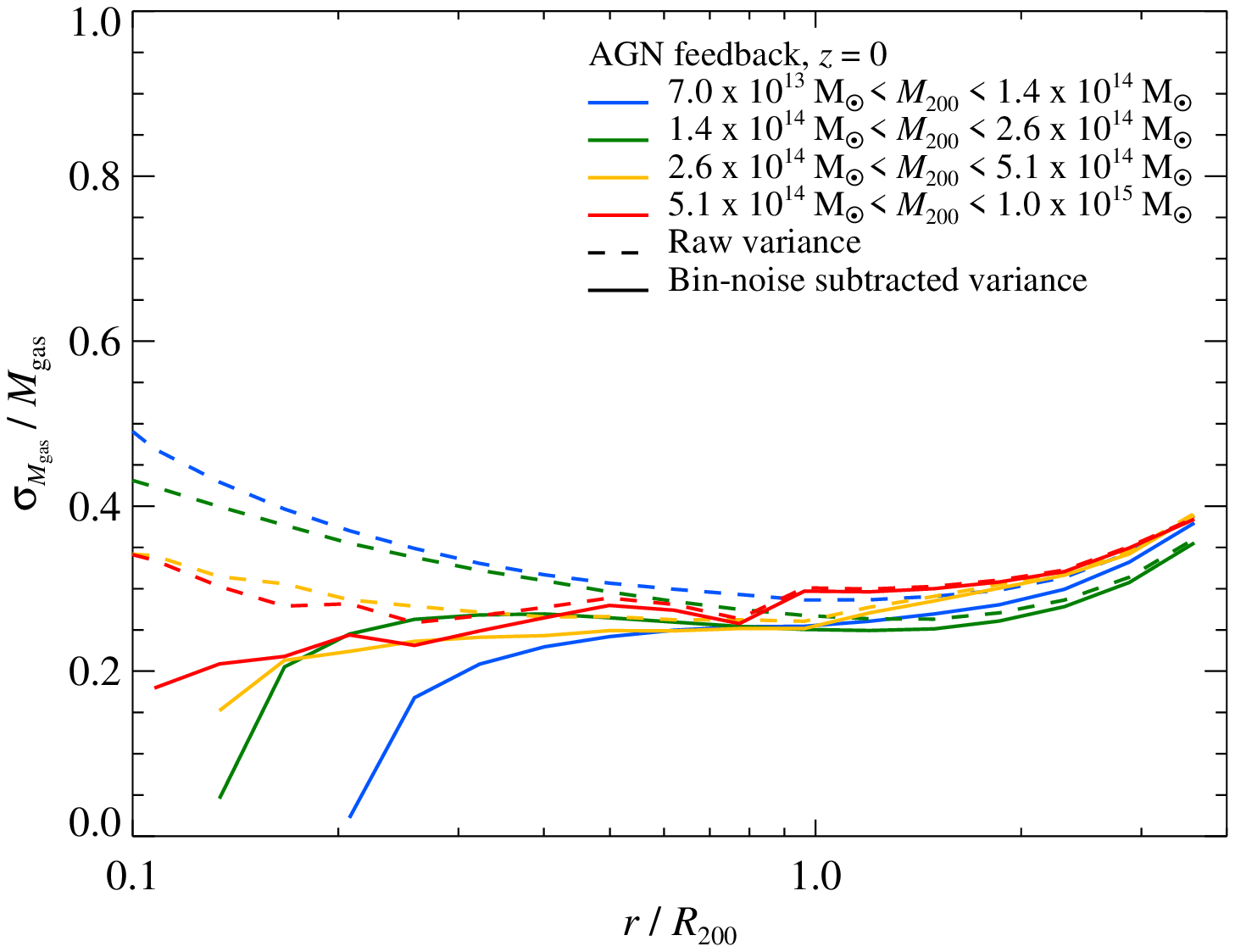}}%
    \resizebox{0.325\hsize}{!}{\includegraphics{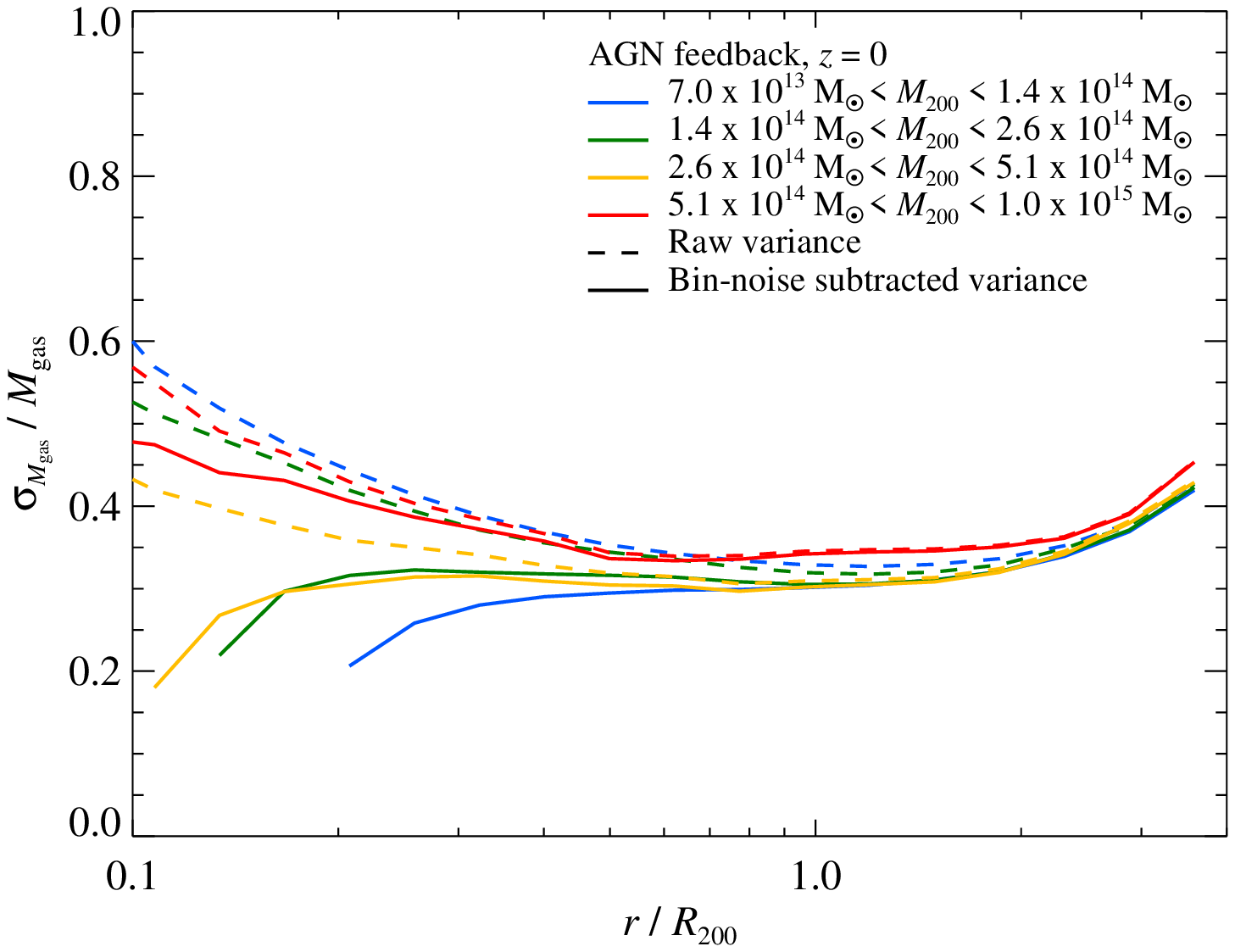}}%
    \resizebox{0.325\hsize}{!}{\includegraphics{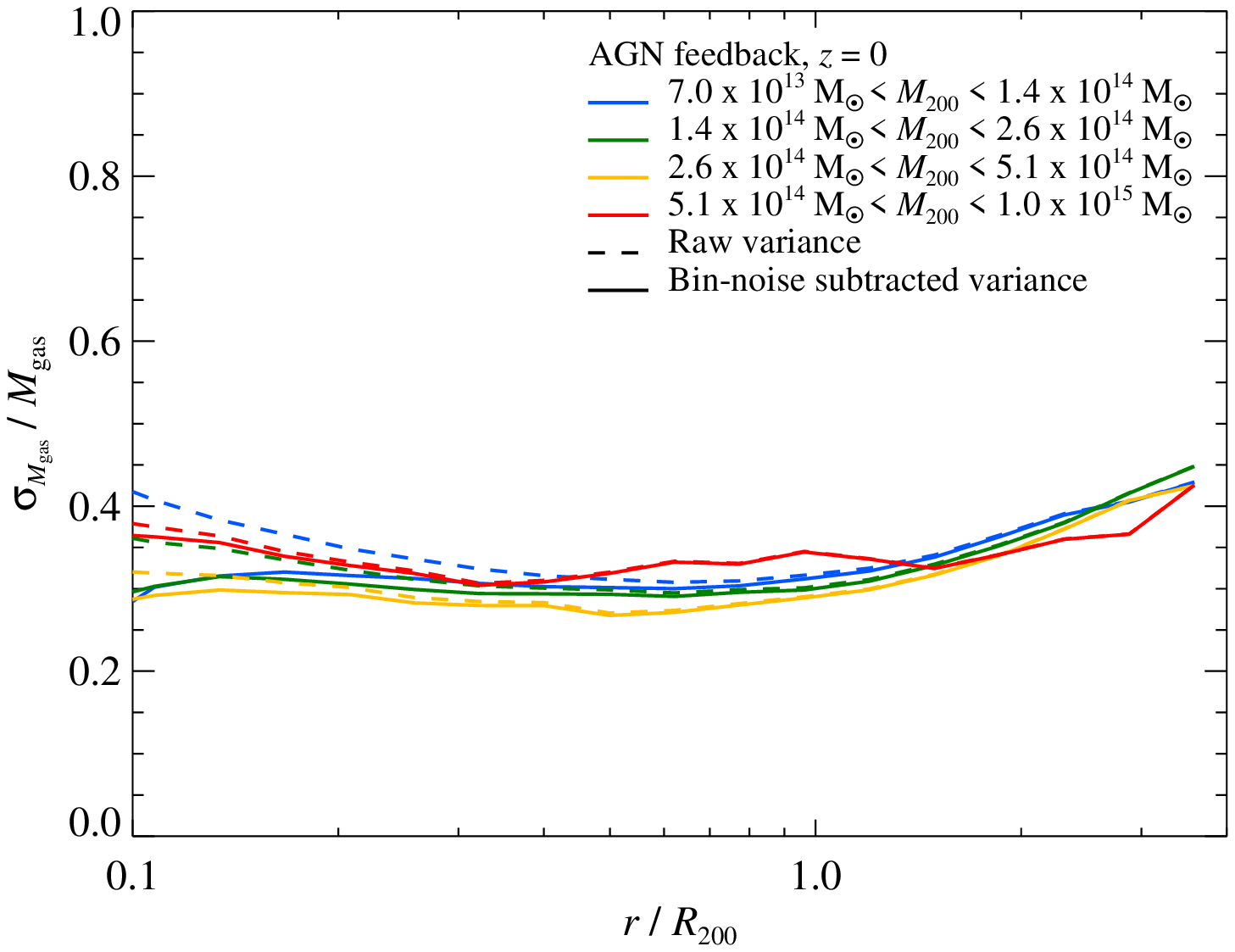}} }\\                  
  \centering{
    \resizebox{0.325\hsize}{!}{\includegraphics{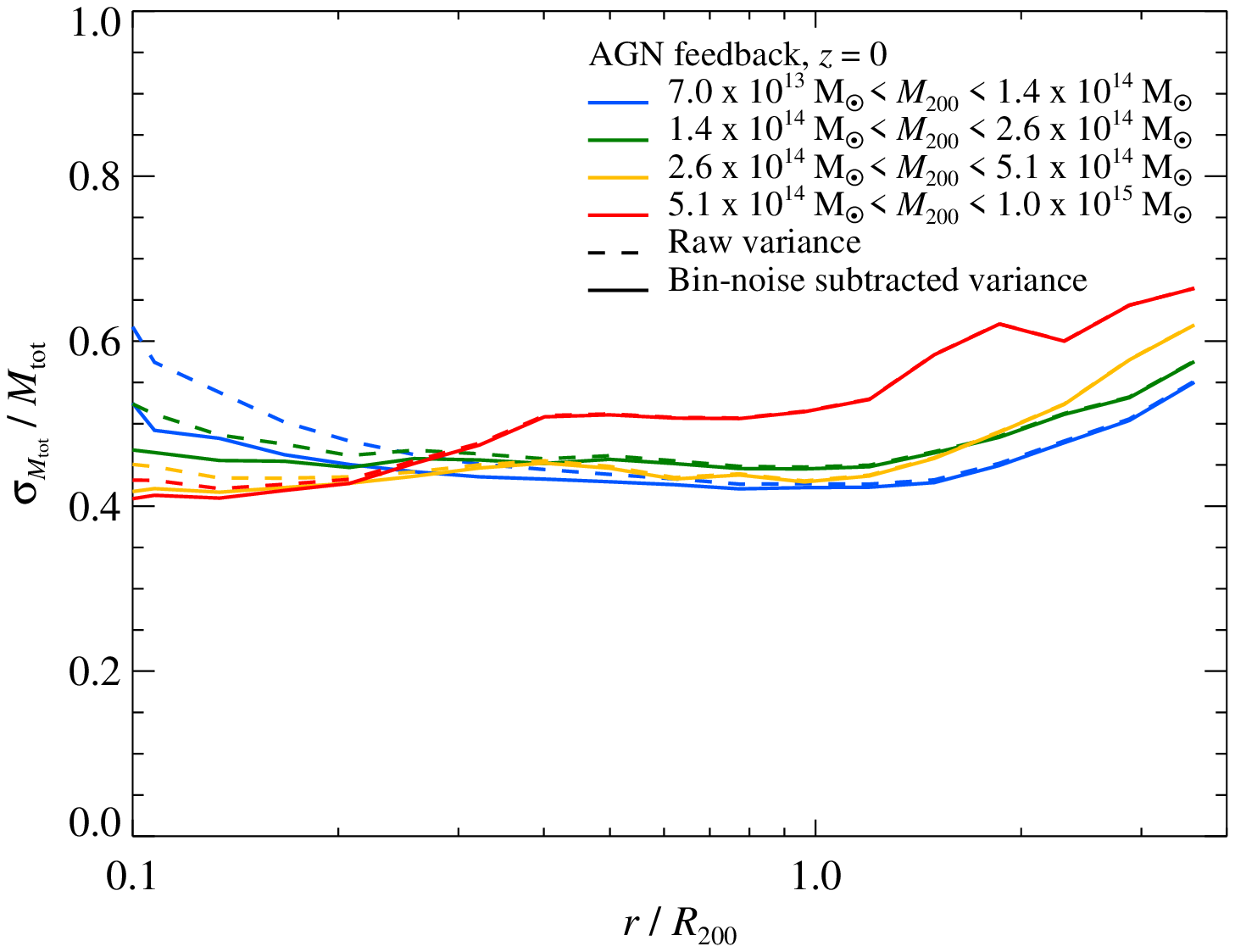}}%
    \resizebox{0.325\hsize}{!}{\includegraphics{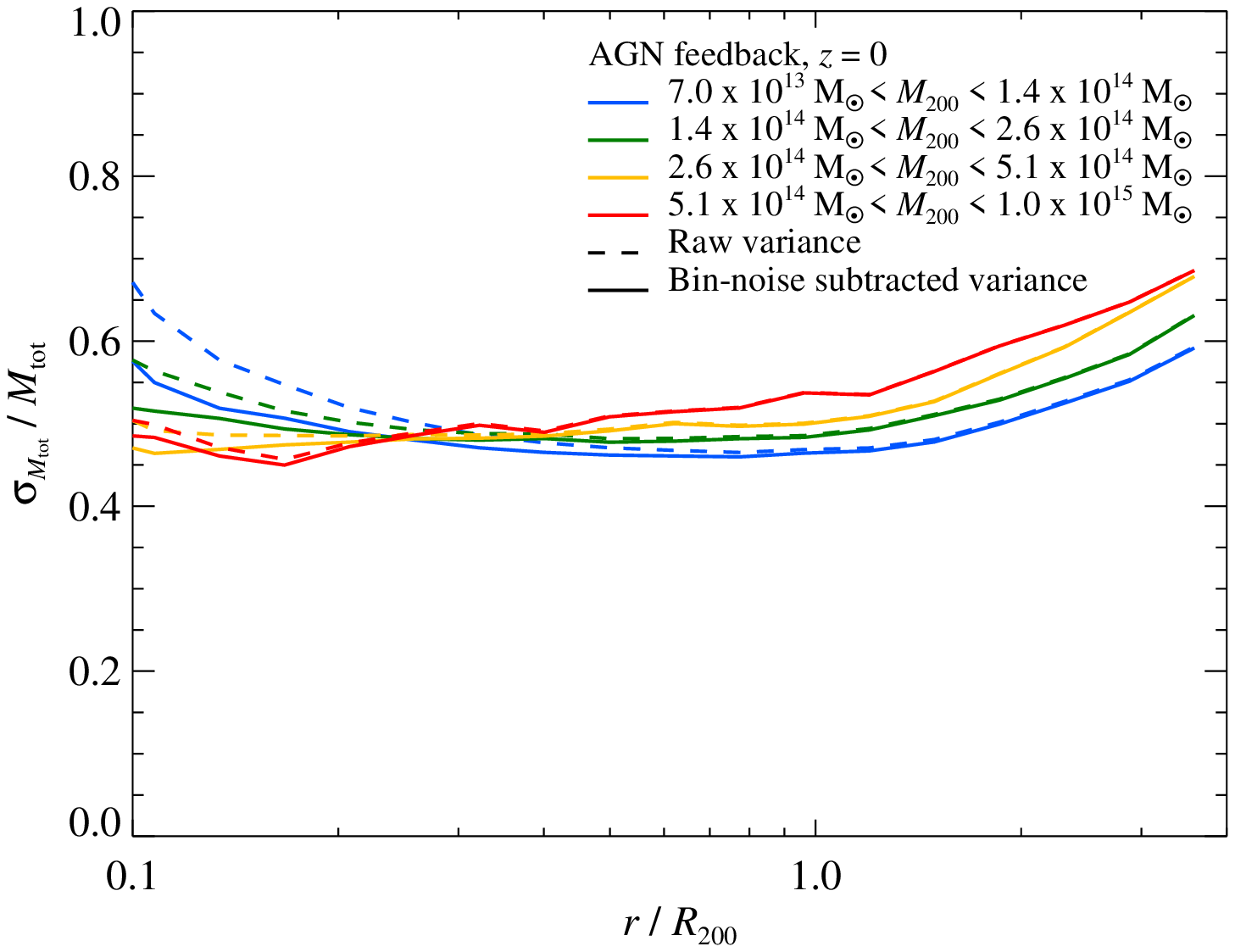}}%
    \resizebox{0.325\hsize}{!}{\includegraphics{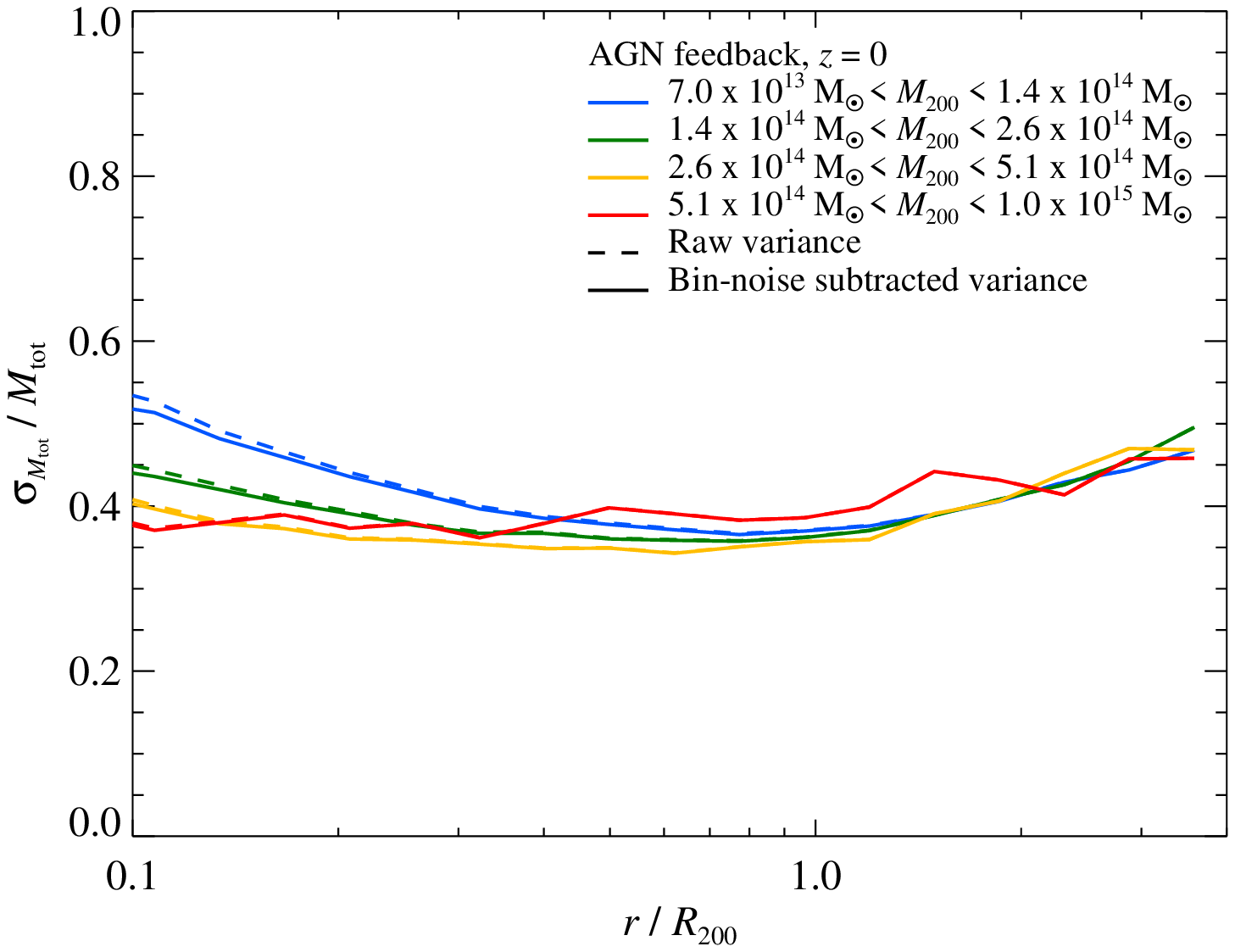}} }\\                  
  \centering{
    \resizebox{0.325\hsize}{!}{\includegraphics{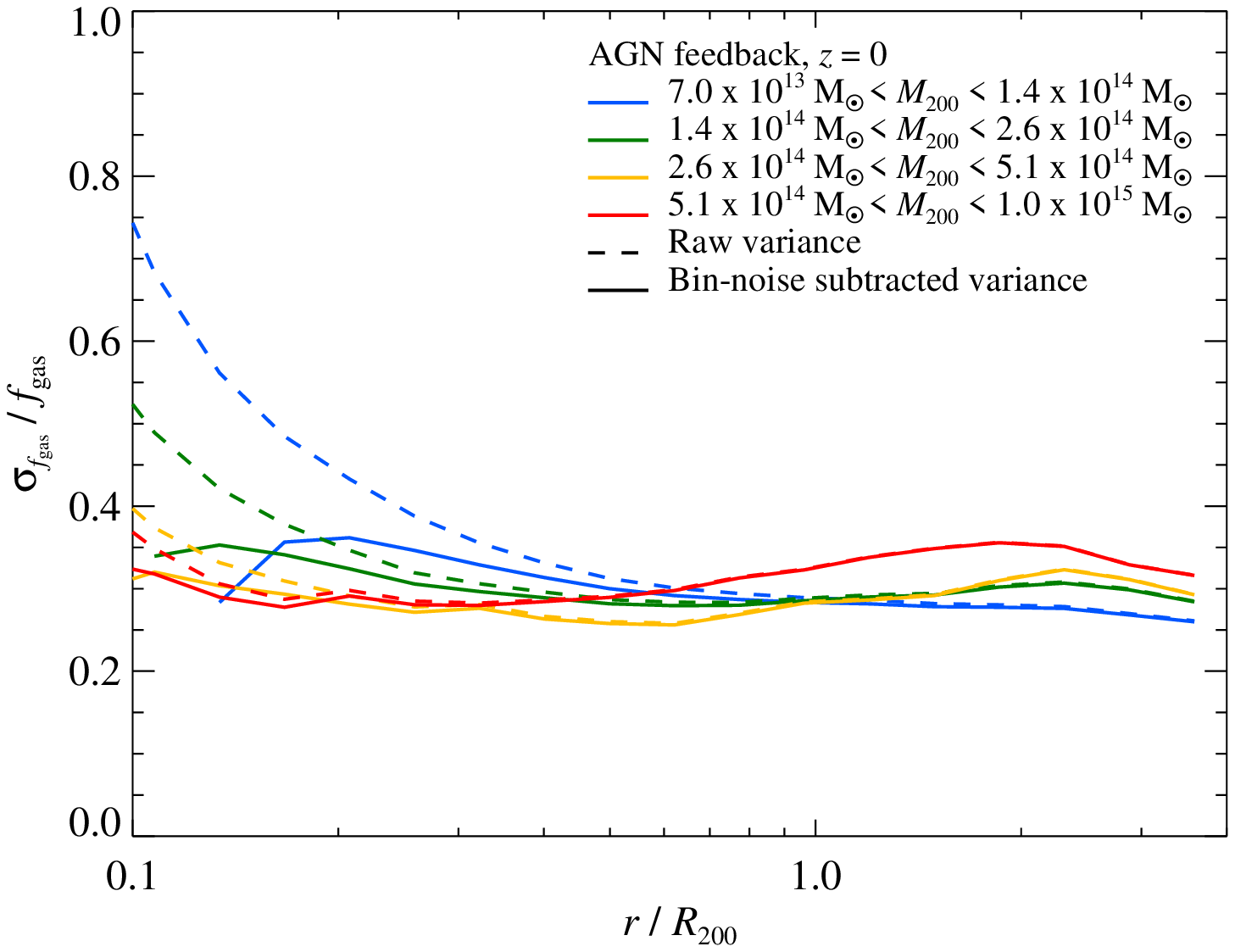}}%
    \resizebox{0.325\hsize}{!}{\includegraphics{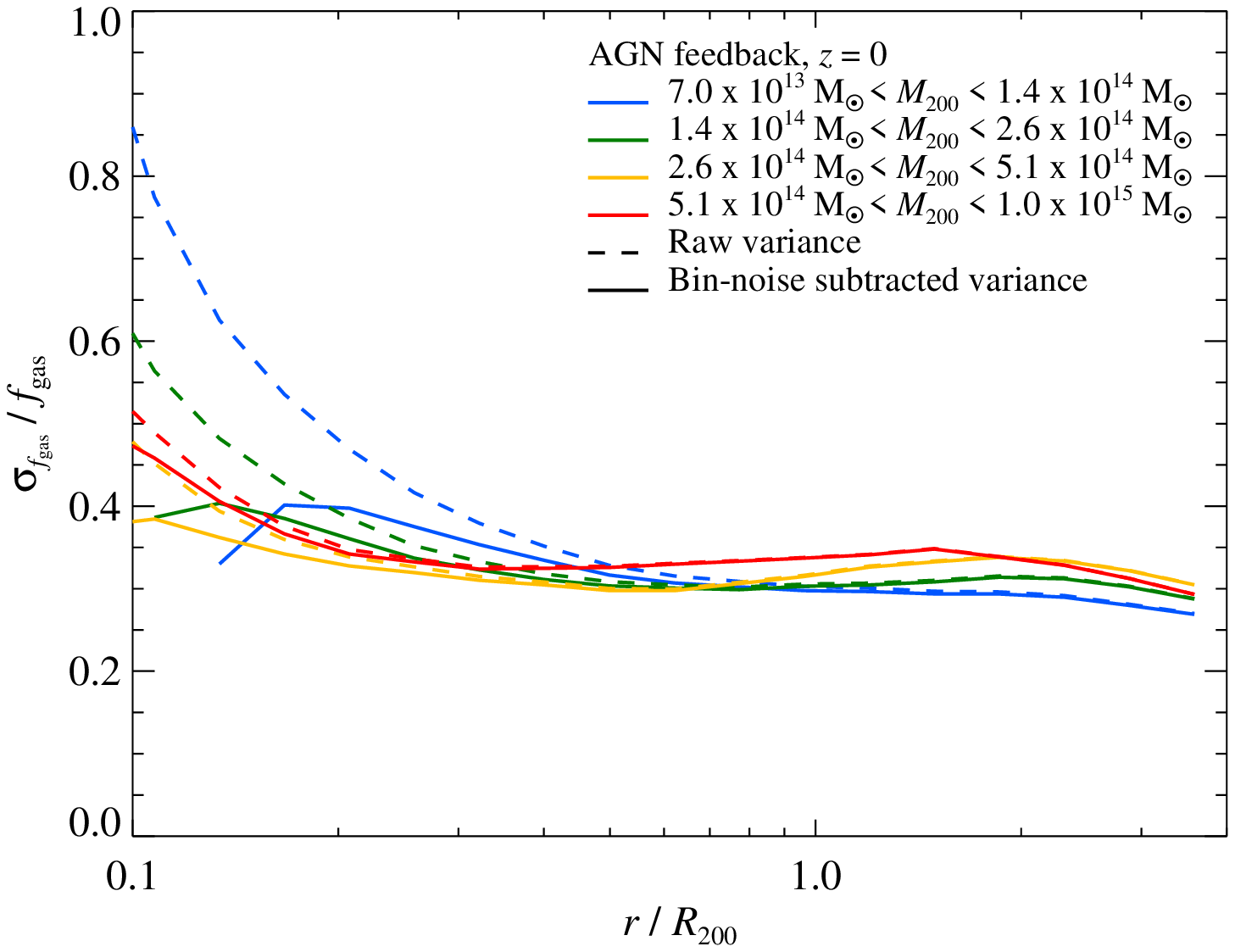}}%
    \resizebox{0.325\hsize}{!}{\includegraphics{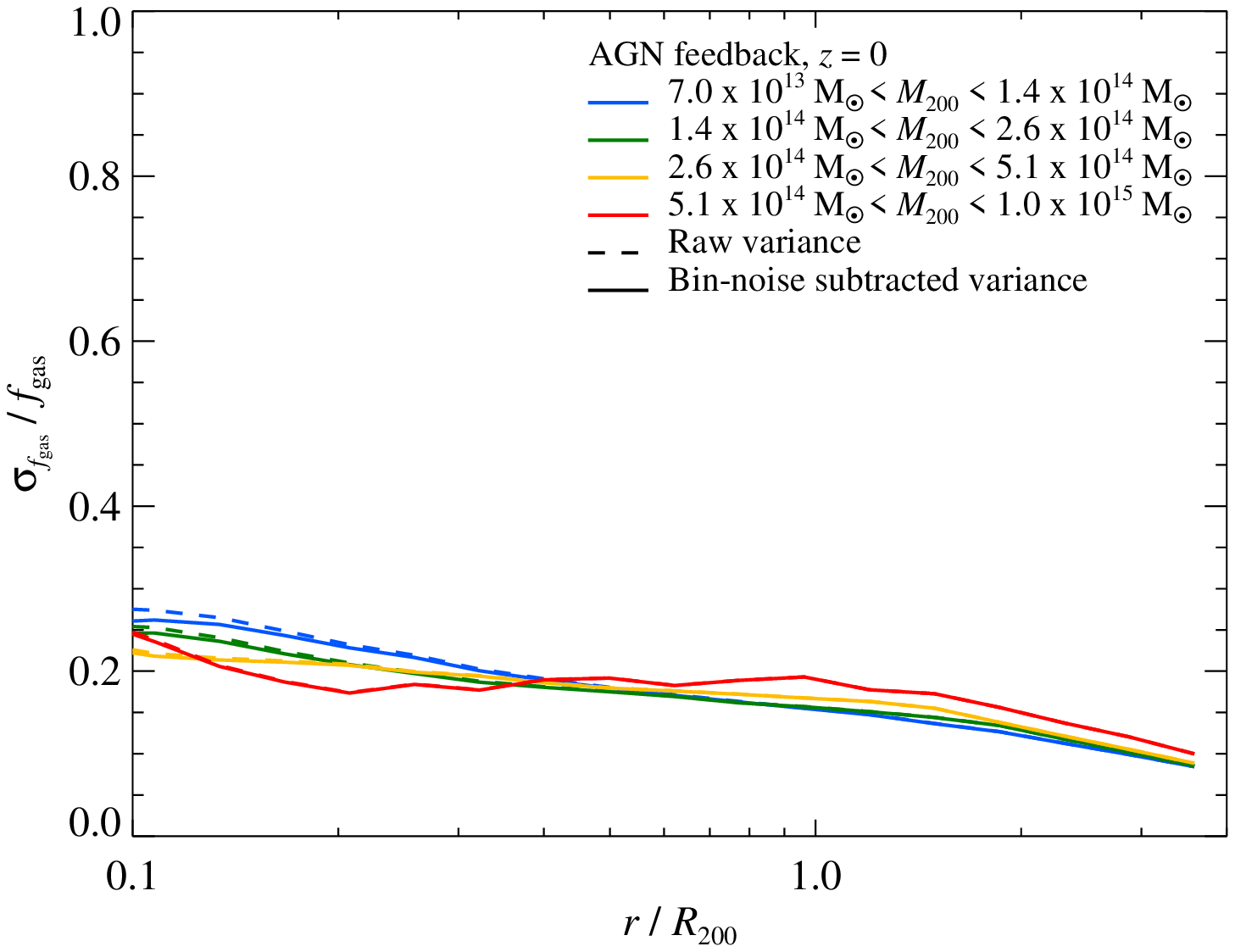}} }\\                  
  \caption{Top panels: we compute the variance of the $M_{\rmn{gas}}$ profiles
    across 48 angular 3D cones using the HEALPix scheme (dashed lines in
    Figure~\ref{fig:ang large}) and across 2D angular sectors (dashed lines in
    Figure~\ref{fig:ang large2D}) for each cluster. Here we show the mean of the
    3D cone variances across a sub-sample of the third most relaxed clusters
    with the lowest ratio of kinetic-to-thermal energy, $\KU$ (left) and across
    all clusters (middle). This is contrasted to the mean of the 2D sector
    variance (right). We define the angular variance of $M_{\rmn{gas}}$,
    $\sigma_{M_{\rmn{gas}}}$, by the difference of the upper and lower
    ($1-\sigma$) percentiles normalized by twice the median of
    $M_{\rmn{gas}}$. We compare the measured variance (dashed) to our estimate
    of the {\em intrinsic} angular variance (solid) where a Poisson noise term
    due to the finite particle number within the cones (respectively sectors)
    has been subtracted in quadrature. Note that in the 3D case this
    over-corrects the intrinsic variance at small radii as the Poisson noise
    estimate caps the noise contribution owing to the true sub-Poissonian
    distribution of SPH particles, which get rearranged as a result of the
    mutual repulsive hydrodynamical forces. Middle and bottom panels: same as
    above, but for $M_{\rmn{tot}}$ and $f_{\rmn{gas}}$, respectively, where
    $M_{\rmn{tot}}$ is calculated by summing up the mass distribution of the
    gaseous, stellar, and DM components in those cones/sectors. Different colors
    correspond to different cluster mass ranges (indicated in the legends).  Due
    to its collisional nature, gas is able to dissipate part of its kinetic
    accretion energy more efficiently than DM, thereby reducing its angular
    variance, in particular for relaxed systems (on the right). This is in
    contrast to the collisionless DM component (dominating the total mass) which
    shows a larger angular variance of its mass distribution.}
\label{fig:ang sigma}
\end{figure*}

\subsection{Cluster sample variance}

It has been suggested that sample variance of clusters could possibly account
for values of $f_{\rmn{gas}}$ larger than the cosmic mean, i.e., when a cluster
experiences an ongoing merger or has close-by cluster companions. To address
such questions, our large sample of clusters is ideal. It consists of 800
clusters with $M_{200}>1\times10^{14}\,\rmn{M}_\odot$ and has 200 clusters with
$kT_\rmn{spec}>2.5$~keV (corresponding to $M_{500}>2.15\times10^{14}\,
\rmn{M}_\odot$). In Figure~\ref{fig:fgas_scatter}, we compare the
cluster-to-cluster scatter of $f_{\rmn{gas}}$ in a sample of all clusters and a
sub-sample of the third most relaxed clusters (as defined by the third of all
clusters with the lowest ratio of kinetic-to-thermal energy, $\KU$, within
$R_{200}$, see Equations \eqref{eq:KU} and \eqref{eq:KU2}). The sample variance
of the true $f_{\rmn{gas}}$ decreases with increasing radius from a relative
1-$\sigma$ scatter of 20\% ($10-15\%$) at $0.1R_{200}$ for our sample of all
(relaxed) clusters down to $\sim5\%$ at $R_{200}$. This demonstrates the smaller
influence of AGN feedback on large scales especially for big clusters where the
gravity of the DM potential starts to dominate.  In contrast, the scatter of the
biased quantity $f_{\rmn{gas,HSE}}$ remains fairly constant at a level of
$\sigma_{f_\rmn{gas}}/f_\rmn{gas} \simeq 0.1 - 0.2$ (depending on the dynamical
state and mass of clusters). Within $R_{200}$, the sample variance $f_\rmn{gas}$
due to the clumping bias is negligible, but it starts to rise steeply outside
$R_{200}$, the innermost locations of accretion shocks (along filaments) that
dissipate kinetic accretion energy and smooth the density distribution interior
to this radius.

The $1-\sigma$ envelope of $f_{\rmn{gas,HSE+clump}}$ starts to rise dramatically
for $r\gtrsim R_{200}$ for our all-cluster sample. This rise is shifted to
somewhat larger radii for our sample of relaxed clusters
(Figure~\ref{fig:fgas_scatter}). However, at $R_{200}$ we obtain an upper
$1-\sigma$ percentile of $f_{\rmn{gas,HSE+clump}}$ for our massive clusters
($M_{200}> 3\times10^{14}\,\rmn{M}_\odot$) that is 30\% above the universal
value for our all-cluster sample and only 10\% for our subsample of relaxed
clusters. Cluster variance alone remains unlikely to support values that are 1.6
times the cosmic mean -- the measured value at $R_{200}$ in Perseus when
accounting for a stellar mass fraction of 12\% \citep[][even when accounting for
the moderate increase of this number by $\sim10\%$ for the most massive clusters
around $M_{200} \sim 10^{15}\,\rmn{M}_\odot$]{2011Sci...331.1576S}. Moreover,
clusters that are lying outside the $1-\sigma$ interval are either merging
systems or are caught in a pre-merger state, properties that can easily be
verified through optical or X-ray observations of the local environment of such
a cluster (unless the merger axis is aligned with the line-of-sight, but such a
configuration is rare and should be easily separable in a statistical sample of
clusters).

\subsection{Angular variance of the mass distribution within a single cluster}

In nearby clusters, the large angular extent of clusters renders it impossible
to cover the entire virial region of a cluster with a few X-ray pointings. Instead, a
limited number of mosaic pointings along radial arms are being observed which
opens the possibility of obtaining a biased answer for $f_{\rmn{gas}}$ if there
is a considerable angular variance. Two effects may play a role. First, a large
mass fraction is being accreted along filaments. These anisotropic ICM features
associated with large-scale structure environment suggest that the
thermalization of the ICM occurs at smaller radii along overdense filamentary
structures (that exert a greater ram pressure on the hot ICM) than along
low-density void regions. Second, since there is a strong internal
baryon-to-dark-matter density bias, a consequence of
collisional-to-collisionless physics, this could in principle also lead to a
significant angular variance depending on the differing efficiencies of gas
thermalization and the relaxation processes of the DM component.

To quantify these considerations, in Figure~\ref{fig:ang large} we show the
profile of $M_{\rmn{gas}}$, $M_{\rmn{tot}}$, and $f_{\rmn{gas}}$ in angular
cones centered on one massive cluster of $M_{200}\approx
1.2\times10^{15}\,\rmn{M}_\odot$. We chose the angular directions of these 48
cones such that they subtend an equal area pixelization on the sphere
\citep[following the HEALPix scheme with
$N_{\rmn{side}}=2$;][]{2005ApJ...622..759G}. The number of cones is a compromise
between obtaining high-enough angular resolution to break the angular smoothing
that is necessarily present in sphericalized profiles and to resolve individual
substructures and cosmic filaments in these cones. At the same time, we aim at
minimizing scatter due to our finite number of particles. To calculate
$M_{\rmn{tot}}$, we sum up the mass distribution of the gaseous, stellar, and DM
components in those cones.\footnote{For the gas component, we sort the volume
  fraction of an SPH smoothing kernel into the corresponding radial bins, but
  neglect the extension of the smoothing kernel in angular directions for
  performance reasons (see Appendix~\ref{sec:SPH-fit}.}

We observe considerable scatter in all three distributions, $M_{\rmn{gas}}$,
$M_{\rmn{tot}}$, and $f_{\rmn{gas}}$, depending on the angular direction. There
are a few enormous outliers in the $M_{\rmn{gas}}$ and $M_{\rmn{tot}}$ cones,
deviating upwards up to a factor of 3 from the median (shown with solid black) in
the case of $M_{\rmn{gas}}$ and even beyond for $M_{\rmn{tot}}$. The right-most
panel in Figure~\ref{fig:ang large} shows that in the most extreme cases, the
cone profiles of $f_{\rmn{gas}}$ can be biased high by a factor of two in such a
massive cluster with $M_{200}\sim10^{15}\,\rmn{M}_\odot$, potentially providing
an explanation of the boosted values of $f_{\rmn{gas}}$ reported in the
north-western arm in Perseus \citep{2011Sci...331.1576S}. Care should be taken
when interpreting these $f_{\rmn{gas}}$ values in cones, since projection effects
may lower these values. However, as demonstrated in Figure~\ref{fig:fgas},
hydrostatic mass bias as well as density clumping may (partially) compensate for
these projection effects.

Only in the $M_{\rmn{tot}}$ cones, those outliers can be seen down to small
radii $r\lesssim 0.1R_{200}$ which is expected as DM substructure only obeys
collisionless physics that allows substructures to penetrate deep into the halo
without too much affecting the dense central subhalo structures by tidal
stripping. In contrast, the gaseous content has been removed by means of shocks
or ram-pressure stripping at larger radii, lowering the variance of
  $M_{\rmn{gas}}$ at smaller radii (see Figure~\ref{fig:ang large}). These
large outliers are absent in the distribution of $f_{\rmn{gas}}$ cones, which
implies a smaller variance (dashed black). This suggests that these outliers are
caused by self-bound substructures that are still holding on to (some) of their
gas at larger cluster radii, i.e., there exists a tight spatial correlation of
the DM and gas distribution (see \citetalias{BBPS4} for a detailed analysis).

Line-of-sight projection smoothes out the largest angular variations in the 3D
distribution. We calculate projected (cylindrical) mass distributions according
to
\begin{eqnarray}
  \Sigma(r_\perp) &=& \int_0^{L} \rho(r_\perp,l) dl, \\
  M_{\rmn{2D}} (<r_\perp) &=& 2 \pi \int_0^{r_\perp} \Sigma(r_\perp') r_\perp' d r_\perp',
\label{eq:M_2D}
\end{eqnarray}
where $L=165\,h^{-1}\,\rmn{Mpc}$ denotes the length of the simulation box.
Before projecting, we smooth the discrete stellar and DM masses by distributing
their individual masses according to an SPH-type smoothing kernel that extends
up to the nearest 48$^\rmn{th}$ neighbouring particle (of the same type). In
Figure \ref{fig:ang large2D}, we show the resulting projected profiles of
$M_{\rmn{gas}}$, $M_{\rmn{tot}}$, and $f_{\rmn{gas}}$ in 12 equally large
angular sectors, centered on the same massive cluster as studied in
Figure~\ref{fig:ang large}. In comparison to the 3D cones, the scatter is
smaller and there are less dramatic outliers in the projected sector
distribution of $M_{\rmn{gas}}$, $M_{\rmn{tot}}$, and $f_{\rmn{gas}}$.  Unlike
the 3D case, the largest 2D outlier sector in this cluster does not exceed far
above the cosmic mean of $f_{\rmn{gas}}$ within $R_{200}$.

In Appendix~\ref{sec:sector-number}, we show that the variance of projected
sectors of $M_{\rmn{gas}}$ and $f_{\rmn{gas}}$ depends only very weakly on the
number of sector elements, hence justifying our choice of 12 elements. Instead,
the scatter of the X-ray luminosity strongly depends on the number of sector
elements, especially in SPH simulations \citep[see Figure 4
of][]{2011MNRAS.413.2305V}. Presumably, this increased scatter solely derives
from the increased clumping factor with radius. The increased scatter in the SPH
simulations in comparison to AMR simulations especially for $r<R_{200}$
\citep{2011MNRAS.413.2305V} may stem from an incomplete clump removal.

\subsection{Variation of the mass distribution with angle across cluster
  samples}

In order to obtain statistically valid answers on how this scatter across
different cones varies among our cluster sample, we first compute the variance
of the profiles of $M_{\rmn{gas}}$, $M_{\rmn{tot}}$, and $f_{\rmn{gas}}$ across
48 angular cones (dashed lines in Figure~\ref{fig:ang large}) for all
clusters. Since the distribution within a single cluster is non-Gaussian and
positively skewed (by the presence of substructures), we define the angular
variance of $M_{\rmn{gas}}$, $\sigma_{M_{\rmn{gas}}}$, by the difference of the
upper and lower $1-\sigma$ percentiles, normalized by twice the median of
$M_{\rmn{gas}}$. In the top panels of Figure~\ref{fig:ang sigma}, we show the
mean of $\sigma_{M_{\rmn{gas}}}$ across our cluster sample (color coded by
cluster mass). The same quantity is being calculated for $M_{\rmn{tot}}$ and
$f_{\rmn{gas}}$ (lower panels of Figure~\ref{fig:ang sigma}) across a sub-sample
of the third most relaxed clusters (left) and across all clusters (middle). Also
shown is the mean of the angular variances $\sigma_{M_{\rmn{gas}}}$,
$\sigma_{M_{\rmn{tot}}}$, and $\sigma_{f_{\rmn{gas}}}$ for the projected 2D maps
(right).

\begin{figure*}
  \begin{minipage}[t]{0.5\hsize}
    \centering{\small $z=0$:}
  \end{minipage}
  \begin{minipage}[t]{0.5\hsize}
    \centering{\small $z=1$:}
  \end{minipage}
  \resizebox{0.5\hsize}{!}{\includegraphics{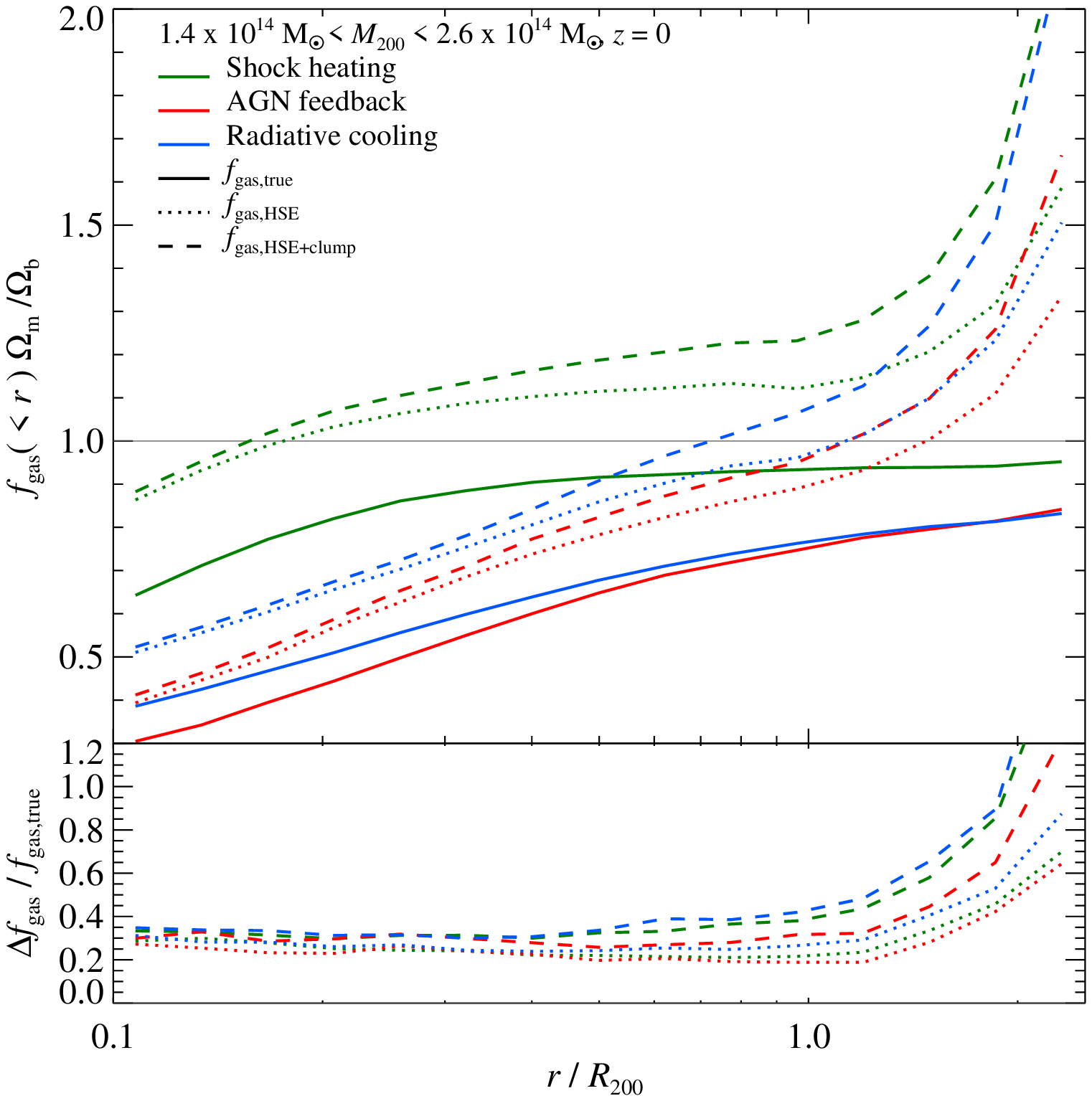}}%
  \resizebox{0.5\hsize}{!}{\includegraphics{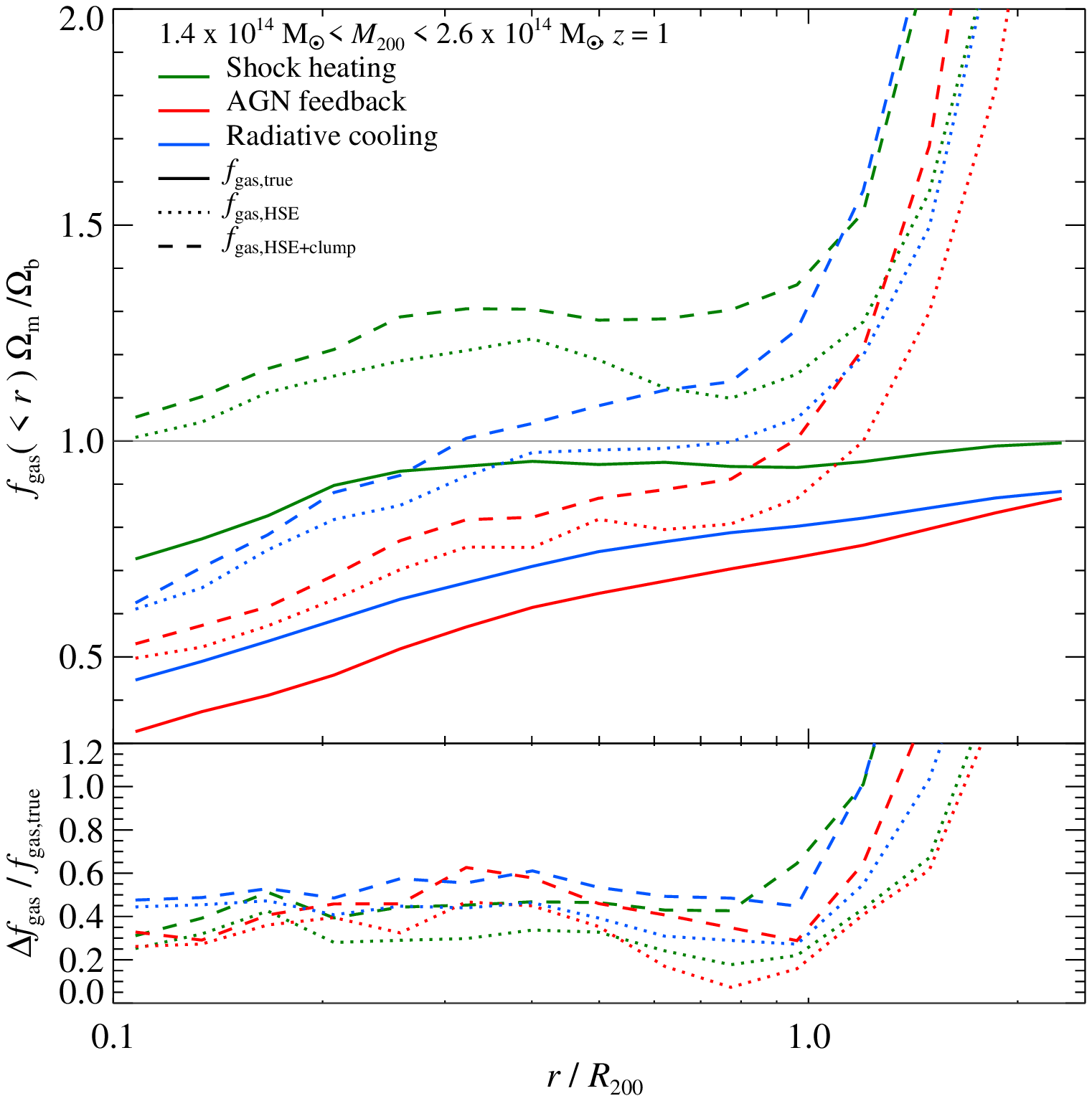}}\\
  \caption{Different levels of sophistication of the simulated physics of the
    ICM imply different profiles for $f_\rmn{gas}$. In the top panels, we show
    the true gas mass fraction $f_{\rmn{gas}} (<r)$ (solid), $f_\rmn{gas,HSE}
    (<r)$ (dotted), and $f_\rmn{gas,HSE+clump} (<r)$ (dashed), all normalized by
    the universal baryon fraction for different physics models at two redshifts,
    $z = 0$ (left) and $z = 1$ (right). In the bottom panels, we show the
    relative differences with respect to the true $f_\rmn{gas}$. There is little
    redshift evolution between the biases at $z = 0$ and $z = 1$, except for
    large radius at which these biases increase dramatically and which moves to
    smaller radii for $z=1$. The HSE bias of the gas mass fraction,
    $f_\rmn{gas,HSE}$, depends less on the simulated physics, whereas the
    $f_\rmn{gas}$ bias that additionally includes clumping,
    $f_\rmn{gas,HSE+clump}$, is marginally stronger in the {\it shock heating}
    and {\it radiative cooling} models in comparison to the {\it AGN feedback
      simulations}.}
\label{fig:fgas_phys}
\end{figure*}

As shown in Figure~\ref{fig:ang sigma}, there is an upturn of the raw 3D
variances (dashed) visible at small radii and in particular for galaxy
groups. This is caused by the noisy distribution of a comparably small number of
SPH particles in the central cluster regions in combination with the smaller
radial bin size of our logarithmic equally-spaced grid towards the center, when
distributing the particle masses among the angular cones. As a result of the
mutual repulsive hydrodynamical forces of neighboring SPH particles, they
maintain a sub-Poissonian distribution with repulsive short distance
correlations on scales of order the corresponding inter-particle
separations. The full characterization of such a distribution with dynamical
evolution of SPH is not possible in terms of known distributions. Hence we
decided to cap this numerical noise contribution by treating it as Poissonian
and subtract a Poissonian noise term in quadrature from the raw 3D angular
variance. We caution that this over-corrects the {\em intrinsic} variance at
small radii, causing an artificial decrease of the bin-noise subtracted variance
(solid). We emphasize that our focus here is on large scales with $r\gtrsim
(0.2-0.5) R_{200}$ (where the range corresponds to large and small clusters,
respectively). Those scales are {\em not} subject to this numerical
limitation. Hence in order to improve upon our analysis in the central regions,
an increase in numerical resolution would be needed. The 2D angular variances in
the central regions are only little affected by the sub-Poissonian variance due
to the sufficiently large number of particles that are projected onto the
central regions.

Inside $R_{200}$, the 3D angular variance of $M_{\rmn{gas}}$ does not strongly
depend on mass and reaches values around 30\% at $R_{200}$ with a moderate
increase at larger radii. In contrast, the 3D angular variance of
$M_{\rmn{tot}}$ is steadily increasing and reaches values around $45-50\%$,
depending on cluster mass (with the upper range indicating larger clusters that
are on average dynamically younger).  These values are only moderately reduced
by $\simeq5$ percentage points for our relaxed cluster sample. The different 3D
angular variances of $M_{\rmn{gas}}$ and $M_{\rmn{tot}}$ can be understood by
the collisional nature of the gas that is able to dissipate part of its kinetic
accretion energy more efficiently (through shocks and ram-pressure stripping of
the gaseous contents of substructures), thereby reducing its 3D angular
variance. This is in contrast to the collisionless DM component (dominating the
total mass) which shows a larger 3D angular variance of its mass distribution.

Interestingly, the Poisson noise-corrected 3D angular variance of
$f_{\rmn{gas}}$ remains almost constant with radius at the level of $30-35\%$
not only within the clusters but also for $r> R_{200}$. Furthermore,
$\sigma_{f_{\rmn{gas}}}/f_{\rmn{gas}}$ depends very weakly on cluster mass.
Both of these properties can be understood by the fact that at these large
radii, most of the accreting substructures have not yet passed the accretion
shock nor are they affected by ram-pressure stripping as they still move through
the dilute warm-hot inter-galactic medium. Hence these self-bound substructures
are still holding on to their gas, implying large spatial correlations of the DM
and gas distribution (see also Figure~\ref{fig:ang large} and
\citetalias{BBPS4}).

The most striking feature of all our Poisson noise-corrected 3D angular
variances is the rather constant behavior with radius, in particular for the 3D
angular variances of $M_{\rmn{gas}}$ and $f_{\rmn{gas}}$, as well as the
moderate radial increase of that of $M_{\rmn{tot}}$. In principle, the mass
distribution in cones allows for finer sampling of the cluster shape beyond the
moment-of-inertia tensor and is sensitive to higher-order moments -- a measure
that would be similar to density clumping for sufficiently high multipoles. In
practice, this is limited by the Poisson noise due to our finite particle
resolution. Moreover in Figure~\ref{fig:ang sigma}, we concentrated on the
radial profiles of the second moment of the 3D angular variance which is
intimately related but not identical to cluster ellipticities as obtained
through eigenvalues of the moment-of-inertia tensor as a function of radius
\citepalias{2012ApJ...758...74B}. The ellipticity of the gas density also shows
a constant behavior inside $R_{200}$ while it is rising at larger radii. In
contrast, the DM ellipticity shows already a moderate increase with radius which
only starts to rise more steeply outside $R_{200}$
\citepalias{2012ApJ...758...74B} -- very similar to our findings for the 3D
angular variance of $M_{\rmn{tot}}$ that is dominated by DM.

When averaged over sufficiently large scales, i.e. at very large radii,
$f_{\rmn{gas}}$ is expected to assume the cosmic mean and the 3D angular
variance $\sigma_{f_{\rmn{gas}}}$ should approach zero. In our case, the
outermost radial bin is centered on $R_i=3.6\,R_{200}$ and has a radial extend
of $\Delta R_i = 0.8\,R_{200}$. The volume associated with that cone bin is
$V_{\rmn{cone,}\,i} = \Delta R_i \,R_i^2 4\pi/48 = (2.9\,\rmn{Mpc})^3$ for a
$10^{15}\,\rmn{M}_\odot$ cluster with $R_{200} = 2.1$~Mpc. This volume is still
small in comparison to the scale associated with the root mean square variance
of linear theory in spheres of radius $8\,\rmn{Mpc}/h$, $\sigma_8 = 0.8$. In
fact, the 2D angular variance $\sigma_{f_{\rmn{gas}}}$ shows the expected
downturn for radii $r\gtrsim2R_{200}$. The volume associated with the outermost
2D radial bin of a cylindrical projection length of $L=165\,h^{-1}\,\rmn{Mpc}$
is $V_{\rmn{sector,}\,i}=L\,R_{200}^2\, (x_{i}^2-x_{i-1}^2) \,\pi/12 =
(11.5\,\rmn{Mpc})^3$ for a $10^{15}\,\rmn{M}_\odot$ cluster and $(x_i,x_{i-1}) =
(4,3.2)$. This volume is 60 times larger than for the corresponding 3D case, but
still somewhat smaller than the volume of a sphere with radius $8\,\rmn{Mpc}/h$.
Why do we then observe a decline in the 2D angular variance?  This is because
the line of sight projection averages over many uncorrelated patches on scales
$\gg 8\,\rmn{Mpc}/h$, which draws $\sigma_{f_{\rmn{gas}}}$ down for larger radii
and brings $f_{\rmn{gas}}$ back to the cosmic mean. In the 3D case, the
considerably smaller volume and the inherent radial correlations of the gas and
DM distributions on these scales (as quantified in \citetalias{BBPS4}) explain
the flat radial profile of $\sigma_{f_{\rmn{gas}}}$ out to $4\,R_{200}$.

All angular variances of $M_{\rmn{gas}}$ and $M_{\rmn{tot}}$ (2D and 3D) show an
increase at the largest radii ($r\gtrsim 2\,R_{200}$). This reflects the
increase of the root mean square variance, $\sigma_R^2 = \bra
\delta^2(\mathbfit{x})\ket = \bra(\delta M/M)^2\ket$, on these scales due to the
dominance of the two-halo term in halo-mass cross-correlation, which corresponds
to the mass autocorrelation function with a mass-dependent offset in amplitude
known as the halo bias factor \citep{2008MNRAS.388....2H}.


\section{Gas and stellar mass fractions in cosmological simulations}
\label{sec:fgas}

Here, we present results on how $f_\rmn{gas}$ varies with simulated physics,
radius, cluster mass, and redshift. These results are summarized in Table
\ref{tab:Sum1} and \ref{tab:Sum2}.

\subsection{How different physics determines $f_\rmn{gas}$ and $f_\rmn{star}$}
\label{sec:physics}

\begin{figure}
\resizebox{\hsize}{!}{\includegraphics{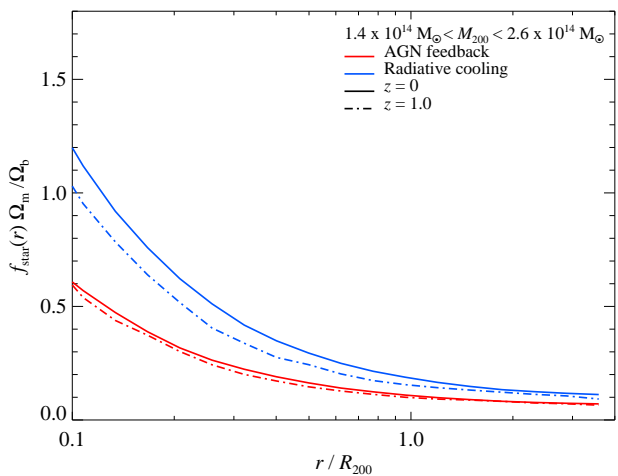}}
\caption{Whether or not we include AGN feedback does affect the median
  profile of the stellar mass fraction, $f_{\rmn{star}}$. We show
  $f_\rmn{star}(<r) \equiv M_{\rmn{star}}(<r) / M_{\rmn{tot}} (<r)$ normalized
  by the universal baryon fraction at $z = 0$ (solid) and $z = 1$
  (dash-dotted). While ongoing cooling increases the central stellar mass
  fraction to unphysical high values in our model with cooling and star
  formation-only, AGN feedback is able to self-regulate the core region and to
  suppress the forming cooling flow.}
\label{fig:fstar_phys}
\end{figure}

\begin{figure*}
  \begin{minipage}[t]{0.5\hsize}
    \centering{\small $z=0$:}
  \end{minipage}
  \begin{minipage}[t]{0.5\hsize}
    \centering{\small $z=1$:}
  \end{minipage}
  \resizebox{0.5\hsize}{!}{\includegraphics{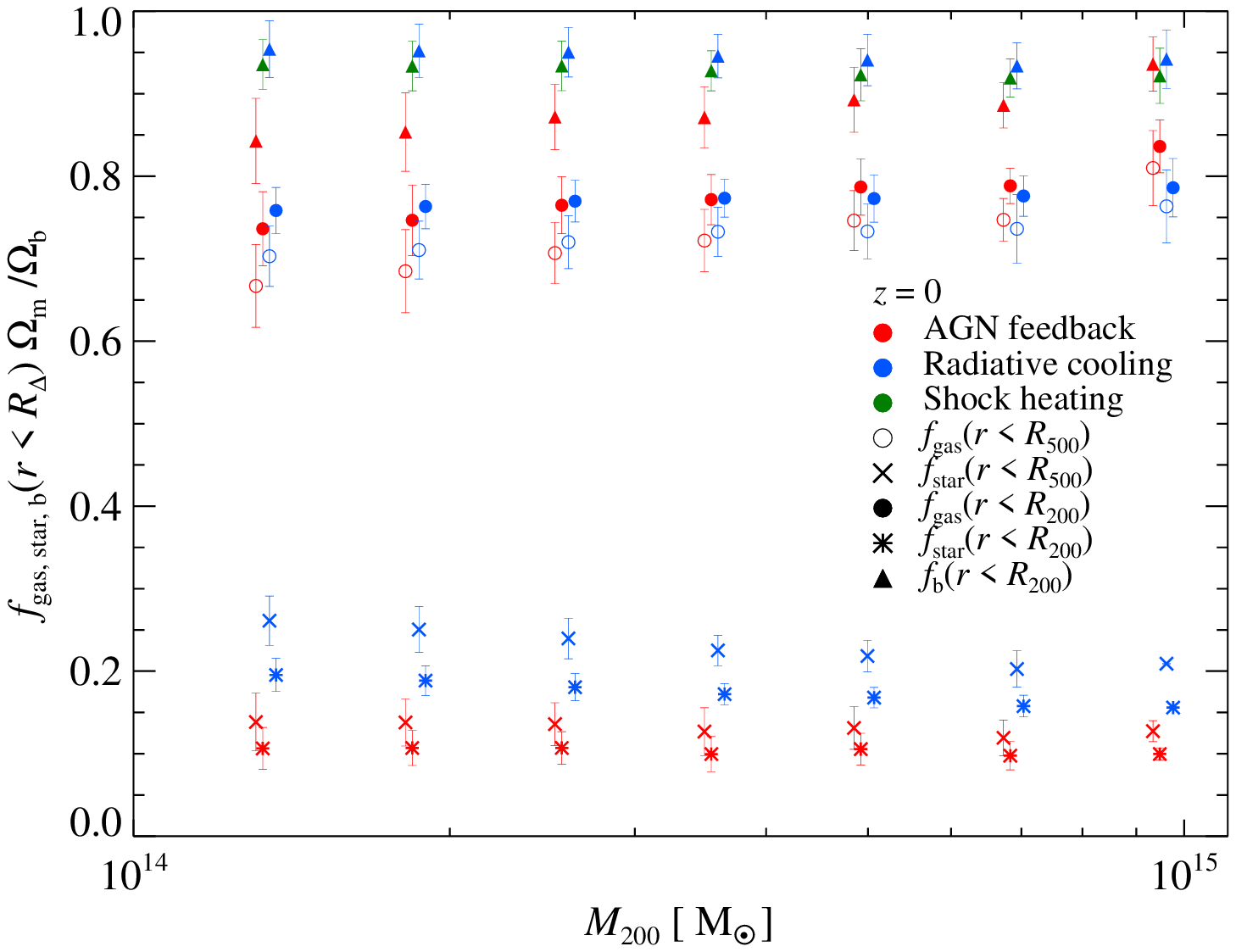}}%
  \resizebox{0.5\hsize}{!}{\includegraphics{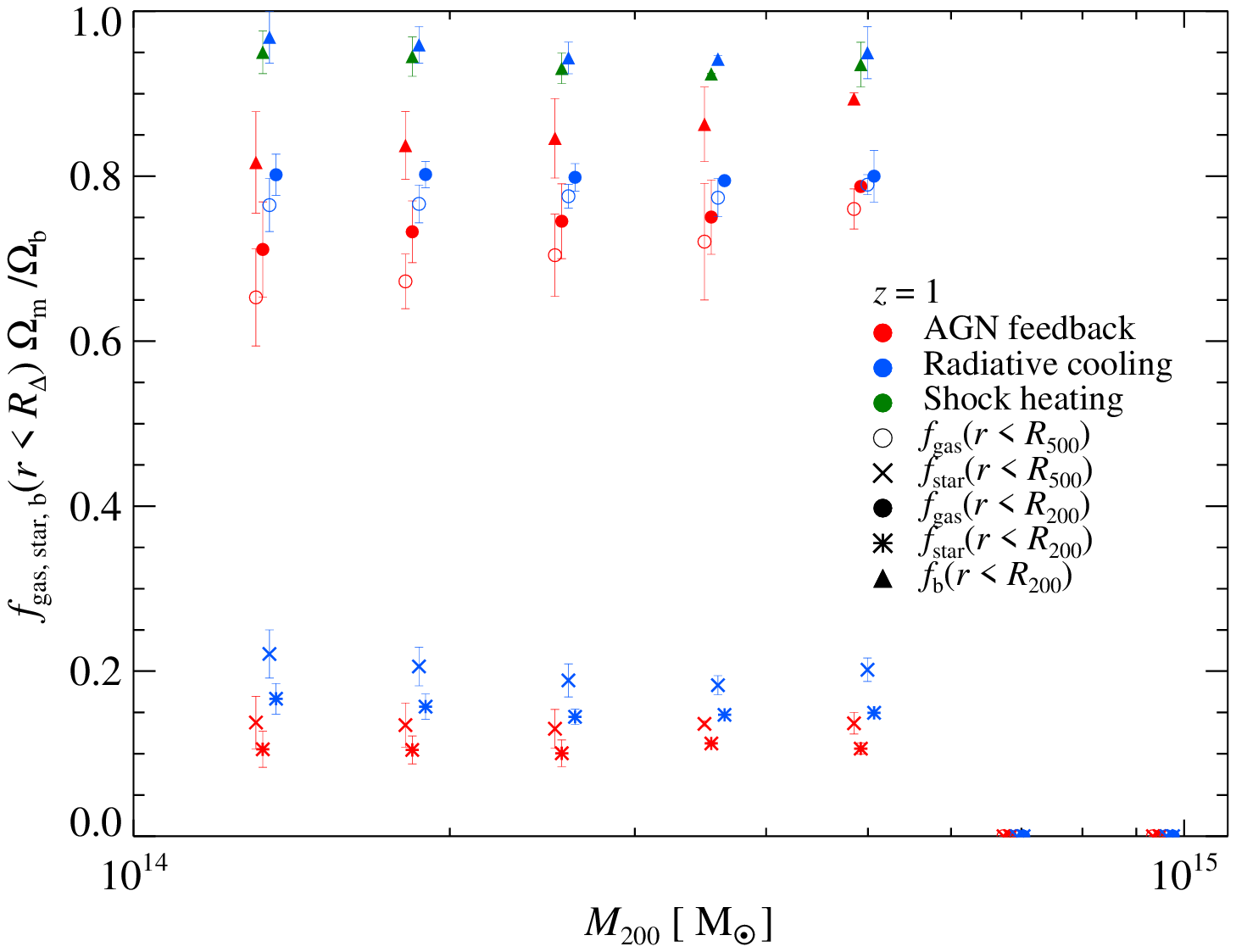}}\\
  \caption{We show gas, stellar, and baryon mass fractions, $f_\rmn{gas}(<r)$,
    $f_\rmn{star}(<r)$, and $f_b(<r)$, (mean and standard deviations), computed
    within a fixed radius ($R_{500}$ and $R_{200}$), as a function of cluster
    mass at $z=0$ (left panel) and $z=1$ (right panel). Different colors
    indicate different simulated physics models: AGN feedback (red), radiative
    cooling (blue), and shock heating-only (green). Note that $f_b$ in our
    shock heating-only and radiative cooling models are almost identical within
    the scatter for the integration radii $R_{500}$ and $R_{200}$. In our
    cooling and star formation model at $z=0$, we find an increasing stellar
    mass fraction towards smaller clusters that is accompanied by a decreasing
    gas mass fraction, particularly within $R_{500}$. This trend of a decreasing
    $f_{\rmn{gas}}$ towards smaller clusters is amplified in our AGN feedback
    model, while the stellar mass fraction remains almost constant across our
    mass range probed. AGN feedback pushes gas out of smaller clusters beyond
    $R_{200}$ which causes a reduction of the baryon mass fraction in comparison
    to our model without AGN feedback. In contrast, the trends of $f_\rmn{star}$
    with cluster mass are almost absent at $z=1$. }
\label{fig:fgas_delta_m}
\end{figure*}

Even though we believe that our AGN feedback model represents the closest
approximation to reality, it is important to compare it to simpler physics
models, which have been extensively studied in earlier literature. Moreover,
there is considerable uncertainty about the correct physics underlying AGN
feedback so that trends relative to our simpler models can be very instructive.
Hence, we compare how different physics models impact the radial profiles of
$f_\rmn{gas}$ (Figure~\ref{fig:fgas_phys}) and $f_\rmn{star}$
(Figure~\ref{fig:fstar_phys}). This is supplemented by a figure demonstrating
the cluster mass dependence of $f_\rmn{gas}$ and $f_\rmn{star}$ within $R_{500}$
and $R_{200}$ for differently simulated physics (Figure~\ref{fig:fgas_delta_m}).

In our shock heating-only model, $f_\rmn{gas}$ approaches a value of
$0.93\pm0.03$ in units of $f_c$ at $r\sim R_{200}$ (which is reduced by one
percentage point for larger clusters). Outside the virial radius, $f_\rmn{gas}$
is continuously rising to asymptote to the cosmic mean only at $4 R_{200}$.
This may be partially due to a transfer of energy from DM to the gas during
group/cluster mergers. In an equal mass collision, SPH simulations show that the
DM can transfer up to 10\% of its energy, and even a 10:1 merger may be able to
reduce the DM energy by a couple of percent \citep{2007MNRAS.376..497M}. There
is also the possibility that some of this gas deficiency may be related to the
SPH technique that we use here. There appears to be a discrepancy to adaptive
mesh-refinement simulations that find higher values of $f_b\approx0.97\pm0.03$
within $R_{\rmn{vir}}$ \citep[][see also Section~\ref{sec:previous}, for a
detailed discussion]{2005ApJ...625..588K}.  At small radii, the shock-heated ICM
provides pressure that causes the inner slope of the gas density profile to be
rather flat, while the collisionless DM exhibits a cusp
\citep[e.g.,][]{2012MNRAS.425.2169G}. As a result, this causes a decreasing gas
fraction inside the DM scaling radius, i.e., for $r\lesssim 0.2 R_{200}$.

Our models with AGN feedback and radiative cooling have similar values of
$f_\rmn{gas}$ at $z=0$, in particular for $r\gtrsim R_{500}$. However, in the
AGN feedback model, $f_\rmn{gas}$ only marginally evolves with $z$, while it
increases for higher $z$ in the case of radiative cooling and star
formation. This also manifests itself in the constant evolution of the
$f_\rmn{star}$ profile in the AGN feedback model. In contrast, the stellar mass
is increasing with time in the radiative model without AGN feedback due to the
well-known overcooling problem in cosmological simulations of galaxy clusters if
cooling is not counteracted by any feedback process
\citep[e.g.,][]{Borgani+2009}.

For $r\gtrsim R_{200}$, the values of $f_\rmn{gas}$ in different radiative
models (as well as for different redshifts) approach each other. However, in
comparison to our shock heating-only model, there is still a gas mass deficit of
10\% at $4 R_{200}$. In the raditive cooling model, the stellar mass fraction
makes up for this deficit (Figure~\ref{fig:fstar_phys}) while in the case of AGN
feedback, stellar mass fractions are reduced by a factor of $\sim2$ in
comparison to our radiative cooling model. (Note that $f_{\rmn{star}}$ in our
radiative cooling models depends on numerical resolution as this model is
physically not converged.) The remaining gas mass has been pushed to even larger
radii through the action of AGN feedback
\citep{2011MNRAS.412.1965M}. Interestingly, these lower values for $f_\rmn{gas}$
in our radiative models in comparison to our non-radiative models imply a
temperature increase in the outer parts \citep[e.g., Figure 4
of][]{2010MNRAS.401.1670F} that is necessary to balance the gravitational
cluster potential. 

At $z=0$, all our models show approximately a similar relative $M_{\rmn{HSE}}$
bias (left bottom panel of Figure~\ref{fig:fgas_phys}) while the relative
differences with respect to the true $f_\rmn{gas}$ are larger at $z=1$ (right
bottom panel of Figure~\ref{fig:fgas_phys}). However, there are some notable
differences. At large radii, the radiative cooling model exhibits the largest
bias, while our AGN feedback model shows the lowest bias. This is presumably due
to AGN outflows depositing heat at these large radii (early on) which manifests
itself in an increased pressure in comparison to the radiative cooling model
\citep{2010ApJ...725...91B}. This increased thermal pressure moves the average
location of accretion shocks to larger radii (since these form at the location
where ram pressure due to infalling gas balances the thermal pressure of the
ICM), which causes a deceleration of the accreting material already further out,
and hence slightly decreases the average kinetic pressure support in the AGN
feedback model. Hence, the bias corrections shown in Figure~\ref{fig:fgas} are
conservative given that the AGN feedback model shows an overall lower bias in
comparison to the other models.

At $z=0$, the bias due to the hydrostatic equilibrium assumption and as a result
of neglecting clumping ($f_\rmn{gas,HSE+clump}$) is rather constant inside
$R_{200}$ and shows values around 30\%. At higher redshifts, $z=1$,
$f_\rmn{gas,HSE+clump}$ increases to a level that fluctuates around
50\%. Outside the virial radius, the $f_\rmn{gas,HSE+clump}$ bias increases
sharply, albeit at a slightly slower rate in our AGN feedback model for reasons
explained above. The location of the sharp increase in the bias moves from
$\sim2 R_{200}$ (at $z=0$) to $\sim R_{200}$ (at $z=1$). This is mostly due to
the definition of $R_{200}$ and can be approximately aligned by changing the
definition to $R_{200m}$, which compares the average interior density to the
mean mass density rather than the critical density of the Universe
\citepalias[Figure 21 of][]{2012ApJ...758...74B}.

\begin{figure*}
  \begin{minipage}[t]{0.5\hsize}
    \centering{\small All clusters:}
  \end{minipage}
  \begin{minipage}[t]{0.5\hsize}
    \centering{\small Relaxed clusters:}
  \end{minipage}
  \resizebox{0.5\hsize}{!}{\includegraphics{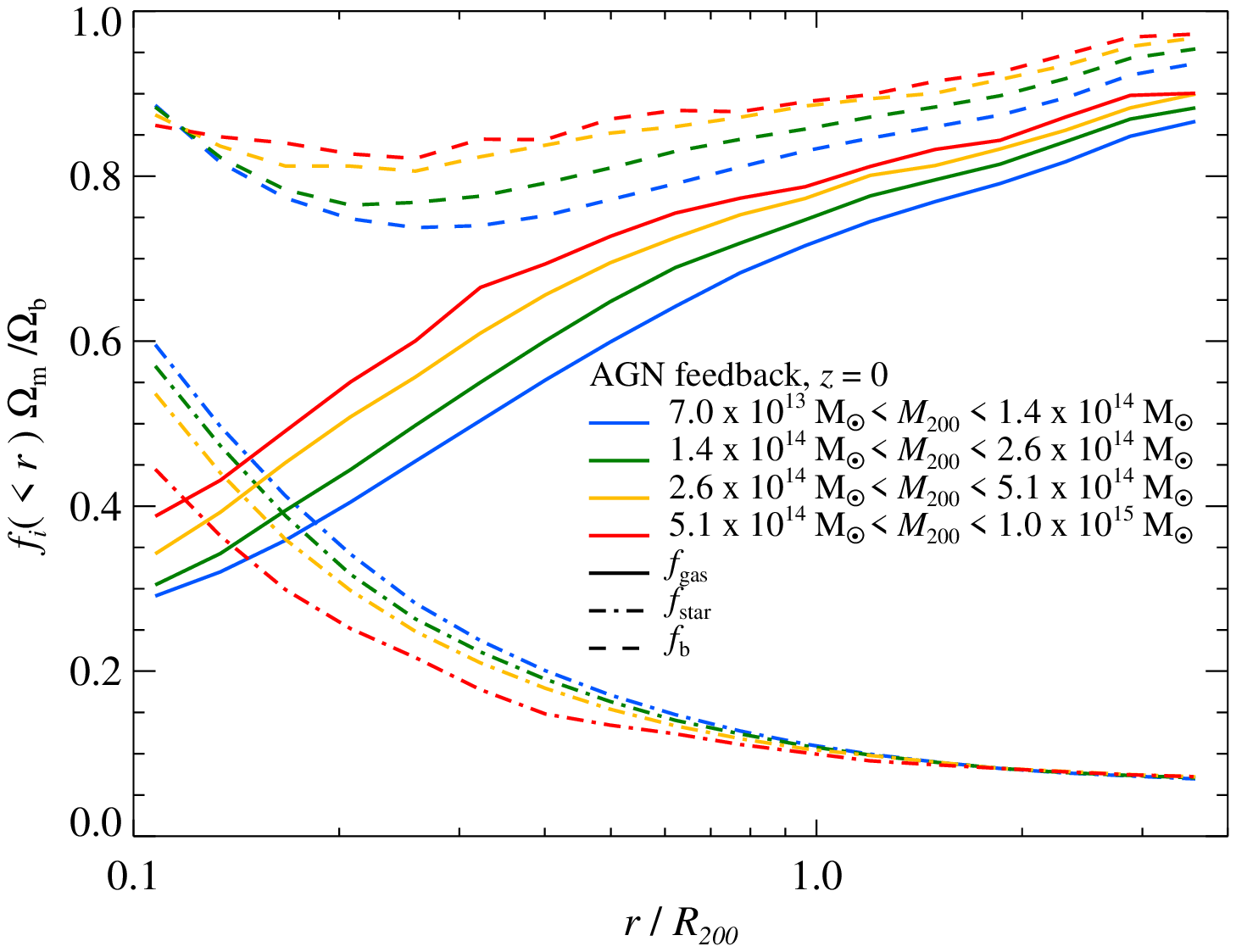}}%
  \resizebox{0.5\hsize}{!}{\includegraphics{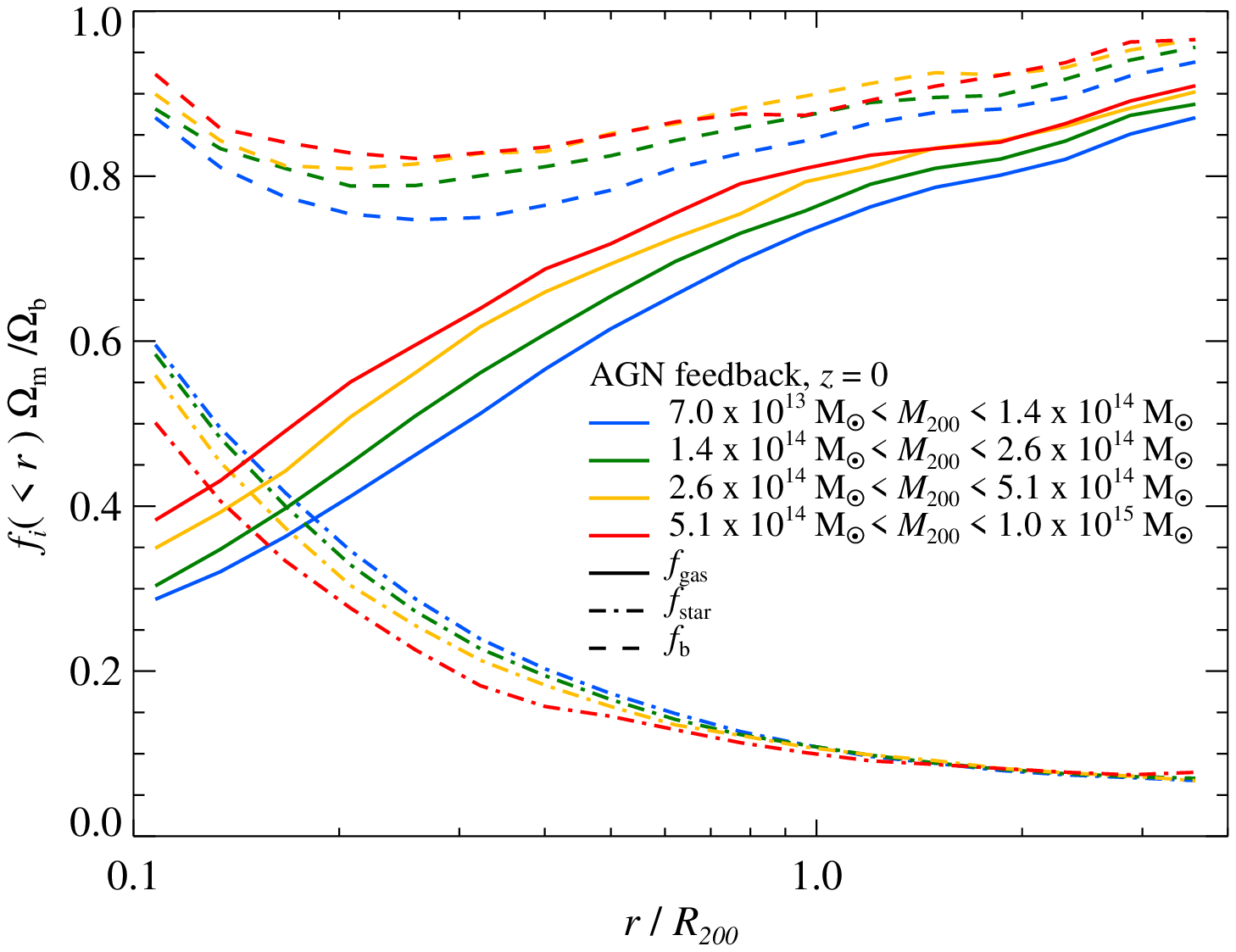}}\\
  \caption{Radial median profiles of the mass fractions in gas ($f_{\rmn{gas}}$,
    solid), stars ($f_{\rmn{star}}$, dash-dotted), and baryons ($f_{b}$,
    dashed). Each profile is normalized by the universal baryon fraction and
    different colors correspond to different cluster mass ranges (indicated in
    the legends).  The profiles in the left panel include all clusters and the
    right panel shows the median of a sub-sample of the third most relaxed
    clusters, selected by the lowest third in total kinetic-to-thermal energy
    ratio, $\KU$. In contrast to our non-equilibrium measures such as clumping
    or kinetic pressure support, the mass fractions in gas, stars, an baryons
    are almost independent on the clusters' dynamical state.}
\label{fig:fbar}
\end{figure*}

\subsection{Gas and stellar mass fractions across cluster masses}

How do the gas and stellar mass fractions, $f_\rmn{gas}$ and $f_\rmn{star}$, as
well as the sum of both, the baryon mass fraction, $f_b$, vary with cluster
mass? In Figure~\ref{fig:fgas_delta_m}, we show $f_\rmn{gas}$ and $f_\rmn{star}$
for our 3 different physics models as a function of cluster mass.  As discussed
in Section~\ref{sec:physics}, our model that only accounts for shock heating
approaches a value of $f_\rmn{gas}/f_c=0.93\pm0.03$ ($1-\sigma$) at $r\sim
R_{200}$ (which is reduced by one percentage point for larger clusters).  These
lower values of $f_\rmn{gas}$ represent the pivot point from which our radiative
models can reduce $f_\rmn{gas}$ furthermore through conversion of baryons into
stars or by expelling it from the cluster region (in the case of our AGN
feedback model).

In our radiative cooling model, $f_\rmn{gas}$ increases moderately with cluster
mass at $R_{500}$ to level off at a value of $f_\rmn{gas}/f_c = 0.79$ for big
clusters. This trend is weakened at $R_{200}$. The opposite behavior is seen
for $f_\rmn{star}$ so that the resulting baryon fraction almost exactly sums to
a value that is seen in our shock heating-only model. At $R_{200}$, the scatter
in $f_\rmn{gas}$ drops by $1-2$ percentage points in comparison to the scatter at
$R_{500}$ for both of our radiative models.

While in our AGN feedback model, the effect of a decreasing $f_\rmn{gas}$
towards lower cluster masses is amplified and the trend of an increasing
$f_\rmn{star}$ towards lower masses is weakened (at $R_{200}$).
Figure~\ref{fig:fbar} shows that these trends are still present at smaller radii
and become progressively stronger towards the center. Interestingly, these effects
partially compensate for the baryon fraction, $f_b$, which attains a much
weaker mass dependence. Comparing the radial profiles of these mass fraction for
all clusters (left panel of Figure~\ref{fig:fbar}) and a sub-sample of the third
most relaxed clusters (right panel) we virtually do not see any differences. In
contrast to our non-equilibrium measures such as clumping or kinetic pressure
support, the mass fractions in gas, stars, and baryons are almost independent on
the clusters' dynamical state. The only exception are the central values for
$f_\rmn{star}$ in massive clusters which are higher by 5\% in our
relaxed-cluster sample in comparison to all clusters.

However, there are two noticeable differences of the mass fractions in our AGN
feedback model in comparison to those in our radiative cooling model
(Figure~\ref{fig:fgas_delta_m}). (1) The baryon fractions do not sum up to the
value of our shock heating-only model. There is a deficit of gas mass at
$R_{200}$ that decreases from 10\% at $M_{200}\simeq 10^{14}\,\rmn{M}_\odot$
down to 5\% at $7\times10^{14}\,\rmn{M}_\odot$. Only for the largest clusters
($M_{200}\simeq 10^{15}\,\rmn{M}_\odot$) does the baryon fraction of the AGN
feedback model correspond to that of the shock heating-only model. As explained
in in Section~\ref{sec:physics}, this is due to the powerful outflows of gas
that deposit considerable amounts of gas beyond $R_{200}$ so that the cosmic
mean is not even reached by $4\,R_{200}$. The only exception to this are the
largest clusters whose gravitational potential provide a strong enough pull on
the gas that was successful in having re-accreted the gas. (2) In our AGN
feedback model at $R_{200}$, $f_\rmn{star}$ remains almost independent of
cluster mass (in contrast to the values at $R_{500}$).

This behavior is very interesting with respect to its implications for the
redshift evolution of $f_\rmn{star}$ on group scales. In order to build up
high-mass clusters with low values of $f_\rmn{star}/f_c \sim 0.1-0.15$, their
progenitors also had to have these low stellar mass fractions in a
hierarchically growing universe since the amount of baryons locked up in
(low-mass) stars remains there throughout a Hubble time in the model of star
formation adopted here \citep{2003MNRAS.339..289S}.\footnote{There can be return
  of stellar mass to the gaseous phase by means of stellar winds which is
  thought to mostly sustain star formation in galactic disks
  \citep{2010ApJ...714L.275M, 2011ApJ...734...48L}. If there is a considerable
  population of intracluster stars \citep{2010MNRAS.406..936P}, stellar mass
  loss could be a potential channel of increasing $f_\rmn{gas}$ at the expense
  of $f_\rmn{star}$, but the low observed mass fraction of intracluster stars
  renders it unlikely to be an important contribution, especially on the
  high-mass cluster end \citep{2003ApJ...591..749L, 2005MNRAS.358..949Z,
    2005ApJ...618..195G, 2007ApJ...666..147G, 2009ApJ...703..982G}.} Since
$f_\rmn{star}$ is increasing towards lower cluster masses in our radiative
cooling model, this would imply that the groups we observe today, do not reflect
the properties of groups at high-$z$ (in terms of $f_\rmn{star}$) that merged to
assemble large clusters today. This is demonstrated by the almost constant
behaviour of $f_\rmn{star}$ as a function of cluster mass at $z=1$ in our
radiative cooling model (right-hand panel of Figure~\ref{fig:fgas_delta_m}). The
values of $f_\rmn{star}$ in groups at $z\simeq1$ is almost at the same level as
$f_\rmn{star}$ for big clusters today.  In contrast, the approximate constancy
of $f_\rmn{star}$ with cluster mass in our AGN feedback models suggests that
groups that are observed today do not need to be different from group-scale
progenitor systems of clusters today.

The increasing value of $f_\rmn{star}$ with time on groups scales in our
radiative cooling model (Figure~\ref{fig:fstar_phys}) also suggests an excess
star formation at late-times which results in too blue colors and too massive
stellar masses at the high-mass end in comparison to observations
\citep{2007MNRAS.380..877S}. We note that the distributions of $f_\rmn{star}$
and $f_\rmn{gas}$ are almost Gaussian (independent of the models), resulting in
almost identical median and mean.\\[1em]

\subsection{Redshift evolution of gas and stellar mass fractions}
\label{sec:redshift}

\begin{figure}
\resizebox{\hsize}{!}{\includegraphics{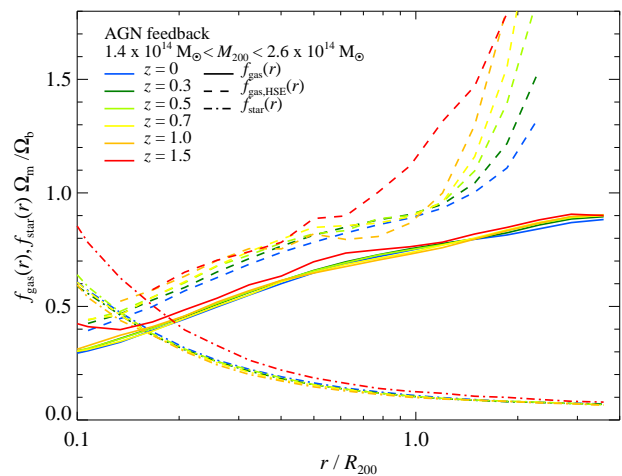}}
\caption{In our AGN feedback model, there is little redshift evolution in the
  median profiles of the gas mass fraction, $f_{\rmn{gas}} (<r)$, (solid) and
  stellar mass fraction, $f_{\rmn{star}} (<r)$ (dash-dotted), when radii are
  scaled to $R_{200}$. Both quantities are normalized by the universal baryon
  fraction and shown as a function of redshift for fixed mass range (different
  colors). In contrast, the profiles $f_{\rmn{gas}} (<r)$ that assume
  hydrostatic equilibrium (dashed) show a larger bias at high redshift due to
  the greater kinetic pressure contribution at earlier times. Those probe on
  average dynamically younger objects that are in the process of assembling.  }
\label{fig:fgas_z}
\end{figure}

\begin{figure}
\resizebox{\hsize}{!}{\includegraphics{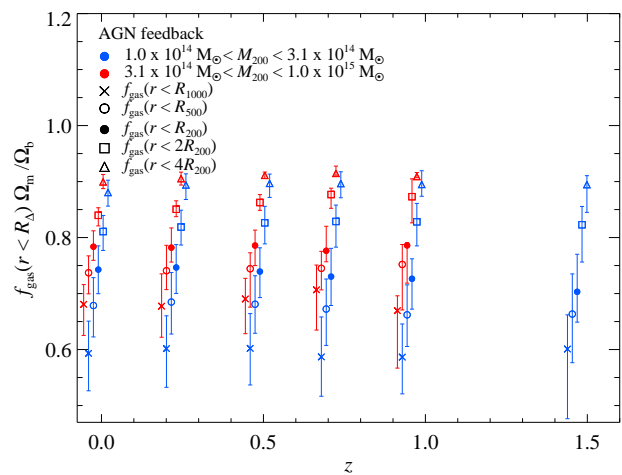}}
\caption{Redshift evolution of the gas mass fractions, $f_\rmn{gas}(<r)$, that
  are computed within different radii ($R_{1000}$, $R_{500}$, $R_{200}$,
  $2\,R_{200}$, and $4\,R_{200}$) for our AGN feedback model. We contrast median
  values for groups ($1\times 10^{14}\,\rmn{M}_\odot<M_{200}< 3.1\times
  10^{14}\,\rmn{M}_\odot$, blue) and clusters ($3.1\times
  10^{14}\,\rmn{M}_\odot<M_{200}< 1\times 10^{15}\,\rmn{M}_\odot$, blue) and
  find very little redshift evolution for $f_\rmn{gas}(<r)$, particularly for
  large clusters. While $f_\rmn{gas}$ increases with increasing radius, the
  baryon mass fraction does not reach the cosmic mean for $r<4\,R_{200}$ since
  the stellar mass fraction is $f_{\rmn{star}}(<4R_{200})\simeq 7\%$.}
\label{fig:fgas_delta_z}
\end{figure}

First, we show the redshift evolution of the median profiles of the stellar and
gas mass fractions in our AGN feedback model at fixed cluster mass, $1.4\times
10^{14}\,\rmn{M}_\odot< M_{200} < 2.6\times 10^{14}\,\rmn{M}_\odot$
(Figure~\ref{fig:fgas_z}). To address the cluster-to-cluster scatter, we
additionally show the redshift evolution of the median and percentiles
containing 68.3\% probability centered on the median (corresponding to
1-$\sigma$ intervals) for small and big clusters separated at the characteristic
mass of $M_{200} = 3.1\times 10^{14}\,\rmn{M}_\odot$
(Figure~\ref{fig:fgas_delta_z}).

For a given cluster-mass bin and with increasing redshift, we are probing
clusters with a higher mass accretion rate on average, i.e., dynamically younger
systems. Back to $z=1$, there is almost no evolution in these mass fractions,
especially for bigger clusters and larger radii. The constancy of
$f_{\rmn{star}}$ implies that the integral of the galaxy mass function increases
at the same rate as the DM mass, which is realized if most of the stellar mass
is already in place by the time it assembles in the cluster halo (since we do
not account for stellar mass loss in our simulations).

How does this finding compare to observations?  Optical observations find an
abundance of red elliptical galaxies within clusters that is the foundation for
the red sequence cluster finding algorithm \citep{2000AJ....120.2148G}. This is
the standard cluster finding method in the optical for $z \leq 1$ and has been
extended to higher redshifts by the Spitzer Adaptation of the Red-sequence
Cluster Survey \citep[SpARCS,][]{2009ApJ...698.1934M} and photometric cluster
finding techniques such as the IRAC Shallow and Deep Cluster Surveys \citep[ISCS
and IDCS,][]{2006ApJ...651..791B, 2012ApJ...761..141M}. In the literature there
is a controversy about the faint-end slope of the red sequence, which translates
into an uncertainty of the built-up of stellar mass in observations across
cosmic time. Some groups find a constant faint-end slope of the red sequence
\citep[e.g.,][]{1998ApJ...492..461S, 2006A&A...448..447A, 2008MNRAS.386.1045A,
  2013arXiv1307.1592D} after accounting for surface brightness selection on red
sequence luminosity function while others observe a decrease of the faint end of
the red sequence \citep{2004ApJ...610L..77D, 2007MNRAS.374..809D,
  2004AJ....128.2677T, 2007ApJ...661...95S, 2008ApJ...673..742G,
  2009ApJ...700.1559R, 2010ApJ...716.1503P, 2011MNRAS.412..246V,
  2012ApJ...755...14R}. Those authors infer an evolving faint end of the red
sequence, which they interpret as a built-up of stellar mass in the red sequence
with a stellar mass growth by factor of two between $z=1.6$ down to 0.6 and
another factor of two since then. This agrees on average with the mass assembly
history of cluster halos after accounting for stellar mass loss.

In our simplified star formation models, we do not account for stellar mass
loss, which can return up to $30\%- 50\%$ to the ambient medium, depending on
the initial mass function. We emphasize that even if our stellar mass
prescription and hence $f_{\rmn{star}}$ may be off by 50\% due to the effect of
stellar mass loss or limiting resolution, this would bias the gas mass fractions
by a factor of at most $\Delta f_{\rmn{star}}f_{\rmn{star}}
/f_{\rmn{gas}}\lesssim0.2$. However, this factor will be substantially reduced
in practice because a larger (smaller) amount of formed stellar mass is partly
balanced by an increased (decreased) inflow of gas from larger radii to maintain
the approximate pressure balance due to the increased gas mass loss to
stars. This renders our results with respect to $f_{\rmn{gas}}$ quite robust.

We demonstrate in Figure~\ref{fig:fgas_delta_z} that there is weak evolution of
$f_{\rmn{gas}}$ (at the level of 1-$\sigma$) for our small clusters at $R_{200}$
which disappears for larger radii or larger clusters. In particular for clusters
with $M_{200} > 3.1\times 10^{14}\,\rmn{M}_\odot$, we see no evolution of the
median $f_\rmn{gas}$ already for radii as small as $R_{500}$. At $z=1.5$, both
$f_{\rmn{star}}$ and $f_{\rmn{gas}}$ are increased towards the center indicating
that those redshifts probe the tail end of the epoch of our AGN feedback which
is most efficient at $z\gtrsim 2$ in blowing gas out (Figure~\ref{fig:fgas_z},
see also \citealt{2011MNRAS.412.1965M}). At $z>1.5$, we run out of statistics
for the mass range under consideration. However, this interpretation is
confirmed by looking at a smaller cluster mass range at an even higher redshift
which shows a further increase in the stellar mass fraction.

This finding is also consistent with the amount of energy injected by AGN
feedback over the simulation time. Approximately one third of total energy is
delivered in the cluster formation phases at $z >2$ (analogous to high-$z$
quasar-like feedback), another third in the redshift range $1<z<2$, and the
final third below $z=1$ (analogous to jet/bubble like feedback). These fractions
depend moderately on the numerical resolution; increasing the resolution
enables to resolve the growth of smaller halos at earlier times and causes a
higher fraction of energy injection at higher redshifts \citep[see][for a
discussion]{{2010ApJ...725...91B}}.

Despite the almost constant evolution of the gas fractions for $r\gtrsim
R_{200}$, there is positive evolution of the biased gas mass fraction
$f_{\rmn{gas,HSE}}$ at these radii due to the evolution of the
kinetic-to-thermal pressure ratio when radii are scaled to $R_{200}$
(Figure~\ref{fig:fgas_z}). Most of this apparent evolution can be absorbed by
redefining $R_{200}$ as the radius at which the mean interior density equals
$\Delta$ times the {\em mean density} of the universe \citepalias[see Figure~21
in][]{2012ApJ...758...74B}. However, this rescaling would then imply a redshift
evolution of $f_{\rmn{gas}}$ with smaller values of $f_{\rmn{gas}}$ at higher
redshifts.

\subsection{Comparison to data}
\label{sec:data}

\begin{figure*}
  \begin{minipage}[t]{0.49\hsize}
    \centering{\small Data comparison of radial $f_\rmn{gas}$ profiles:}
  \end{minipage}
  \begin{minipage}[t]{0.49\hsize}
    \centering{\small Data comparison of enclosed mass fractions:}
  \end{minipage}
  \centering{
    \resizebox{0.49\hsize}{!}{\includegraphics{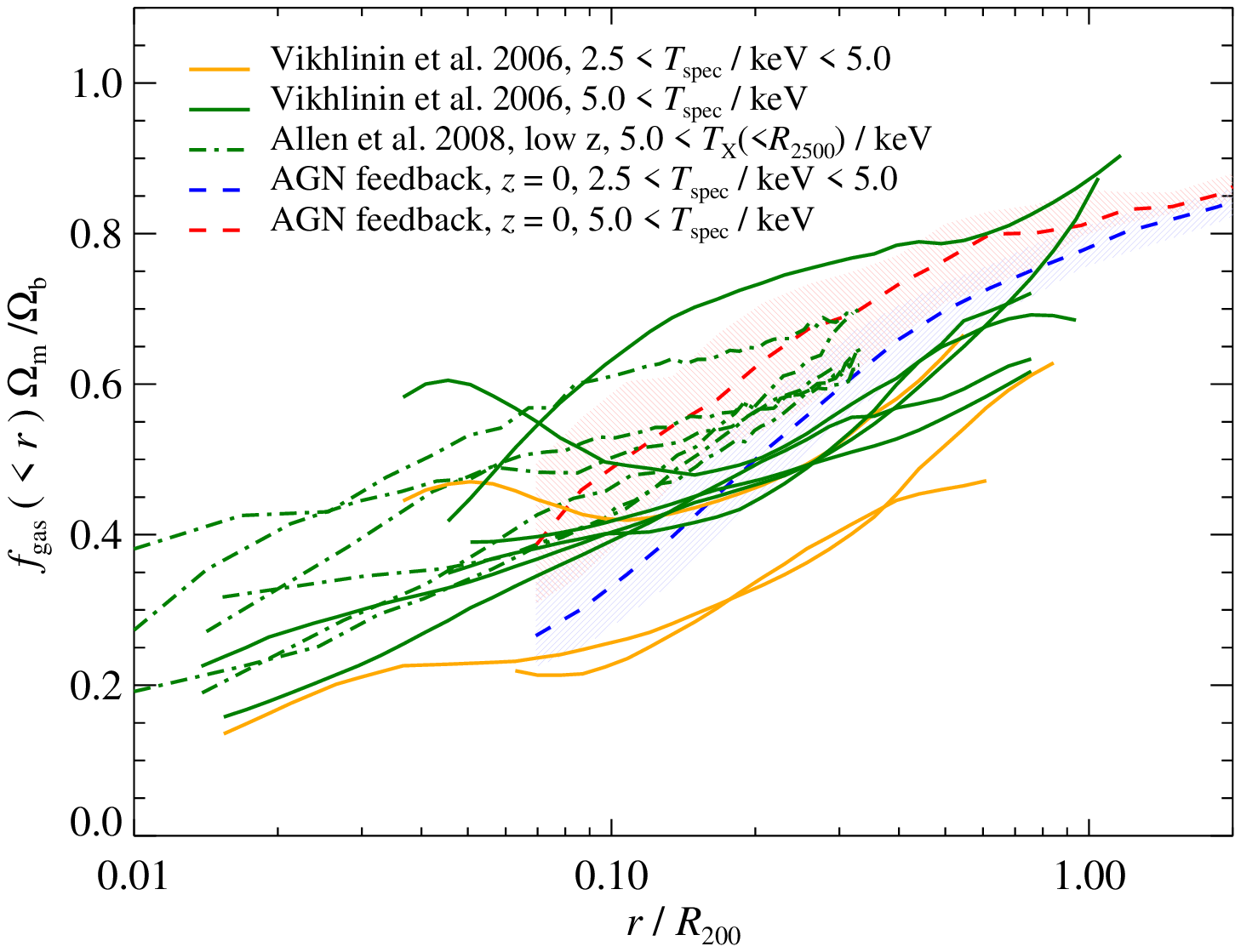}}%
    \resizebox{0.49\hsize}{!}{\includegraphics{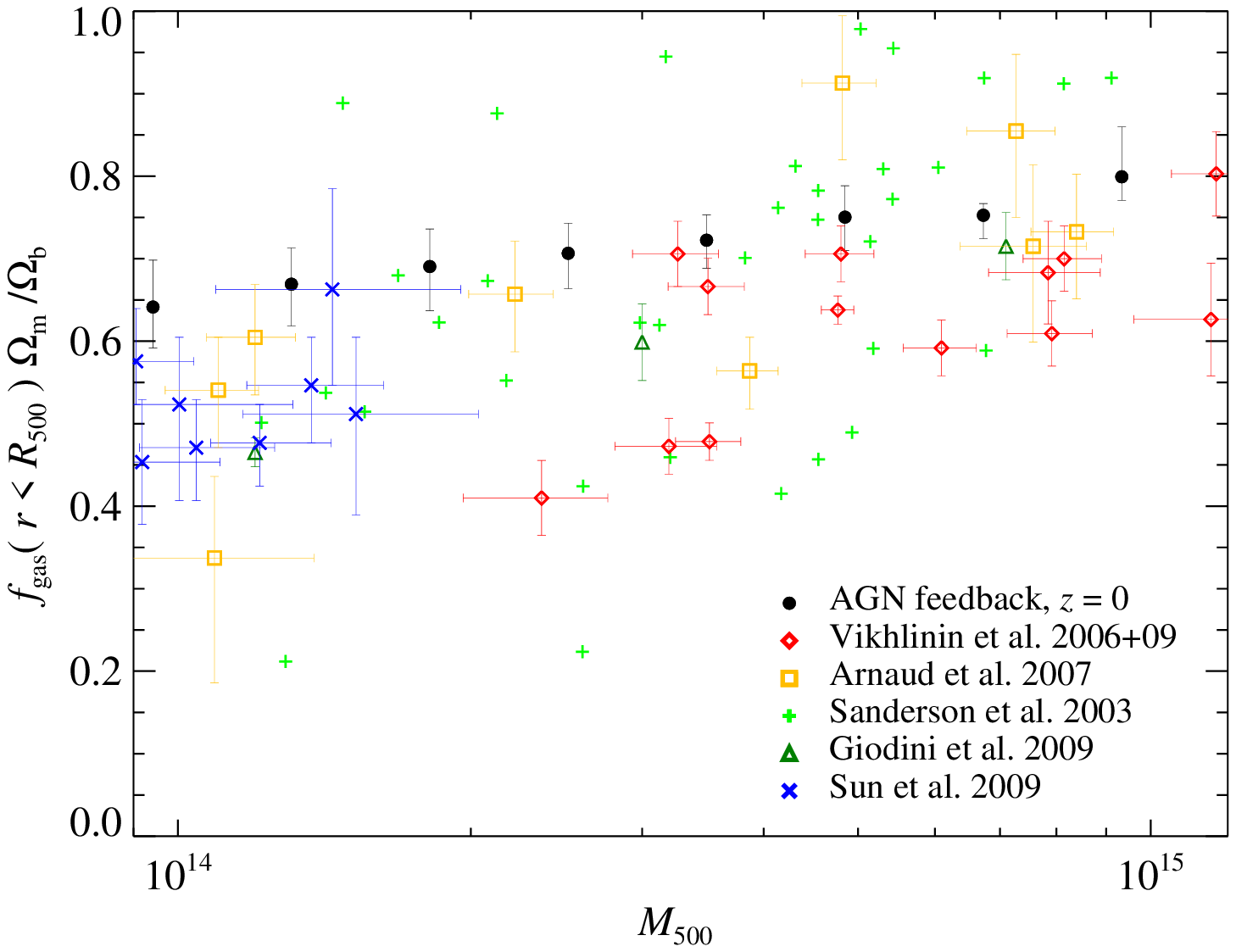}}}\\
  \centering{
    \resizebox{0.49\hsize}{!}{\includegraphics{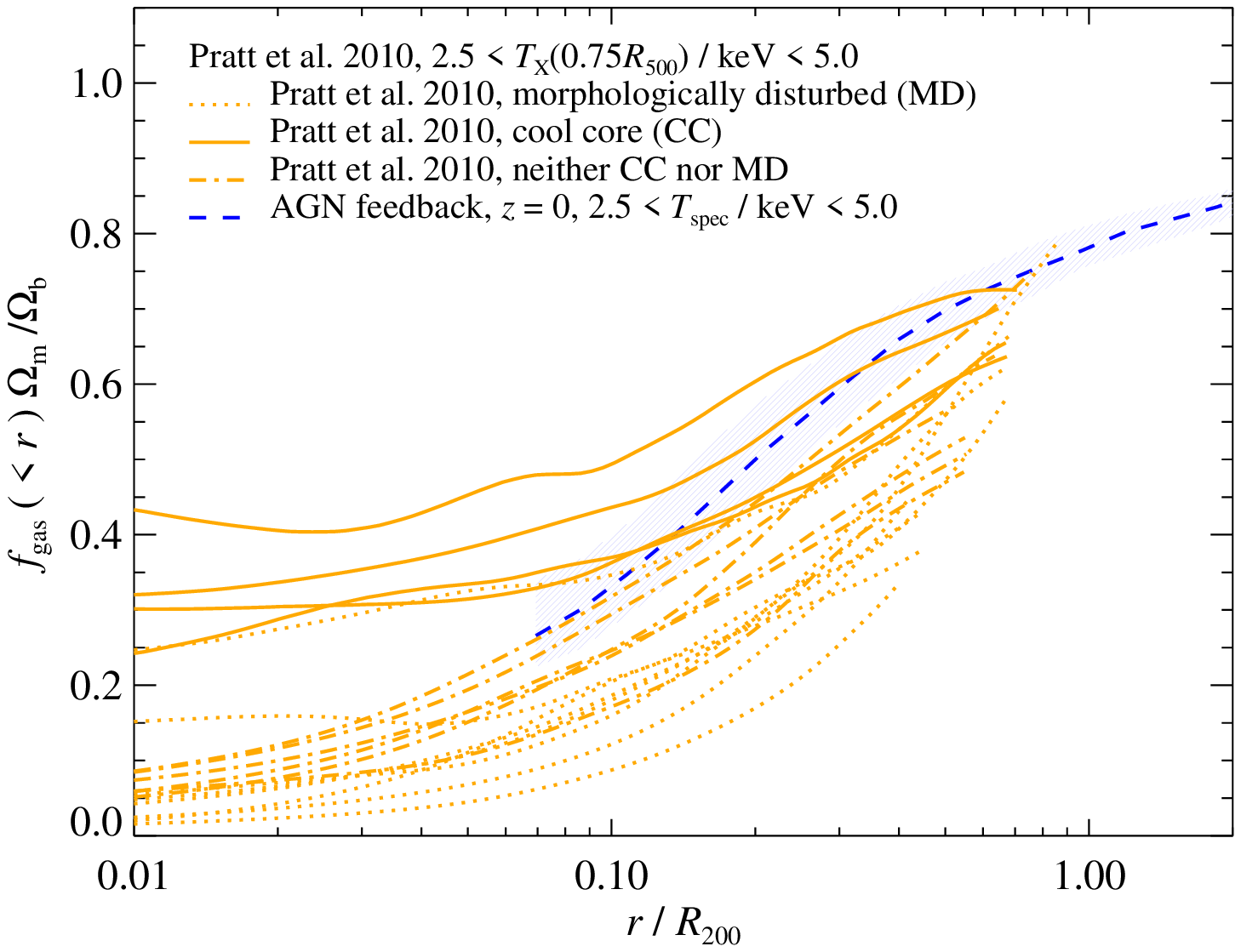}}%
    \resizebox{0.49\hsize}{!}{\includegraphics{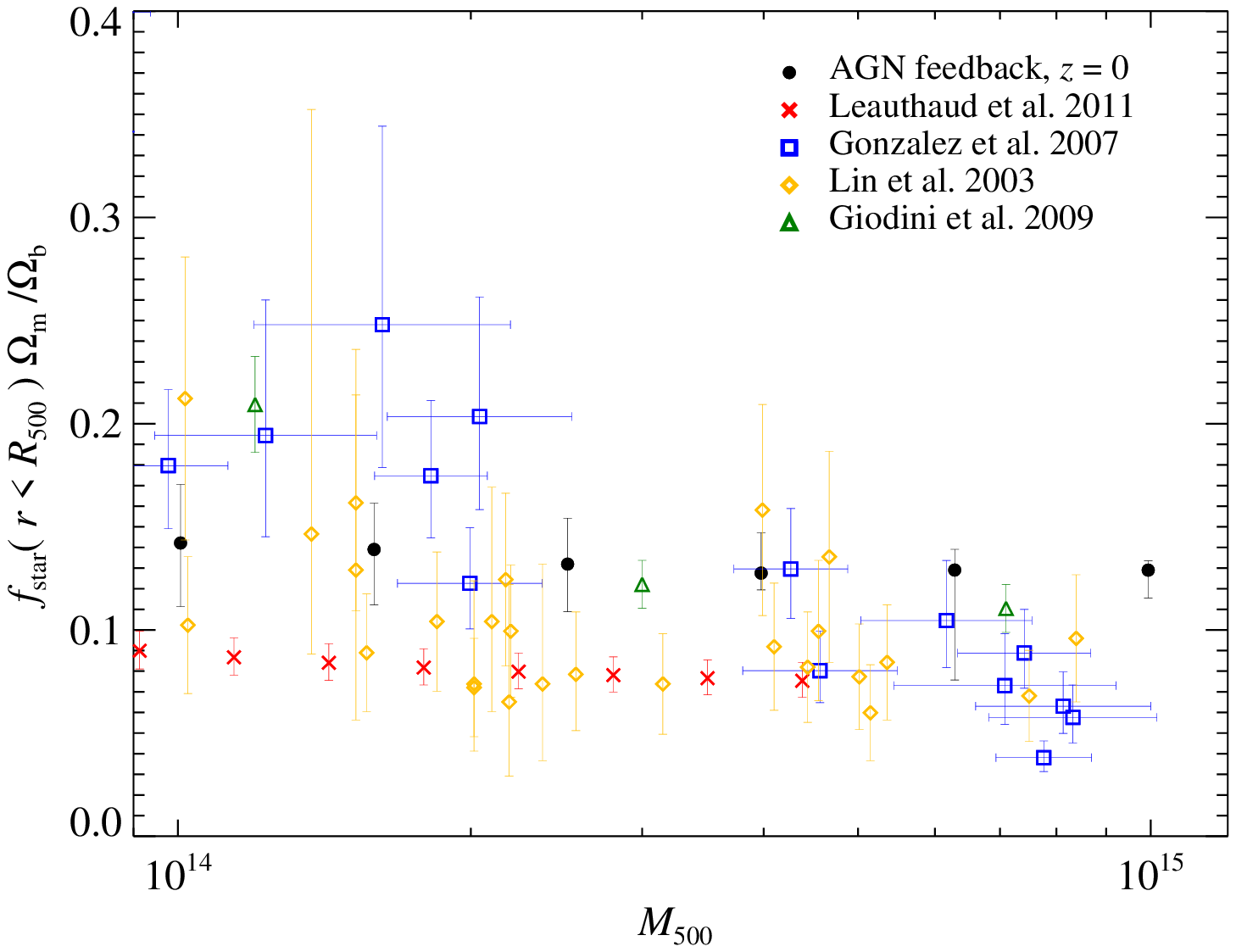}}}\\
  \centering{
    \resizebox{0.49\hsize}{!}{\includegraphics{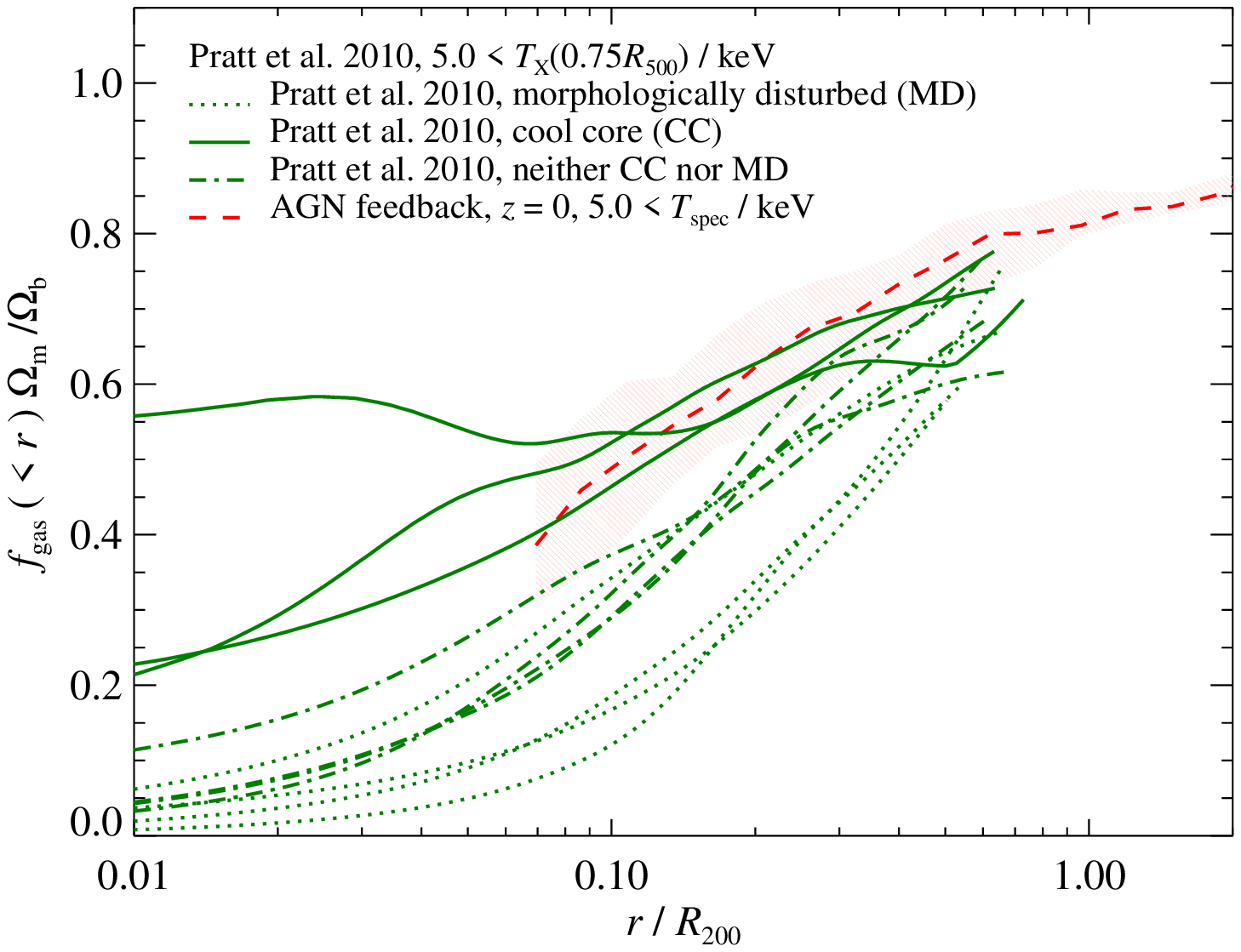}}%
    \resizebox{0.49\hsize}{!}{\includegraphics{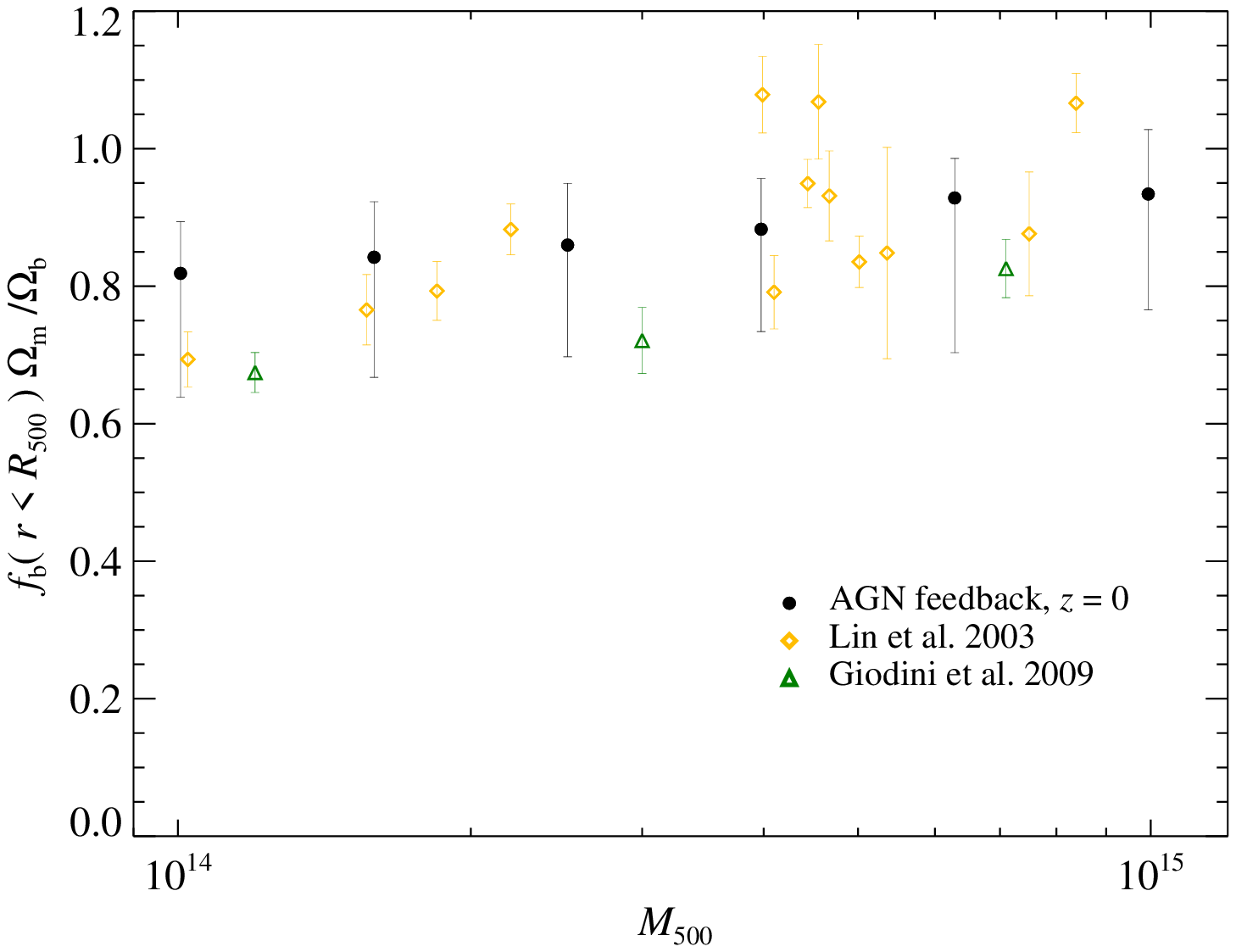}}}\\
  \caption{Comparison of our AGN feedback model with data. In the left panels,
    we compare radial $f_\rmn{gas}$ profiles of two cluster temperature bins
    that are separated by spectroscopic temperature $T_{\rmn{spec}}=5$~keV
    (using the $T_{\rmn{spec}}-M_{500}$ relation by \citet{2009ApJ...692.1033V}
    for translating our $M_{500}$ values). The top panel shows Chandra data from
    \citet{2006ApJ...640..691V} and \citet{2008MNRAS.383..879A}. The middle and
    bottom panels shows the {\em XMM-Newton} data of the REXCESS cluster sample
    \citep{2010A&A...511A..85P}, separated by spectroscopic temperature
    $T_{\rmn{spec}}=5$~keV, and differentiated by dynamical state.  Right
    panels: we compare the cluster mass dependence of enclosed gas mass
    fractions (top), stellar mass fractions (middle), and consistent treatments
    of the baryon mass fractions, i.e., $f_b$ obtained from the same objects,
    respectively (bottom). We use $f_\rmn{gas}$ data by
    \citet{2003MNRAS.340..989S, 2006ApJ...640..691V, 2009ApJ...692.1033V,
      2007A&A...474L..37A, 2009ApJ...703..982G, 2009ApJ...693.1142S} and
    $f_\rmn{star}$ data by \citet{2003ApJ...591..749L, 2007ApJ...666..147G,
      2009ApJ...703..982G, 2011ApJ...738...45L}. Overall, our AGN feedback model
    successfully reproduces observed values and trends, albeit with slightly
    higher values of $f_\rmn{gas}$ at the mass scales of groups. While the
    scatter of our simulated distribution is noticeably smaller, we caution that
    we did not include measurement biases that would substantially add to the
    simulated scatter.}
\label{fig:data}
\end{figure*}

In Figure~\ref{fig:data}, we compare our AGN feedback model with data. We
compare radial $f_\rmn{gas}$ profiles of clusters that are sorted in two bins
according to their spectroscopic temperatures, $2.5<
kT_{\rmn{spec}}/\rmn{keV}<5.0$ and $kT_{\rmn{spec}}> 5.0\,\rmn{keV}$. We use the
$T_{\rmn{spec}}-M_{500}$ relation by \citet{2009ApJ...692.1033V} to translate
our $M_{500}$ values to spectroscopic temperatures. We adopt a 15\% correction
to the X-ray-inferred $M_\rmn{HSE}$ estimates \citep{2006ApJ...650..128K}, which
is valid for the respective observational sample selection criterion. Thus, the
spectroscopic temperatures 2.5 keV and 5 keV correspond to $\sim 1.7\times
10^{14}\, \mathrm{M}_{\odot}$ and $\sim 5.0 \times 10^{14}\,
\mathrm{M}_{\odot}$, respectively.

The left panels of Figure~\ref{fig:data} show gas mass fractions from individual
X-ray cluster observations that combine surface brightness and temperature maps
of deep {\em Chandra} \citep{2006ApJ...640..691V, 2008MNRAS.383..879A} and {\em
  XMM-Newton} data of a representative sample of nearby X-ray galaxy clusters
\citep[REXCESS,][]{2010A&A...511A..85P}.\footnote{We separate the {\em
    XMM-Newton} clusters additionally into cool core and morphologically
  disturbed systems. The cool core definition is based on scaled central
  density, and the morphologically disturbed definition is based on centroid
  shift, $\bra w\ket$, following \citet{2009A&A...498..361P}. The two clusters
  that are classified as both cool core and morphologically disturbed systems,
  are presented as being morphologically disturbed.}  There is generally good
agreement of our model clusters populating the high-temperature bin, especially
with the {\em Chandra} data. In contrast, our simulated small-mass clusters tend
to have on average slightly too high $f_\rmn{gas}$ values. However, these {\em
  Chandra} data represent a biased sample since \citet{2006ApJ...640..691V} and
\citet{2008MNRAS.383..879A} targeted extremely relaxed cool core clusters for
their mass measurements. REXCESS, on the other hand, should be unbiased for
X-ray selected systems. Indeed, our model $f_\rmn{gas}$ profiles compare
similarly well to the REXCESS cool core systems while we overproduce the
apparent $f_\rmn{gas}$ estimates in morphologically disturbed systems
(independent of cluster mass respectively spectroscopic temperature). It is
unclear whether the cause for the discrepancy is on the data or simulation
side. There is the possibility that our simulations fail to reproduce the
non-cool core systems while they apparently match the $f_\rmn{gas}$
characteristics of cool core systems. However, statistically, we find no
evidence that the true $f_\rmn{gas}$ is considerably changed as a result of the
cluster's dynamical state (Figure~\ref{fig:fgas}). On the other side, it remains
to be seen whether spherical profiles of these inherently non-spherical merging
systems bias the derived values for $f_\rmn{gas}$ low.

\begin{figure*}
  \centering{
    \resizebox{0.49\hsize}{!}{\includegraphics{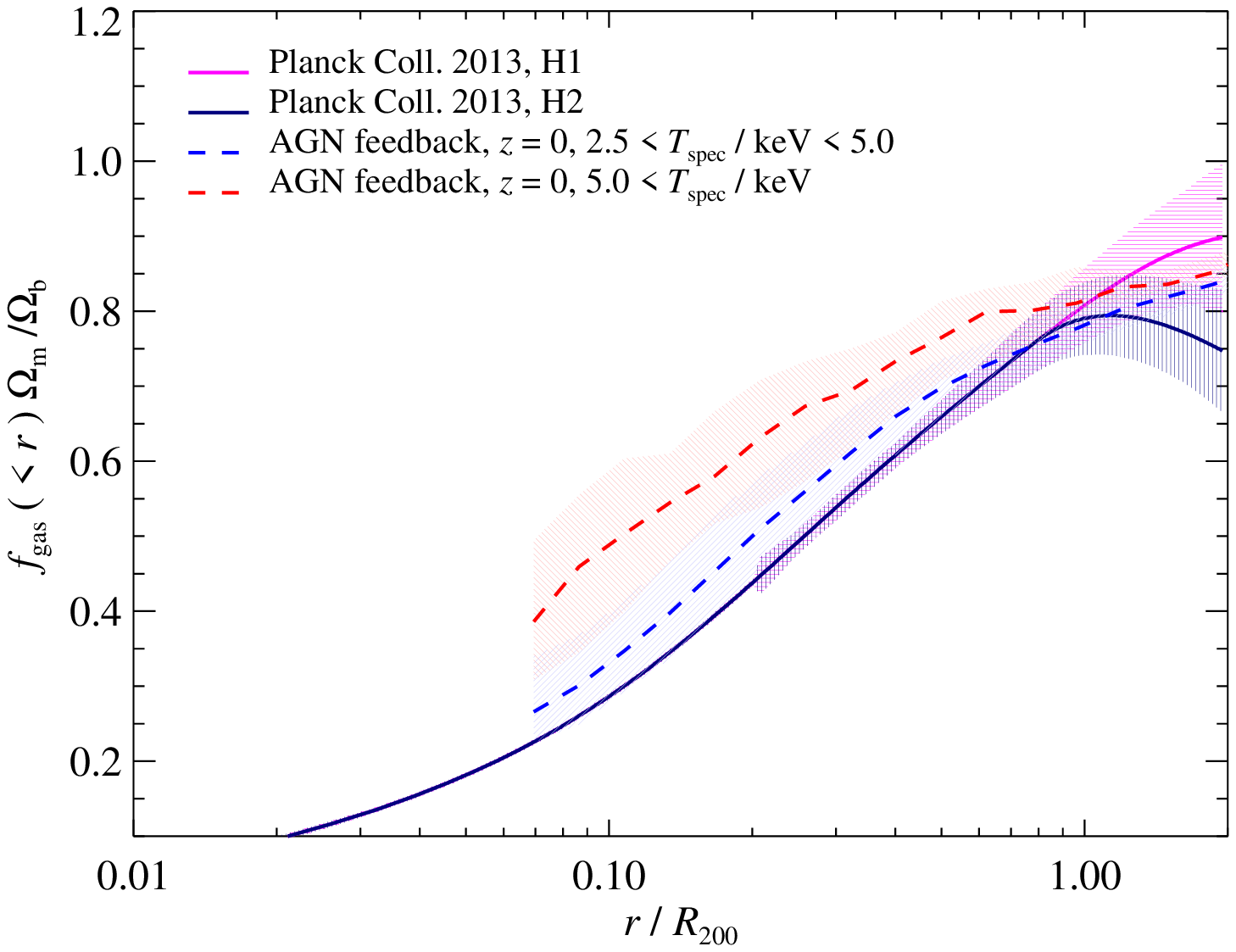}}%
    \resizebox{0.49\hsize}{!}{\includegraphics{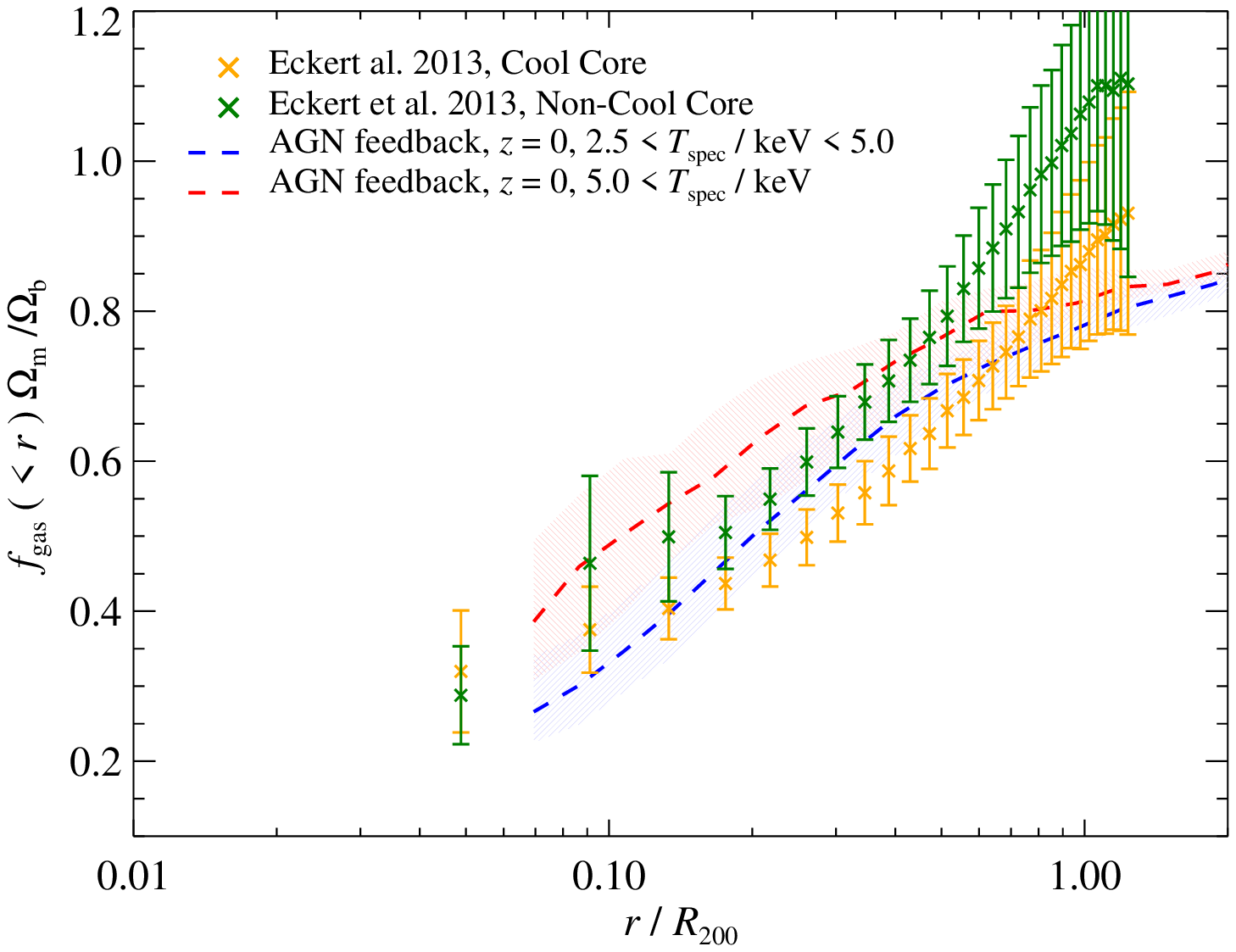}}}
  \caption{Comparison of radial $f_\rmn{gas}$ profiles in our AGN feedback
      model and data that combine different X-ray and {\em Planck}
      Sunyaev-Zel'dovich observations of massive clusters
      ($kT_{\rmn{spec}}\gtrsim5~$keV). Left panel: the gas density profile is
      obtained by combining the pressure profile derived from {\em Planck} data
      with spectroscopic temperature profiles interior to $R_{500}$ from {\em
        XMM-Newton} data \citep{2013A&A...550A.131P}.  Two differing hypotheses
      (H1 and H2) have been adopted to extrapolate those beyond $R_{500}$,
      bounding the plausible range. H1 extrapolates the average best fit model
      across the sample from X-ray spectroscopy and H2 assumes a constant
      temperature beyond $R_{500}$ that is equal to the average temperature
      measured in the last radial bin across the sample. Enclosed gas fractions
      are derived by dividing $M_{\rmn{gas}}(<r)$ with the total enclosed mass,
      assuming a generalized NFW model and a $M_{500}-Y_X$ scaling
      relation. Right panel: $f_\rmn{gas}$ is derived by combining {\em
        Planck}-inferred pressure profiles and {\em ROSAT} X-ray data
      \citep{2013A&A...551A..23E} while assuming hydrostatic equilibrium and
      neglecting the density clumping bias. }
\label{fig:data2}
\end{figure*}

In Figure~\ref{fig:data2}, we compare our models with $f_\rmn{gas}$ estimates
from the combination of {\em Planck} Sunyaev-Zel'dovich data and different X-ray
observations of massive clusters ($kT_{\rmn{spec}}\gtrsim5~$keV). Depending on
the availability of X-ray spectra, different approaches have been pursued in the
literature. The {\em Planck} collaboration used {\em XMM-Newton} data interior
to $R_{500}$ with spectroscopic temperature profiles \citep{2013A&A...550A.131P}
and extrapolated these beyond $R_{500}$ while assuming two hypotheses that are
likely bounding the true profile.\footnote{Note that we use an updated version
  of $f_\rmn{gas}(<r)$ as presented by the \citet{Planck_erratum}.}  In
combination with deprojected {\em Planck} pressure profiles, this yields gas
density profiles and---through integration---$M_{\rmn{gas}}(<r)$. The total mass
profile is assumed to follow a generic NFW model \citep{1997ApJ...490..493N},
which is normalised to $M_{500}$ from the $M_{500}-Y_X$ scaling relation
\citep{2007A&A...474L..37A,2010A&A...517A..92A}. The resulting $f_\rmn{gas}$
profile is somewhat lower than that of our AGN simulations for $r<R_{500}$ while
it agrees with the simulation model at larger radii. Since our simulated
pressure profiles match the observed ones \citep{2013A&A...550A.131P}, the
reason for the discrepancy could either lie in the assumptions used to derive
$f_\rmn{gas}$ or that our simulations yield too high values for $f_\rmn{gas}$.
Possibilities for observational biases include incomplete accounting for
multi-temperature phases in cluster outskirts and biases of the total mass
profile. $M_{\rmn{tot}}$ could be biased by neglecting the substantial scatter
among individual DM density profiles as well as not accounting for the
covariances between the different parameters adopted for the DM profile such as
mass-concentration and $M_{500}-Y_X$ scaling relation.

Using {\em ROSAT} data (that lacks spectroscopic X-ray information) of a sample
of 18 clusters that also has {\em Planck} Sunyaev-Zel'dovich data,
\citet{2013A&A...551A..23E} follow a different methodology. This uses the
hydrostatic equilibrium mass,
\begin{equation}
  \label{eq:MHSE2}
  M_{\rmn{HSE}}(<r) = -\frac{r^2}{G\rho_{\rmn{clump}}}\frac{d P_{\rmn{th}}}{d r},
\end{equation}
where $\rho_{\rmn{clump}}$ derives from the X-ray emission measure and is
subject to residual density clumping (that escaped the masking in the X-ray
analysis). The thermal pressure gradient is obtained from the {\em Planck}
best-fit generalized NFW models for the pressure, which does not depend on
clumping. The gas mass fraction is derived via $f_{\rmn{gas}}(<r) =
M_{\rmn{gas,clump}}(<r)/M_{\rmn{HSE}}(<r)$, where $M_{\rmn{gas,clump}}(<r)$ is
given by Equation~(\ref{eq:M_clump}). Hence, the gas mass fraction obtained by
this method scales approximately with the (residual) clumping factor,
\begin{equation}
  \label{eq:SZ_fgas_bias}
  f_{\rmn{gas}}(<r)\propto
  \sqrt{C_{2,\rho}(r)} \int_0^r \sqrt{C_{2,\rho}(r')} \rho_{\rmn{true}}(r')r'^2dr',
\end{equation}
rather than the square root as in the case of pure X-ray based analysis (where
the pseudo pressure scales also with the residual $\sqrt{C_{2,\rho}}$, canceling
the clumping factor dependence in $M_{\rmn{HSE}}$). Additionally,
$f_{\rmn{gas}}$ is biased high owing to the hydrostatic mass bias, i.e. by
neglecting the kinetic pressure support (see Figure~\ref{fig:fgas} or
\citet{2006MNRAS.369.2013R, 2011A&A...529A..17V, 2012ApJ...751..121N}) and the
non-thermal pressure contributed by a potential relativistic particle population
in cluster outskirts \citep{2007MNRAS.378..385P, 2010MNRAS.409..449P,
  2012MNRAS.421.3375V}. As before for the individual {\em Chandra} and {\em
  XMM-Newton} clusters, our simulation models provide a good match to the median
of six cool core systems by \citet{2013A&A...551A..23E}. However, their median
profile of twelve non-cool core clusters shows larger values than predicted by
our simulations for radii $\gtrsim0.7\,R_{200}$. This is especially remarkable
since there is a substantial discrepancy between the non-cool core systems in
the sample by \citet{2013A&A...551A..23E} and the morphologically disturbed
systems of \citet{2010A&A...511A..85P}, some of which may be explained by the
different dependencies on the clumping factor as discussed here. We caution that
more careful X-ray mocks (exactly quantifying the various biases discussed here)
are needed to draw final conclusions regarding the direct comparison of our
simulated models with observations.

In the right-hand panels of Figure~\ref{fig:data}, we compare the cluster mass
dependence of enclosed gas mass fractions (top), stellar mass fractions
(middle), and baryon mass fractions (bottom). Note that we only use those
compilations of $f_b$ versus mass where the gas and stellar masses have been
obtained from the same clusters, respectively. Overall, our AGN feedback model
successfully reproduces observed values and trends, albeit with slightly higher
values of $f_\rmn{gas}$ (and to a lesser extent $f_b$) at the mass scales of
groups in comparison to X-ray observations \citep[cf.][]{2007A&A...474L..37A,
  2009ApJ...693.1142S}. While the scatter of our simulated distribution is
noticeably smaller, we caution that we did not include measurement biases that
would substantially add to the simulated scatter as demonstrated in
Section~\ref{sec:biases}.

There could be several explanations for an overshooting of our simulated
$f_\rmn{gas}$ values at group scales. Those could include numerical resolution,
shortcomings of the AGN modeling, or missing physics in the
simulations. Increasing resolution would enable us to resolve the formation of
smaller objects earlier on which may allow a more vigorous blow-out of the gas
already in halos in the mass range $10^{12}\lesssim M_{200}/\rmn{M}_\odot
\lesssim10^{13}$. Those are just above the knee of the galaxy luminosity
function and hence have a decreasing efficiency of star formation per unit halos
mass in more massive halos which is usually attributed to AGN feedback
\citep[e.g.,][]{2012MNRAS.422.2816B, 2013MNRAS.428.2966P}.

As laid out in Section~\ref{sec:sims}, we use an extremely low-resolution
incarnation of AGN feedback that enables us to simulate large volumes and
cluster samples, which are needed to provide precise predictions for the next
generation X-ray and SZ surveys. Our AGN feedback prescription couples the
thermal energy feedback to the global SFR of the central cluster region. While
this model is able to arrest overcooling in massive clusters, it may be too
simplified for groups. The observed (low-frequency) radio emission that
spatially coincides with X-ray cavities suggests the presence of non-thermal
components composed of cosmic rays and magnetic fields in these lobes. Modelling
AGN feedback into cosmic ray rather than thermal energy \citep{Sijacki+2008} or
exploring kinetic jet feedback that thermalizes at larger radii
\citep{Dubois+2010} may be viable alternatives to our simulated scenario, but
would also require better spatial resolution. If cosmic rays are injected into
the ICM by the break-up of radio bubbles due to Kelvin-Helmholtz instabilities,
they can stream at the Alfv{\'e}n velocity with respect to the plasma rest frame
and heat the surrounding thermal plasma at a rate that can balance that of
radiative cooling {\em on average} (as demonstrated for the Virgo cluster),
thereby providing a physical solution to the ``cluster cooling flow problem''
\citep{1991ApJ...377..392L, 2008MNRAS.384..251G, 2013arXiv1303.5443P}.

Finally, our model may be missing important physics that could additionally add
entropy to the centers of groups, such as the recently suggested
\citep{2012ApJ...752...22B,2012ApJ...752...23C,2012ApJ...752...24P,2012MNRAS.423..149P}
blazar heating model that proposes the bolometric luminosity of TeV blazars as
an additional heating source of the inter-galactic medium. Phenomenologically,
highly biased regions that first turn-around to form groups and clusters should
be heated first. This would imply an evolving pre-heated entropy floor and cause
an additional decrease of $f_\rmn{gas}$ in late-forming groups that are observed
on a sufficiently short time scale after formation so that the central group
medium had not have time to enter the strongly cooling regime
\citep{2012ApJ...752...24P}.

\subsection{Comparison to previous work}
\label{sec:previous}

Given the importance of cluster X-ray observations for cosmological parameter
estimation, there has been a comparably large body of work concerning gas
fractions in cosmological cluster simulations. Early work on $f_b$ in
non-radiative simulations showed a strong systematic discrepancy between
different numerical methods. This is surprising since all codes should solve the
same equations for the gas and collisionless DM physics without any ambiguity in
formulating a subgrid description of star formation and associated feedback
processes. SPH simulations of ten massive clusters in a $\Lambda$CDM cosmology
derived values of $f_b\approx0.87$ and 0.83 within $R_{\rmn{vir}}$ and
$0.5R_{\rmn{vir}}\approx R_{500}$, respectively, and show only a very weak
evolution of $f_b$ from $z=1$ to 0 \citep{1998ApJ...503..569E}. A number
of further studies that all employ the energy formulation of SPH confirm these
findings \citep{Frenk+1999, 2001ApJ...555..597B, 2002MNRAS.336..527M}. In
contrast, cluster studies that employ the adaptive mesh refinement (AMR)
technique for solving the hydrodynamic equations find higher values of
$f_b\approx0.97\pm0.03$ and $0.94\pm0.03$ within $R_{\rmn{vir}}$ and
$0.5R_{\rmn{vir}}\approx R_{500}$, respectively
\citep{2005ApJ...625..588K}. Cluster simulations with the entropy-conserving SPH
code {\sc GADGET} yield values of $f_b\approx0.92-0.93$, higher by about 5
percentage points than the energy-conserving formulation of SPH but still
smaller than the AMR results \citep[see][which is also in line with our
findings]{2004MNRAS.355.1091K, 2005ApJ...625..588K}. All the different methods
find no significant redshift evolution of the baryon fraction in non-radiative
simulations.

Simulations with radiative cooling and star formation necessarily yield larger
discrepancies in the resulting $f_b$ values owing to the larger uncertainties of
implementing subgrid star formation and feedback processes. Common to all of
these approaches is that too many stars condense out of the ICM, which results
in the ``overcooling problem'', i.e., $f_\rmn{gas}$ values that are too small
and $f_\rmn{star}$ values that are too high in comparison to observations. SPH
simulations (employing the energy formulation) with radiative cooling or
feedback yield values of $f_b\approx0.85-0.9$ for the cluster mass range
considered here \citep[see Figure 3 of][]{2002MNRAS.336..527M}. For smaller
galaxy clusters (within a factor of two of mass
$M_{200}=10^{14}h^{-1}\,\rmn{M}_\odot$), entropy-conserving SPH simulations by
\citet{2004MNRAS.355.1091K} and AMR simulations by \citet{2005ApJ...625..588K}
find similar average radial profiles of $f_\rmn{gas}$. However, those results
differ for larger clusters and the AMR results obtain substantially higher
$f_\rmn{gas}$ values.  \citet{2006MNRAS.365.1021E} presented an analysis for a
sample of galaxy clusters simulated with the entropy-conserving {\sc GADGET} SPH
code that followed radiative cooling and star formation (essentially similar to
our radiative cooling model). They find total {\em baryon} fractions of
$f_b(<R_\rmn{vir})\approx 0.93-0.95$, similar to our values and those by
\citet{2005ApJ...625..588K} on cluster-mass scales. The {\em gas} fractions
found in \citet{2006MNRAS.365.1021E} and more recently in
\citet{2013MNRAS.429..323S} of $0.75-0.8$ agree with our values (for the
radiative cooling model), but are higher than the corresponding values by
\citet{2005ApJ...625..588K} because of the different implementations of cooling
and star formation.

More recently, baryon fractions have been studied in AGN feedback models.
\citet{2008ApJ...687L..53P} presented mock X-ray observations of a sample of
zoomed cluster simulations which account for radiatively cooling gas of
primordial composition, star formation, and thermal energy feedback by black
holes that grow through Bondi-Hoyle accretion. These simulations produce gas
mass fractions that match the observations and are also in agreement with our
AGN feedback results for larger clusters ($M_{500}\gtrsim2\times
10^{14}\,\rmn{M}_\odot$). However, our smaller systems appear to have slightly
larger values of $f_\rmn{gas}$ due to the coarser resolution of our cosmological
simulations in comparison to those zoomed simulations that are able to resolve
AGN feedback at earlier times in galaxies above $L^*$. In contrast, our stellar
mass fractions on cluster scales match the observed values, but are smaller by
about a factor of three in comparison to the AGN feedback simulation by
\citet{2010MNRAS.406..936P} (see their Figure 5). Improving the simulated
physics of these AGN feedback models, \citet{2013MNRAS.431.1487P} follow the
detailed chemical evolution of the IGM and self-consistently account for
metallicity-dependent radiative cooling (thus improving the simplifying
assumptions of collisional ionization equilibrium and solar relative abundances,
which were underlying earlier work by \citealt{2010MNRAS.401.1670F}). On
cluster-mass scales, they find slightly smaller gas fractions than ours and
those by \citet{2008ApJ...687L..53P}. Additionally, they have similarly large
stellar mass fractions as \citet{2010MNRAS.406..936P}, which appear to be in
conflict with the data. Common to all that body of work (including ours) is that
it estimates the black hole accretion rate and hence the amount of feedback
energy directly from simulated hydrodynamical quantities (i.e., entropy or
density) by means of a subgrid accretion model, which ensures a substantial
predictive power of this approach. In contrast, \citet{2011MNRAS.413..691Y}
derive their feedback energy from a semi-analytical model of galaxy formation
that they couple to hydrodynamical simulations. By construction, their approach
suffers less from resolution effects in comparison to ours and produces a strong
cluster-mass dependence of $f_\rmn{gas}(<R_\Delta)$ for various characteristic
cluster radii, in agreement with the observations, particularly on group
scales. However, their models appear to have problems in reproducing the larger
$f_\rmn{gas}$ values of some of the observed $f_\rmn{gas}$ profiles in large
clusters with $T_\rmn{spec}>5 \,\rmn{keV}$ \citep{2006ApJ...640..691V,
  2008MNRAS.383..879A}.

Summarizing, it appears that there are still systematic differences between
Lagrangian SPH and Eulerian AMR codes in computing baryon fraction in
non-radiative simulations. Those however are reduced to the level of about five
percentage points when AMR results are compared to those obtained with the
entropy-conserving {\sc GADGET} SPH code. Since this is the baseline model out
of which radiative cooling starts to condense cool gas and to form stars, more
work is needed to fully understand the reason of the discrepancy. Purely
radiative models, which only account for star-formation feedback, suffer from
the over-cooling problem and produce stellar and gas mass fractions in conflict
with observations. Instead, a growing body of AGN feedback models demonstrates
that the overcooling problem can be weakened or even circumvented, depending on
the detailed implementation of cooling and feedback. Moreover, since that type
of physics causes larger changes in $f_b$ and $f_\rmn{gas}$ in comparison
to the systematic differences owing to the numerical method, this holds the
promise that the simulations are starting to produce robust and predictive
results (for a certain implementation of AGN feedback) so that a comparison to
observations is a meaningful and well-posed task.


\section{Conclusions}
\label{sec:conclusions}

In this paper, we performed a study of gas and stellar mass fractions
in clusters with the goal to address (1) the bias of $f_\rmn{gas}$ measurements
in recent {\em Suzaku} X-ray measurements of the ICM beyond the virial radius
and (2) provide a theoretical framework for using gas masses as a tightly
correlated and {\em unbiased} proxy for the halo masses. This property lies at
the heart of using X-ray-inferred gas masses to derive precision cosmological
parameters and is of critical importance for SZ surveys as $f_\rmn{gas}$ sets
the normalization of the $Y-M$ scaling relation and the SZ power spectrum scales
with the square of $f_\rmn{gas}$.

\subsection{Biases in X-ray inferred gas mass fractions}

We identified two main sources of bias for X-ray inferred values of
$f_\rmn{gas}$ in addition to sample variance of clusters and angular orientation
variance within clusters; all of which can be addressed with cosmological
simulations of a large statistical sample of clusters. (1) Adopting hydrostatic
equilibrium masses, i.e., neglecting the non-thermal pressure, biases the
inferred masses low by $20-25\%$ on average (or $15-20\%$ if we considered only
a sample of the third most relaxed clusters). Hence $f_\rmn{gas}$ is biased high
by the same factor that increases dramatically for even larger radii due to the
larger relative contribution of kinetic pressure. (2) The presence of gas
density clumping biases the X-ray surface brightness and hence the inferred
density and gas masses high. Hence, the resulting values for $f_\rmn{gas}$ are
also biased high at a level of $\sim10-20\%$ within $R_{200}$ (slightly
mass-dependent and at the low-end for relaxed clusters). For the most massive
clusters with $M_{200} \gtrsim 3\times 10^{14}\,\rmn{M}_\odot$, it increases
even more steeply outside $R_{200}$ in comparison with the bias by HSE masses.

While these two processes represent an upward measurement bias of $f_\rmn{gas}$
in comparison to its cosmic value at $R_{200}$, we also consider the sample
variance of the true $f_\rmn{gas}$ as well as its biased measurement across
clusters and within clusters for different angular directions. Interestingly,
the cluster-to-cluster scatter of the true $f_{\rmn{gas}}$ decreases for larger
radii to values around 5\% at $R_{200}$. In contrast, the cluster sample
variance of the biased measurement of $f_{\rmn{gas}}$ due to the assumption of
HSE masses and neglecting density clumping remains fairly constant as a function
of radius at a level of $\sigma_{f_\rmn{gas}}/f_\rmn{gas} \simeq 0.1 - 0.2$
(depending on the dynamical state and mass of clusters) within $R_{200}$ and
even starts to increase for larger radii due to the strong clumping term.

These different measurement biases are all manifestations of two underlying
physical processes which both lead to significant angular variance of the (gas
and total) mass distributions. First, there is a strong internal
baryon-to-dark-matter density bias, a consequence of
collisional-to-collisionless physics as a result of the differing efficiencies
of gas thermalization and the relaxation processes of the DM component. Second,
hierarchical structure formation implies that cluster are the latest object that
had have time to form until now which manifests itself in a considerable
anisotropy of the mass distributions, in particular around the virial radius
where the filamentary cosmic web connects to the cluster interiors. The
accretion shock of gas that accretes along filaments forms further inwards,
leaving a larger level of clumping and kinetic pressure contribution already
around $R_{200}$. Simultaneously, this causes a considerable measure of
ellipticity (as measured by the eigenvalues of the moment-of-inertia tensor)
throughout the clusters \citepalias{2012ApJ...758...74B}.

Equivalently, computing the mass profiles in 48 cones, whose footprints
partition the sphere, we find anisotropic gas and total mass distributions that
imply an angular variance of $f_{\rmn{gas}}$ at the level of 30\% (almost
independent of radius and cluster mass).  Since the angular variance as well as
the outliers of the gas and DM mass distribution in angular cones is larger than
the angular variance of $f_{\rmn{gas}}$, this implies spatial correlations of
the DM and gas distribution through self-bound substructures that are still
holding on to (some) of their gas, especially outside $R_{200}$. We also
demonstrated that in the most extreme cases, the cone profiles of
$f_{\rmn{gas}}$ can be biased high by a factor of two in massive clusters
($M_{200}\sim10^{15}\,\rmn{M}_\odot$). This factor is lowered by considering
projection effects. The 3D angular variance drops from $30-35\%$ at $R_{200}$
down to $\sim15-20\%$ in case of the projected (2D) variance. However,
hydrostatic mass bias as well as density clumping may (partially) compensate for
these projection effects.  If the anisotropic mass distribution is indeed
responsible for some of the high $f_\rmn{gas}$ values inferred from {\em Suzaku}
measurements along individual radial arms, it predicts a significantly lower
level of $f_\rmn{gas}$ along other directions. This prediction can be tested by
increasing the number of radial arms in a given cluster.

We caution that the magnitude of the angular variance of the mass distributions
is not independent of the cluster variance, and hence is not additive. Instead,
both variances are complementary ways of characterizing the non-equilibrium
processes in clusters (anisotropy, kinetic pressure contribution, clumping) that
become increasingly complicated at the virial radius and beyond
\citepalias{BBPS4}. We also emphasize that the distributions of mass profiles in
angular cones is positively skewed (there are more extreme positive outliers in
$M_\rmn{gas}$ and $M_\rmn{tot}$ cone profiles) which is expected from the
presence of substructures. In particular, the outliers with high values of the
mass profiles in individual cones likely correspond to filaments that channel
the accreting substructures to the cluster \citepalias{BBPS4}.

Alternatively the increased level of X-ray surface brightness towards the virial
region may be due to instrumental or observational effects, such as of
unresolved X-ray point sources, point-spread function leakage from masked point
sources, or stray light from the bright inner cluster core.

\subsection{Gas and stellar mass fractions in cosmological simulations} 

We find little redshift dependence of the radial profile of $f_\rmn{gas}$ in our
AGN feedback model, if radii are scaled to $R_{200}$. In particular, this is the
case for big clusters with $M_{200} \gtrsim 3\times 10^{14}\,\rmn{M}_\odot$
or for larger containment radii $r\gtrsim R_{500}$, or for both. The redshift
independence of $f_\rmn{gas}$ depends critically on the ability to scale the
radial variable with $R_{200}$ as the $f_\rmn{gas}$ profile increases steadily
with radius. If the redshift-dependent systematics of inferring $R_{200}$ can be
controlled, this provides a good theoretical basis of using the gas mass
fraction to do precision cosmology. The physical reason behind this argument is
the deep gravitational potential of big clusters ($M_{200} \gtrsim 3\times
10^{14}\,\rmn{M}_\odot$). This makes it impossible for AGNs to expel gas today,
in particular as the bubble enthalpy is poorly coupled to the thermal energy of
the surrounding ICM which results in little heating and entropy generation of
the central ICM on a buoyancy time scale \citep{2012ApJ...752...24P}. Moreover,
gas that was pushed beyond the virial radius by AGN feedback early-on gets
pulled in during assembly of these big systems.

Similarly, we find almost no evolution of $f_{\rmn{star}}$ back to $z=1$ in our
AGN feedback model. This implies that the stellar mass increases at the same
rate as the DM mass, which is realized if most of the stellar mass is already in
place by the time it assembles in the cluster halo (since we do not account for
stellar mass loss in our simulations). This appears to agree with observations
of the stellar mass assembly in clusters, although there is considerable
uncertainty about the faint-end slope of the red sequence of cluster galaxies.

Cluster mergers may transfer energy from the DM component to the gas,
potentially reducing $f_\rmn{gas}$ at the virial radius from its cosmic
value. While our entropy-conserving SPH simulations suggest a value of
$f_\rmn{gas}\simeq 0.93$ when scaled with the cosmic value, AMR simulations find
somewhat larger values of $f_\rmn{gas}\simeq 0.97$
\citep{2005ApJ...625..588K}. Radiative physics furthermore reduces $f_\rmn{gas}$
by locking baryons up into stars, causing a steeper slope of $f_\rmn{gas}$
towards the center. AGN feedback (or its quasar-like appearance) is most
effective early-on during the assembly time of groups at $z\gtrsim2$ and expels
gas far beyond $R_{200}$ \citep{2011MNRAS.412.1965M}. This causes a much steeper
dependence of $f_\rmn{gas}$ on cluster mass. At the same time, AGN feedback (in
the incarnation that we modeled) is able to arrest overcooling and to drop the
stellar mass by a factor of two.\footnote{This suppression factor would be
  larger if we had run at higher resolution since models without AGN feedback do
  not produce numerically converged results (at the resolution that we could
  afford running our cosmological boxes).} As a result, our AGN feedback model
keeps $f_\rmn{star}$ almost constant as a function of redshift (for
$z\lesssim1$) and mass, in particular for larger systems. This together implies
a significantly increasing baryon fraction, $f_\rmn{b}$, with cluster mass
from 0.85 (at $M_{200} \simeq 10^{14}\,\rmn{M}_\odot$) to 0.94 (at $M_{200}
\simeq 10^{15}\,\rmn{M}_\odot$). Those scalings of $f_\rmn{gas}$ and
$f_\rmn{b}$ need to be taken into account (e.g., through a marginalization
process) when using X-ray and SZ surveys of cluster to constrain cosmological
parameters. We emphasize that AGN feedback produces large effects on various
thermodynamic quantities, including significant changes of stellar and gas mass
fractions. While the qualitative results of our AGN feedback model are in line
with approaches by other groups and produces robust predictions of the
interesting parameter space preferred by observations, the detailed scalings and
profiles may, however, depend on the particular model realization and could vary
for different implementations.

\acknowledgments 

We thank Volker Springel for enlightening discussions and careful reading the
manuscript and thank Stefano Andreon, Dominique Eckert, Daisuke Nagai, Etienne
Pointecouteau, Gabriel Pratt, and Greg Rudnick, Simon D.M. White for useful
discussions. We thank our referee for providing an insightful report.
C.P. gratefully acknowledges financial support of the Klaus Tschira
Foundation. Research in Canada is supported by NSERC and CIFAR. Simulations were
run on SCINET and CITA's Sunnyvale high-performance computing clusters. SCINET
is funded and supported by CFI, NSERC, Ontario, ORF-RE and UofT deans.  We also
thank KITP for their hospitality during the 2011 galaxy cluster workshop. KITP
is supported by National Science Foundation under Grant No. NSF PHY05-51164.

\bibliography{bibtex/nab}

\begin{thebibliography}{141}
\expandafter\ifx\csname natexlab\endcsname\relax\def\natexlab#1{#1}\fi

\bibitem[{{Akamatsu} {et~al.}(2011){Akamatsu}, {Hoshino}, {Ishisaki}, {Ohashi},
  {Sato}, {Takei}, \& {Ota}}]{2011PASJ...63S1019A}
{Akamatsu}, H., {Hoshino}, A., {Ishisaki}, Y., {Ohashi}, T., {Sato}, K.,
  {Takei}, Y., \& {Ota}, N. 2011, \pasj, 63, 1019

\bibitem[{{Allen} {et~al.}(2008){Allen}, {Rapetti}, {Schmidt}, {Ebeling},
  {Morris}, \& {Fabian}}]{2008MNRAS.383..879A}
{Allen}, S.~W., {Rapetti}, D.~A., {Schmidt}, R.~W., {Ebeling}, H., {Morris},
  R.~G., \& {Fabian}, A.~C. 2008, \mnras, 383, 879

\bibitem[{{Allen} {et~al.}(2004){Allen}, {Schmidt}, {Ebeling}, {Fabian}, \&
  {van Speybroeck}}]{2004MNRAS.353..457A}
{Allen}, S.~W., {Schmidt}, R.~W., {Ebeling}, H., {Fabian}, A.~C., \& {van
  Speybroeck}, L. 2004, \mnras, 353, 457

\bibitem[{{Allen} {et~al.}(2002){Allen}, {Schmidt}, \&
  {Fabian}}]{2002MNRAS.334L..11A}
{Allen}, S.~W., {Schmidt}, R.~W., \& {Fabian}, A.~C. 2002, \mnras, 334, L11

\bibitem[{{Andreon}(2006)}]{2006A&A...448..447A}
{Andreon}, S. 2006, \aap, 448, 447

\bibitem[{{Andreon}(2008)}]{2008MNRAS.386.1045A}
---. 2008, \mnras, 386, 1045

\bibitem[{{Arnaud} {et~al.}(2007){Arnaud}, {Pointecouteau}, \&
  {Pratt}}]{2007A&A...474L..37A}
{Arnaud}, M., {Pointecouteau}, E., \& {Pratt}, G.~W. 2007, \aap, 474, L37

\bibitem[{{Arnaud} {et~al.}(2010){Arnaud}, {Pratt}, {Piffaretti},
  {B{\"o}hringer}, {Croston}, \& {Pointecouteau}}]{2010A&A...517A..92A}
{Arnaud}, M., {Pratt}, G.~W., {Piffaretti}, R., {B{\"o}hringer}, H., {Croston},
  J.~H., \& {Pointecouteau}, E. 2010, \aap, 517, A92

\bibitem[{{Battaglia} {et~al.}(2012{\natexlab{a}}){Battaglia}, {Bond},
  {Pfrommer}, \& {Sievers}}]{2012ApJ...758...74B}
{Battaglia}, N., {Bond}, J.~R., {Pfrommer}, C., \& {Sievers}, J.~L.
  2012{\natexlab{a}}, \apj, 758, 74

\bibitem[{{Battaglia} {et~al.}(2012{\natexlab{b}}){Battaglia}, {Bond},
  {Pfrommer}, \& {Sievers}}]{2012ApJ...758...75B}
---. 2012{\natexlab{b}}, \apj, 758, 75

\bibitem[{{Battaglia} {et~al.}(2013{\natexlab{a}}){Battaglia}, {Bond},
  {Pfrommer}, \& {Sievers}}]{BBPS4}
---. 2013{\natexlab{a}}, in prep.

\bibitem[{{Battaglia} {et~al.}(2013{\natexlab{b}}){Battaglia}, {Bond},
  {Pfrommer}, \& {Sievers}}]{BBPS5}
---. 2013{\natexlab{b}}, in prep.

\bibitem[{{Battaglia} {et~al.}(2010){Battaglia}, {Bond}, {Pfrommer}, {Sievers},
  \& {Sijacki}}]{2010ApJ...725...91B}
{Battaglia}, N., {Bond}, J.~R., {Pfrommer}, C., {Sievers}, J.~L., \& {Sijacki},
  D. 2010, \apj, 725, 91

\bibitem[{{Battye} \& {Weller}(2003)}]{2003PhRvD..68h3506B}
{Battye}, R.~A., \& {Weller}, J. 2003, \prd, 68, 083506

\bibitem[{{Bautz} {et~al.}(2009){Bautz}, {Miller}, {Sanders}, {Arnaud},
  {Mushotzky}, {Porter}, {Hayashida}, {Henry}, {Hughes}, {Kawaharada},
  {Makashima}, {Sato}, \& {Tamura}}]{2009PASJ...61.1117B}
{Bautz}, M.~W. {et~al.} 2009, \pasj, 61, 1117

\bibitem[{{Bialek} {et~al.}(2001){Bialek}, {Evrard}, \&
  {Mohr}}]{2001ApJ...555..597B}
{Bialek}, J.~J., {Evrard}, A.~E., \& {Mohr}, J.~J. 2001, \apj, 555, 597

\bibitem[{{Booth} \& {Schaye}(2009)}]{Booth+2009}
{Booth}, C.~M., \& {Schaye}, J. 2009, \mnras, 398, 53

\bibitem[{{Borgani} \& {Kravtsov}(2009)}]{Borgani+2009}
{Borgani}, S., \& {Kravtsov}, A. 2009, arXiv:0906.4370

\bibitem[{{Bower} {et~al.}(2012){Bower}, {Benson}, \&
  {Crain}}]{2012MNRAS.422.2816B}
{Bower}, R.~G., {Benson}, A.~J., \& {Crain}, R.~A. 2012, \mnras, 422, 2816

\bibitem[{{Broderick} {et~al.}(2012){Broderick}, {Chang}, \&
  {Pfrommer}}]{2012ApJ...752...22B}
{Broderick}, A.~E., {Chang}, P., \& {Pfrommer}, C. 2012, \apj, 752, 22

\bibitem[{{Brodwin} {et~al.}(2006){Brodwin}, {Brown}, {Ashby}, {Bian}, {Brand},
  {Dey}, {Eisenhardt}, {Eisenstein}, {Gonzalez}, {Huang}, {Jannuzi},
  {Kochanek}, {McKenzie}, {Murray}, {Pahre}, {Smith}, {Soifer}, {Stanford},
  {Stern}, \& {Elston}}]{2006ApJ...651..791B}
{Brodwin}, M. {et~al.} 2006, \apj, 651, 791

\bibitem[{{Carlstrom} {et~al.}(2011){Carlstrom}, {Ade}, {Aird}, {Benson},
  {Bleem}, {Busetti}, {Chang}, {Chauvin}, {Cho}, {Crawford}, {Crites}, {Dobbs},
  {Halverson}, {Heimsath}, {Holzapfel}, {Hrubes}, {Joy}, {Keisler}, {Lanting},
  {Lee}, {Leitch}, {Leong}, {Lu}, {Lueker}, {Luong-van}, {McMahon}, {Mehl},
  {Meyer}, {Mohr}, {Montroy}, {Padin}, {Plagge}, {Pryke}, {Ruhl}, {Schaffer},
  {Schwan}, {Shirokoff}, {Spieler}, {Staniszewski}, {Stark}, {Tucker},
  {Vanderlinde}, {Vieira}, \& {Williamson}}]{2011PASP..123..568C}
{Carlstrom}, J.~E. {et~al.} 2011, \pasp, 123, 568

\bibitem[{{Chang} {et~al.}(2012){Chang}, {Broderick}, \&
  {Pfrommer}}]{2012ApJ...752...23C}
{Chang}, P., {Broderick}, A.~E., \& {Pfrommer}, C. 2012, \apj, 752, 23

\bibitem[{{Churazov} {et~al.}(2012){Churazov}, {Vikhlinin}, {Zhuravleva},
  {Schekochihin}, {Parrish}, {Sunyaev}, {Forman}, {B{\"o}hringer}, \&
  {Randall}}]{2012MNRAS.421.1123C}
{Churazov}, E. {et~al.} 2012, \mnras, 421, 1123

\bibitem[{{De Lucia} {et~al.}(2004){De Lucia}, {Poggianti},
  {Arag{\'o}n-Salamanca}, {Clowe}, {Halliday}, {Jablonka}, {Milvang-Jensen},
  {Pell{\'o}}, {Poirier}, {Rudnick}, {Saglia}, {Simard}, \&
  {White}}]{2004ApJ...610L..77D}
{De Lucia}, G. {et~al.} 2004, \apjl, 610, L77

\bibitem[{{De Lucia} {et~al.}(2007){De Lucia}, {Poggianti},
  {Arag{\'o}n-Salamanca}, {White}, {Zaritsky}, {Clowe}, {Halliday}, {Jablonka},
  {von der Linden}, {Milvang-Jensen}, {Pell{\'o}}, {Rudnick}, {Saglia}, \&
  {Simard}}]{2007MNRAS.374..809D}
---. 2007, \mnras, 374, 809

\bibitem[{{De Propris} {et~al.}(2013){De Propris}, {Phillipps}, \&
  {Bremer}}]{2013arXiv1307.1592D}
{De Propris}, R., {Phillipps}, S., \& {Bremer}, M. 2013, arXiv:1307.1592

\bibitem[{{Dubois} {et~al.}(2010){Dubois}, {Devriendt}, {Slyz}, \&
  {Teyssier}}]{Dubois+2010}
{Dubois}, Y., {Devriendt}, J., {Slyz}, A., \& {Teyssier}, R. 2010, \mnras, 409,
  985

\bibitem[{{Duffy} {et~al.}(2008){Duffy}, {Schaye}, {Kay}, \& {Dalla
  Vecchia}}]{2008MNRAS.390L..64D}
{Duffy}, A.~R., {Schaye}, J., {Kay}, S.~T., \& {Dalla Vecchia}, C. 2008,
  \mnras, 390, L64

\bibitem[{{Eckert} {et~al.}(2013){Eckert}, {Ettori}, {Molendi}, {Vazza}, \&
  {Paltani}}]{2013A&A...551A..23E}
{Eckert}, D., {Ettori}, S., {Molendi}, S., {Vazza}, F., \& {Paltani}, S. 2013,
  \aap, 551, A23

\bibitem[{{Eckert} {et~al.}(2012){Eckert}, {Vazza}, {Ettori}, {Molendi},
  {Nagai}, {Lau}, {Roncarelli}, {Rossetti}, {Snowden}, \&
  {Gastaldello}}]{2012A&A...541A..57E}
{Eckert}, D. {et~al.} 2012, \aap, 541, A57

\bibitem[{{Eke} {et~al.}(1998){Eke}, {Navarro}, \&
  {Frenk}}]{1998ApJ...503..569E}
{Eke}, V.~R., {Navarro}, J.~F., \& {Frenk}, C.~S. 1998, \apj, 503, 569

\bibitem[{{En{\ss}lin} {et~al.}(2007){En{\ss}lin}, {Pfrommer}, {Springel}, \&
  {Jubelgas}}]{2007A&A...473...41E}
{En{\ss}lin}, T.~A., {Pfrommer}, C., {Springel}, V., \& {Jubelgas}, M. 2007,
  \aap, 473, 41

\bibitem[{{Ettori} {et~al.}(2006){Ettori}, {Dolag}, {Borgani}, \&
  {Murante}}]{2006MNRAS.365.1021E}
{Ettori}, S., {Dolag}, K., {Borgani}, S., \& {Murante}, G. 2006, \mnras, 365,
  1021

\bibitem[{{Ettori} {et~al.}(2009){Ettori}, {Morandi}, {Tozzi}, {Balestra},
  {Borgani}, {Rosati}, {Lovisari}, \& {Terenziani}}]{2009A&A...501...61E}
{Ettori}, S., {Morandi}, A., {Tozzi}, P., {Balestra}, I., {Borgani}, S.,
  {Rosati}, P., {Lovisari}, L., \& {Terenziani}, F. 2009, \aap, 501, 61

\bibitem[{{Evrard}(1990)}]{1990ApJ...363..349E}
{Evrard}, A.~E. 1990, \apj, 363, 349

\bibitem[{{Fabjan} {et~al.}(2010){Fabjan}, {Borgani}, {Tornatore}, {Saro},
  {Murante}, \& {Dolag}}]{2010MNRAS.401.1670F}
{Fabjan}, D., {Borgani}, S., {Tornatore}, L., {Saro}, A., {Murante}, G., \&
  {Dolag}, K. 2010, \mnras, 401, 1670

\bibitem[{{Fowler} {et~al.}(2007){Fowler}, {Niemack}, {Dicker}, {Aboobaker},
  {Ade}, {Battistelli}, {Devlin}, {Fisher}, {Halpern}, {Hargrave}, {Hincks},
  {Kaul}, {Klein}, {Lau}, {Limon}, {Marriage}, {Mauskopf}, {Page}, {Staggs},
  {Swetz}, {Switzer}, {Thornton}, \& {Tucker}}]{2007ApOpt..46.3444F}
{Fowler}, J.~W. {et~al.} 2007, \ao, 46, 3444

\bibitem[{{Frenk} {et~al.}(1999){Frenk}, {White}, {Bode}, {Bond}, {Bryan},
  {Cen}, {Couchman}, {Evrard}, {Gnedin}, {Jenkins}, {Khokhlov}, {Klypin},
  {Navarro}, {Norman}, {Ostriker}, {Owen}, {Pearce}, {Pen}, {Steinmetz},
  {Thomas}, {Villumsen}, {Wadsley}, {Warren}, {Xu}, \& {Yepes}}]{Frenk+1999}
{Frenk}, C.~S. {et~al.} 1999, \apj, 525, 554

\bibitem[{{Galli} {et~al.}(2012){Galli}, {Bartlett}, \&
  {Melchiorri}}]{2012PhRvD..86d3516G}
{Galli}, S., {Bartlett}, J.~G., \& {Melchiorri}, A. 2012, \prd, 86, 043516

\bibitem[{{Gao} {et~al.}(2012){Gao}, {Navarro}, {Frenk}, {Jenkins}, {Springel},
  \& {White}}]{2012MNRAS.425.2169G}
{Gao}, L., {Navarro}, J.~F., {Frenk}, C.~S., {Jenkins}, A., {Springel}, V., \&
  {White}, S.~D.~M. 2012, \mnras, 425, 2169

\bibitem[{{George} {et~al.}(2009){George}, {Fabian}, {Sanders}, {Young}, \&
  {Russell}}]{2009MNRAS.395..657G}
{George}, M.~R., {Fabian}, A.~C., {Sanders}, J.~S., {Young}, A.~J., \&
  {Russell}, H.~R. 2009, \mnras, 395, 657

\bibitem[{{Gilbank} {et~al.}(2008){Gilbank}, {Yee}, {Ellingson}, {Gladders},
  {Loh}, {Barrientos}, \& {Barkhouse}}]{2008ApJ...673..742G}
{Gilbank}, D.~G., {Yee}, H.~K.~C., {Ellingson}, E., {Gladders}, M.~D., {Loh},
  Y.-S., {Barrientos}, L.~F., \& {Barkhouse}, W.~A. 2008, \apj, 673, 742

\bibitem[{{Giodini} {et~al.}(2009){Giodini}, {Pierini}, {Finoguenov}, {Pratt},
  {Boehringer}, {Leauthaud}, {Guzzo}, {Aussel}, {Bolzonella}, {Capak}, {Elvis},
  {Hasinger}, {Ilbert}, {Kartaltepe}, {Koekemoer}, {Lilly}, {Massey},
  {McCracken}, {Rhodes}, {Salvato}, {Sanders}, {Scoville}, {Sasaki}, {Smolcic},
  {Taniguchi}, {Thompson}, \& {COSMOS Collaboration}}]{2009ApJ...703..982G}
{Giodini}, S. {et~al.} 2009, \apj, 703, 982

\bibitem[{{Gladders} \& {Yee}(2000)}]{2000AJ....120.2148G}
{Gladders}, M.~D., \& {Yee}, H.~K.~C. 2000, \aj, 120, 2148

\bibitem[{{Gonzalez} {et~al.}(2005){Gonzalez}, {Zabludoff}, \&
  {Zaritsky}}]{2005ApJ...618..195G}
{Gonzalez}, A.~H., {Zabludoff}, A.~I., \& {Zaritsky}, D. 2005, \apj, 618, 195

\bibitem[{{Gonzalez} {et~al.}(2007){Gonzalez}, {Zaritsky}, \&
  {Zabludoff}}]{2007ApJ...666..147G}
{Gonzalez}, A.~H., {Zaritsky}, D., \& {Zabludoff}, A.~I. 2007, \apj, 666, 147

\bibitem[{{G{\'o}rski} {et~al.}(2005){G{\'o}rski}, {Hivon}, {Banday},
  {Wandelt}, {Hansen}, {Reinecke}, \& {Bartelmann}}]{2005ApJ...622..759G}
{G{\'o}rski}, K.~M., {Hivon}, E., {Banday}, A.~J., {Wandelt}, B.~D., {Hansen},
  F.~K., {Reinecke}, M., \& {Bartelmann}, M. 2005, \apj, 622, 759

\bibitem[{{Guo} \& {Oh}(2008)}]{2008MNRAS.384..251G}
{Guo}, F., \& {Oh}, S.~P. 2008, \mnras, 384, 251

\bibitem[{{Hayashi} \& {White}(2008)}]{2008MNRAS.388....2H}
{Hayashi}, E., \& {White}, S.~D.~M. 2008, \mnras, 388, 2

\bibitem[{{Hoshino} {et~al.}(2010){Hoshino}, {Henry}, {Sato}, {Akamatsu},
  {Yokota}, {Sasaki}, {Ishisaki}, {Ohashi}, {Bautz}, {Fukazawa}, {Kawano},
  {Furuzawa}, {Hayashida}, {Tawa}, {Hughes}, {Kokubun}, \&
  {Tamura}}]{2010PASJ...62..371H}
{Hoshino}, A. {et~al.} 2010, \pasj, 62, 371

\bibitem[{{Hu}(2003)}]{2003PhRvD..67h1304H}
{Hu}, W. 2003, \prd, 67, 081304

\bibitem[{{Humphrey} {et~al.}(2012){Humphrey}, {Buote}, {Brighenti}, {Flohic},
  {Gastaldello}, \& {Mathews}}]{2012ApJ...748...11H}
{Humphrey}, P.~J., {Buote}, D.~A., {Brighenti}, F., {Flohic}, H.~M.~L.~G.,
  {Gastaldello}, F., \& {Mathews}, W.~G. 2012, \apj, 748, 11

\bibitem[{{Jubelgas} {et~al.}(2008){Jubelgas}, {Springel}, {En{\ss}lin}, \&
  {Pfrommer}}]{2008A&A...481...33J}
{Jubelgas}, M., {Springel}, V., {En{\ss}lin}, T., \& {Pfrommer}, C. 2008, \aap,
  481, 33

\bibitem[{{Kawaharada} {et~al.}(2010){Kawaharada}, {Okabe}, {Umetsu},
  {Takizawa}, {Matsushita}, {Fukazawa}, {Hamana}, {Miyazaki}, {Nakazawa}, \&
  {Ohashi}}]{2010ApJ...714..423K}
{Kawaharada}, M. {et~al.} 2010, \apj, 714, 423

\bibitem[{{Kay} {et~al.}(2004){Kay}, {Thomas}, {Jenkins}, \&
  {Pearce}}]{2004MNRAS.355.1091K}
{Kay}, S.~T., {Thomas}, P.~A., {Jenkins}, A., \& {Pearce}, F.~R. 2004, \mnras,
  355, 1091

\bibitem[{{Khedekar} {et~al.}(2010){Khedekar}, {Majumdar}, \&
  {Das}}]{2010PhRvD..82d1301K}
{Khedekar}, S., {Majumdar}, S., \& {Das}, S. 2010, \prd, 82, 041301

\bibitem[{{Kravtsov} {et~al.}(2005){Kravtsov}, {Nagai}, \&
  {Vikhlinin}}]{2005ApJ...625..588K}
{Kravtsov}, A.~V., {Nagai}, D., \& {Vikhlinin}, A.~A. 2005, \apj, 625, 588

\bibitem[{{Kravtsov} {et~al.}(2006){Kravtsov}, {Vikhlinin}, \&
  {Nagai}}]{2006ApJ...650..128K}
{Kravtsov}, A.~V., {Vikhlinin}, A., \& {Nagai}, D. 2006, \apj, 650, 128

\bibitem[{{LaRoque} {et~al.}(2006){LaRoque}, {Bonamente}, {Carlstrom}, {Joy},
  {Nagai}, {Reese}, \& {Dawson}}]{2006ApJ...652..917L}
{LaRoque}, S.~J., {Bonamente}, M., {Carlstrom}, J.~E., {Joy}, M.~K., {Nagai},
  D., {Reese}, E.~D., \& {Dawson}, K.~S. 2006, \apj, 652, 917

\bibitem[{{Lau} {et~al.}(2009){Lau}, {Kravtsov}, \&
  {Nagai}}]{2009ApJ...705.1129L}
{Lau}, E.~T., {Kravtsov}, A.~V., \& {Nagai}, D. 2009, \apj, 705, 1129

\bibitem[{{Leauthaud} {et~al.}(2011){Leauthaud}, {Tinker}, {Behroozi}, {Busha},
  \& {Wechsler}}]{2011ApJ...738...45L}
{Leauthaud}, A., {Tinker}, J., {Behroozi}, P.~S., {Busha}, M.~T., \&
  {Wechsler}, R.~H. 2011, \apj, 738, 45

\bibitem[{{Leitner} \& {Kravtsov}(2011)}]{2011ApJ...734...48L}
{Leitner}, S.~N., \& {Kravtsov}, A.~V. 2011, \apj, 734, 48

\bibitem[{{Lin} {et~al.}(2003){Lin}, {Mohr}, \&
  {Stanford}}]{2003ApJ...591..749L}
{Lin}, Y.-T., {Mohr}, J.~J., \& {Stanford}, S.~A. 2003, \apj, 591, 749

\bibitem[{{Loewenstein} {et~al.}(1991){Loewenstein}, {Zweibel}, \&
  {Begelman}}]{1991ApJ...377..392L}
{Loewenstein}, M., {Zweibel}, E.~G., \& {Begelman}, M.~C. 1991, \apj, 377, 392

\bibitem[{{Majumdar} \& {Mohr}(2003)}]{2003ApJ...585..603M}
{Majumdar}, S., \& {Mohr}, J.~J. 2003, \apj, 585, 603

\bibitem[{{Majumdar} \& {Mohr}(2004)}]{2004ApJ...613...41M}
---. 2004, \apj, 613, 41

\bibitem[{{Mancone} {et~al.}(2012){Mancone}, {Baker}, {Gonzalez}, {Ashby},
  {Stanford}, {Brodwin}, {Eisenhardt}, {Snyder}, {Stern}, \&
  {Wright}}]{2012ApJ...761..141M}
{Mancone}, C.~L. {et~al.} 2012, \apj, 761, 141

\bibitem[{{Mantz} {et~al.}(2010{\natexlab{a}}){Mantz}, {Allen}, {Ebeling},
  {Rapetti}, \& {Drlica-Wagner}}]{2010MNRAS.406.1773M}
{Mantz}, A., {Allen}, S.~W., {Ebeling}, H., {Rapetti}, D., \& {Drlica-Wagner},
  A. 2010{\natexlab{a}}, \mnras, 406, 1773

\bibitem[{{Mantz} {et~al.}(2010{\natexlab{b}}){Mantz}, {Allen}, \&
  {Rapetti}}]{2010MNRAS.406.1805M}
{Mantz}, A., {Allen}, S.~W., \& {Rapetti}, D. 2010{\natexlab{b}}, \mnras, 406,
  1805

\bibitem[{{Mantz} {et~al.}(2010{\natexlab{c}}){Mantz}, {Allen}, {Rapetti}, \&
  {Ebeling}}]{2010MNRAS.406.1759M}
{Mantz}, A., {Allen}, S.~W., {Rapetti}, D., \& {Ebeling}, H.
  2010{\natexlab{c}}, \mnras, 406, 1759

\bibitem[{{Martig} \& {Bournaud}(2010)}]{2010ApJ...714L.275M}
{Martig}, M., \& {Bournaud}, F. 2010, \apjl, 714, L275

\bibitem[{{McCarthy} {et~al.}(2007){McCarthy}, {Bower}, {Balogh}, {Voit},
  {Pearce}, {Theuns}, {Babul}, {Lacey}, \& {Frenk}}]{2007MNRAS.376..497M}
{McCarthy}, I.~G. {et~al.} 2007, \mnras, 376, 497

\bibitem[{{McCarthy} {et~al.}(2011){McCarthy}, {Schaye}, {Bower}, {Ponman},
  {Booth}, {Vecchia}, \& {Springel}}]{2011MNRAS.412.1965M}
{McCarthy}, I.~G., {Schaye}, J., {Bower}, R.~G., {Ponman}, T.~J., {Booth},
  C.~M., {Vecchia}, C.~D., \& {Springel}, V. 2011, \mnras, 412, 1965

\bibitem[{{McCarthy} {et~al.}(2010){McCarthy}, {Schaye}, {Ponman}, {Bower},
  {Booth}, {Dalla Vecchia}, {Crain}, {Springel}, {Theuns}, \&
  {Wiersma}}]{McCarthy+2010}
{McCarthy}, I.~G. {et~al.} 2010, \mnras, 406, 822

\bibitem[{{Miller} {et~al.}(2012){Miller}, {Bautz}, {George}, {Mushotzky},
  {Davis}, \& {Henry}}]{2012AIPC.1427...13M}
{Miller}, E.~D., {Bautz}, M., {George}, J., {Mushotzky}, R., {Davis}, D., \&
  {Henry}, J.~P. 2012, in American Institute of Physics Conference Series, Vol.
  1427, American Institute of Physics Conference Series, ed. R.~{Petre},
  K.~{Mitsuda}, \& L.~{Angelini}, 13--20

\bibitem[{{Mohr} {et~al.}(1993){Mohr}, {Fabricant}, \&
  {Geller}}]{1993ApJ...413..492M}
{Mohr}, J.~J., {Fabricant}, D.~G., \& {Geller}, M.~J. 1993, \apj, 413, 492

\bibitem[{{Muanwong} {et~al.}(2002){Muanwong}, {Thomas}, {Kay}, \&
  {Pearce}}]{2002MNRAS.336..527M}
{Muanwong}, O., {Thomas}, P.~A., {Kay}, S.~T., \& {Pearce}, F.~R. 2002, \mnras,
  336, 527

\bibitem[{{Muzzin} {et~al.}(2009){Muzzin}, {Wilson}, {Yee}, {Hoekstra},
  {Gilbank}, {Surace}, {Lacy}, {Blindert}, {Majumdar}, {Demarco}, {Gardner},
  {Gladders}, \& {Lonsdale}}]{2009ApJ...698.1934M}
{Muzzin}, A. {et~al.} 2009, \apj, 698, 1934

\bibitem[{{Nagai} \& {Lau}(2011)}]{2011ApJ...731L..10N}
{Nagai}, D., \& {Lau}, E.~T. 2011, \apjl, 731, L10

\bibitem[{{Navarro} {et~al.}(1997){Navarro}, {Frenk}, \&
  {White}}]{1997ApJ...490..493N}
{Navarro}, J.~F., {Frenk}, C.~S., \& {White}, S.~D.~M. 1997, \apj, 490, 493

\bibitem[{{Nelson} {et~al.}(2012){Nelson}, {Rudd}, {Shaw}, \&
  {Nagai}}]{2012ApJ...751..121N}
{Nelson}, K., {Rudd}, D.~H., {Shaw}, L., \& {Nagai}, D. 2012, \apj, 751, 121

\bibitem[{{Papovich} {et~al.}(2010){Papovich}, {Momcheva}, {Willmer},
  {Finkelstein}, {Finkelstein}, {Tran}, {Brodwin}, {Dunlop}, {Farrah}, {Khan},
  {Lotz}, {McCarthy}, {McLure}, {Rieke}, {Rudnick}, {Sivanandam}, {Pacaud}, \&
  {Pierre}}]{2010ApJ...716.1503P}
{Papovich}, C. {et~al.} 2010, \apj, 716, 1503

\bibitem[{{Pfrommer}(2013)}]{2013arXiv1303.5443P}
{Pfrommer}, C. 2013, arXiv:1303.5443

\bibitem[{{Pfrommer} {et~al.}(2012){Pfrommer}, {Chang}, \&
  {Broderick}}]{2012ApJ...752...24P}
{Pfrommer}, C., {Chang}, P., \& {Broderick}, A.~E. 2012, \apj, 752, 24

\bibitem[{{Pfrommer} {et~al.}(2007){Pfrommer}, {En{\ss}lin}, {Springel},
  {Jubelgas}, \& {Dolag}}]{2007MNRAS.378..385P}
{Pfrommer}, C., {En{\ss}lin}, T.~A., {Springel}, V., {Jubelgas}, M., \&
  {Dolag}, K. 2007, \mnras, 378, 385

\bibitem[{{Pfrommer} {et~al.}(2006){Pfrommer}, {Springel}, {En{\ss}lin}, \&
  {Jubelgas}}]{2006MNRAS.367..113P}
{Pfrommer}, C., {Springel}, V., {En{\ss}lin}, T.~A., \& {Jubelgas}, M. 2006,
  \mnras, 367, 113

\bibitem[{{Pinzke} \& {Pfrommer}(2010)}]{2010MNRAS.409..449P}
{Pinzke}, A., \& {Pfrommer}, C. 2010, \mnras, 409, 449

\bibitem[{{Planck Collaboration} {et~al.}(2013{\natexlab{a}}){Planck
  Collaboration}, {Ade}, {Aghanim}, {Armitage-Caplan}, {Arnaud}, {Ashdown},
  {Atrio-Barandela}, {Aumont}, {Baccigalupi}, {Banday}, \&
  et~al.}]{2013arXiv1303.5076P}
{Planck Collaboration} {et~al.} 2013{\natexlab{a}}, arXiv:1303.5076

\bibitem[{{Planck Collaboration} {et~al.}(2013{\natexlab{b}}){Planck
  Collaboration}, {Ade}, {Aghanim}, {Arnaud}, {Ashdown}, {Atrio-Barandela},
  {Aumont}, {Baccigalupi}, {Balbi}, {Banday}, \& et~al.}]{2013A&A...550A.131P}
---. 2013{\natexlab{b}}, \aap, 550, A131

\bibitem[{{Planck Collaboration} {et~al.}(2013{\natexlab{c}}){Planck
  Collaboration}, {Ade}, {Aghanim}, {Arnaud}, {Ashdown}, {Atrio-Barandela},
  {Aumont}, {Baccigalupi}, {Balbi}, {Banday}, \& et~al.}]{Planck_erratum}
---. 2013{\natexlab{c}}, subm.

\bibitem[{{Planelles} {et~al.}(2013){Planelles}, {Borgani}, {Dolag}, {Ettori},
  {Fabjan}, {Murante}, \& {Tornatore}}]{2013MNRAS.431.1487P}
{Planelles}, S., {Borgani}, S., {Dolag}, K., {Ettori}, S., {Fabjan}, D.,
  {Murante}, G., \& {Tornatore}, L. 2013, \mnras, 431, 1487

\bibitem[{{Poole} {et~al.}(2007){Poole}, {Babul}, {McCarthy}, {Fardal},
  {Bildfell}, {Quinn}, \& {Mahdavi}}]{2007MNRAS.380..437P}
{Poole}, G.~B., {Babul}, A., {McCarthy}, I.~G., {Fardal}, M.~A., {Bildfell},
  C.~J., {Quinn}, T., \& {Mahdavi}, A. 2007, \mnras, 380, 437

\bibitem[{{Pratt} {et~al.}(2010){Pratt}, {Arnaud}, {Piffaretti},
  {B{\"o}hringer}, {Ponman}, {Croston}, {Voit}, {Borgani}, \&
  {Bower}}]{2010A&A...511A..85P}
{Pratt}, G.~W. {et~al.} 2010, \aap, 511, A85

\bibitem[{{Pratt} {et~al.}(2009){Pratt}, {Croston}, {Arnaud}, \&
  {B{\"o}hringer}}]{2009A&A...498..361P}
{Pratt}, G.~W., {Croston}, J.~H., {Arnaud}, M., \& {B{\"o}hringer}, H. 2009,
  \aap, 498, 361

\bibitem[{{Puchwein} {et~al.}(2012){Puchwein}, {Pfrommer}, {Springel},
  {Broderick}, \& {Chang}}]{2012MNRAS.423..149P}
{Puchwein}, E., {Pfrommer}, C., {Springel}, V., {Broderick}, A.~E., \& {Chang},
  P. 2012, \mnras, 423, 149

\bibitem[{{Puchwein} {et~al.}(2008){Puchwein}, {Sijacki}, \&
  {Springel}}]{2008ApJ...687L..53P}
{Puchwein}, E., {Sijacki}, D., \& {Springel}, V. 2008, \apjl, 687, L53

\bibitem[{{Puchwein} \& {Springel}(2013)}]{2013MNRAS.428.2966P}
{Puchwein}, E., \& {Springel}, V. 2013, \mnras, 428, 2966

\bibitem[{{Puchwein} {et~al.}(2010){Puchwein}, {Springel}, {Sijacki}, \&
  {Dolag}}]{2010MNRAS.406..936P}
{Puchwein}, E., {Springel}, V., {Sijacki}, D., \& {Dolag}, K. 2010, \mnras,
  406, 936

\bibitem[{{Rapetti} {et~al.}(2009){Rapetti}, {Allen}, {Mantz}, \&
  {Ebeling}}]{2009MNRAS.400..699R}
{Rapetti}, D., {Allen}, S.~W., {Mantz}, A., \& {Ebeling}, H. 2009, \mnras, 400,
  699

\bibitem[{{Rapetti} {et~al.}(2010){Rapetti}, {Allen}, {Mantz}, \&
  {Ebeling}}]{2010MNRAS.406.1796R}
---. 2010, \mnras, 406, 1796

\bibitem[{{Rasia} {et~al.}(2006){Rasia}, {Ettori}, {Moscardini}, {Mazzotta},
  {Borgani}, {Dolag}, {Tormen}, {Cheng}, \& {Diaferio}}]{2006MNRAS.369.2013R}
{Rasia}, E. {et~al.} 2006, \mnras, 369, 2013

\bibitem[{{Rasia} {et~al.}(2012){Rasia}, {Meneghetti}, {Martino}, {Borgani},
  {Bonafede}, {Dolag}, {Ettori}, {Fabjan}, {Giocoli}, {Mazzotta}, {Merten},
  {Radovich}, \& {Tornatore}}]{2012NJPh...14e5018R}
---. 2012, New Journal of Physics, 14, 055018

\bibitem[{{Rasia} {et~al.}(2004){Rasia}, {Tormen}, \&
  {Moscardini}}]{2004MNRAS.351..237R}
{Rasia}, E., {Tormen}, G., \& {Moscardini}, L. 2004, \mnras, 351, 237

\bibitem[{{Reichardt} {et~al.}(2012){Reichardt}, {Shaw}, {Zahn}, {Aird},
  {Benson}, {Bleem}, {Carlstrom}, {Chang}, {Cho}, {Crawford}, {Crites}, {de
  Haan}, {Dobbs}, {Dudley}, {George}, {Halverson}, {Holder}, {Holzapfel},
  {Hoover}, {Hou}, {Hrubes}, {Joy}, {Keisler}, {Knox}, {Lee}, {Leitch},
  {Lueker}, {Luong-Van}, {McMahon}, {Mehl}, {Meyer}, {Millea}, {Mohr},
  {Montroy}, {Natoli}, {Padin}, {Plagge}, {Pryke}, {Ruhl}, {Schaffer},
  {Shirokoff}, {Spieler}, {Staniszewski}, {Stark}, {Story}, {van Engelen},
  {Vanderlinde}, {Vieira}, \& {Williamson}}]{2012ApJ...755...70R}
{Reichardt}, C.~L. {et~al.} 2012, \apj, 755, 70

\bibitem[{{Reiprich} {et~al.}(2009){Reiprich}, {Hudson}, {Zhang}, {Sato},
  {Ishisaki}, {Hoshino}, {Ohashi}, {Ota}, \& {Fujita}}]{2009A&A...501..899R}
{Reiprich}, T.~H. {et~al.} 2009, \aap, 501, 899

\bibitem[{{Roncarelli} {et~al.}(2013){Roncarelli}, {Ettori}, {Borgani},
  {Dolag}, {Fabjan}, \& {Moscardini}}]{2013MNRAS.432.3030R}
{Roncarelli}, M., {Ettori}, S., {Borgani}, S., {Dolag}, K., {Fabjan}, D., \&
  {Moscardini}, L. 2013, \mnras, 432, 3030

\bibitem[{{Roncarelli} {et~al.}(2006){Roncarelli}, {Ettori}, {Dolag},
  {Moscardini}, {Borgani}, \& {Murante}}]{2006MNRAS.373.1339R}
{Roncarelli}, M., {Ettori}, S., {Dolag}, K., {Moscardini}, L., {Borgani}, S.,
  \& {Murante}, G. 2006, \mnras, 373, 1339

\bibitem[{{Rudnick} {et~al.}(2009){Rudnick}, {von der Linden}, {Pell{\'o}},
  {Arag{\'o}n-Salamanca}, {Marchesini}, {Clowe}, {De Lucia}, {Halliday},
  {Jablonka}, {Milvang-Jensen}, {Poggianti}, {Saglia}, {Simard}, {White}, \&
  {Zaritsky}}]{2009ApJ...700.1559R}
{Rudnick}, G. {et~al.} 2009, \apj, 700, 1559

\bibitem[{{Rudnick} {et~al.}(2012){Rudnick}, {Tran}, {Papovich}, {Momcheva}, \&
  {Willmer}}]{2012ApJ...755...14R}
{Rudnick}, G.~H., {Tran}, K.-V., {Papovich}, C., {Momcheva}, I., \& {Willmer},
  C. 2012, \apj, 755, 14

\bibitem[{{Sadat} {et~al.}(2005){Sadat}, {Blanchard}, {Vauclair}, {Lumb},
  {Bartlett}, {Romer}, {Bernard}, {Boer}, {Marty}, {Nevalainen}, {Burke},
  {Collins}, \& {Nichol}}]{2005A&A...437...31S}
{Sadat}, R. {et~al.} 2005, \aap, 437, 31

\bibitem[{{Sanders} \& {Fabian}(2012)}]{2012MNRAS.421..726S}
{Sanders}, J.~S., \& {Fabian}, A.~C. 2012, \mnras, 421, 726

\bibitem[{{Sanderson} {et~al.}(2003){Sanderson}, {Ponman}, {Finoguenov},
  {Lloyd-Davies}, \& {Markevitch}}]{2003MNRAS.340..989S}
{Sanderson}, A.~J.~R., {Ponman}, T.~J., {Finoguenov}, A., {Lloyd-Davies},
  E.~J., \& {Markevitch}, M. 2003, \mnras, 340, 989

\bibitem[{{Sato} {et~al.}(2012){Sato}, {Sasaki}, {Matsushita}, {Sakuma},
  {Sato}, {Fujita}, {Okabe}, {Fukazawa}, {Ichikawa}, {Kawaharada}, {Nakazawa},
  {Ohashi}, {Ota}, {Takizawa}, \& {Tamura}}]{2012PASJ...64...95S}
{Sato}, T. {et~al.} 2012, \pasj, 64, 95

\bibitem[{{Sembolini} {et~al.}(2013){Sembolini}, {Yepes}, {De Petris},
  {Gottl{\"o}ber}, {Lamagna}, \& {Comis}}]{2013MNRAS.429..323S}
{Sembolini}, F., {Yepes}, G., {De Petris}, M., {Gottl{\"o}ber}, S., {Lamagna},
  L., \& {Comis}, B. 2013, \mnras, 429, 323

\bibitem[{{Shaw} {et~al.}(2010){Shaw}, {Nagai}, {Bhattacharya}, \&
  {Lau}}]{2010ApJ...725.1452S}
{Shaw}, L.~D., {Nagai}, D., {Bhattacharya}, S., \& {Lau}, E.~T. 2010, \apj,
  725, 1452

\bibitem[{{Sievers} {et~al.}(2013){Sievers}, {Hlozek}, {Nolta}, {Acquaviva},
  {Addison}, {Ade}, {Aguirre}, {Amiri}, {Appel}, {Barrientos}, {Battistelli},
  {Battaglia}, {Bond}, {Brown}, {Burger}, {Calabrese}, {Chervenak}, {Crichton},
  {Das}, {Devlin}, {Dicker}, {Bertrand Doriese}, {Dunkley}, {D{\"u}nner},
  {Essinger-Hileman}, {Faber}, {Fisher}, {Fowler}, {Gallardo}, {Gordon},
  {Gralla}, {Hajian}, {Halpern}, {Hasselfield}, {Hern{\'a}ndez-Monteagudo},
  {Hill}, {Hilton}, {Hilton}, {Hincks}, {Holtz}, {Huffenberger}, {Hughes},
  {Hughes}, {Infante}, {Irwin}, {Jacobson}, {Johnstone}, {Baptiste Juin},
  {Kaul}, {Klein}, {Kosowsky}, {Lau}, {Limon}, {Lin}, {Louis}, {Lupton},
  {Marriage}, {Marsden}, {Martocci}, {Mauskopf}, {McLaren}, {Menanteau},
  {Moodley}, {Moseley}, {Netterfield}, {Niemack}, {Page}, {Page}, {Parker},
  {Partridge}, {Plimpton}, {Quintana}, {Reese}, {Reid}, {Rojas}, {Sehgal},
  {Sherwin}, {Schmitt}, {Spergel}, {Staggs}, {Stryzak}, {Swetz}, {Switzer},
  {Thornton}, {Trac}, {Tucker}, {Uehara}, {Visnjic}, {Warne}, {Wilson},
  {Wollack}, {Zhao}, \& {Zuncke}}]{2013arXiv1301.0824S}
{Sievers}, J.~L. {et~al.} 2013, arXiv:1301.0824

\bibitem[{{Sijacki} {et~al.}(2008){Sijacki}, {Pfrommer}, {Springel}, \&
  {En{\ss}lin}}]{Sijacki+2008}
{Sijacki}, D., {Pfrommer}, C., {Springel}, V., \& {En{\ss}lin}, T.~A. 2008,
  \mnras, 387, 1403

\bibitem[{{Sijacki} \& {Springel}(2006)}]{Sijacki+2006}
{Sijacki}, D., \& {Springel}, V. 2006, \mnras, 366, 397

\bibitem[{{Sijacki} {et~al.}(2007){Sijacki}, {Springel}, {Di Matteo}, \&
  {Hernquist}}]{2007MNRAS.380..877S}
{Sijacki}, D., {Springel}, V., {Di Matteo}, T., \& {Hernquist}, L. 2007,
  \mnras, 380, 877

\bibitem[{{Simionescu} {et~al.}(2011){Simionescu}, {Allen}, {Mantz}, {Werner},
  {Takei}, {Morris}, {Fabian}, {Sanders}, {Nulsen}, {George}, \&
  {Taylor}}]{2011Sci...331.1576S}
{Simionescu}, A. {et~al.} 2011, Science, 331, 1576

\bibitem[{{Springel}(2005)}]{2005MNRAS.364.1105S}
{Springel}, V. 2005, \mnras, 364, 1105

\bibitem[{{Springel} \& {Hernquist}(2003)}]{2003MNRAS.339..289S}
{Springel}, V., \& {Hernquist}, L. 2003, \mnras, 339, 289

\bibitem[{{Stanford} {et~al.}(1998){Stanford}, {Eisenhardt}, \&
  {Dickinson}}]{1998ApJ...492..461S}
{Stanford}, S.~A., {Eisenhardt}, P.~R., \& {Dickinson}, M. 1998, \apj, 492, 461

\bibitem[{{Stott} {et~al.}(2007){Stott}, {Smail}, {Edge}, {Ebeling}, {Smith},
  {Kneib}, \& {Pimbblet}}]{2007ApJ...661...95S}
{Stott}, J.~P., {Smail}, I., {Edge}, A.~C., {Ebeling}, H., {Smith}, G.~P.,
  {Kneib}, J.-P., \& {Pimbblet}, K.~A. 2007, \apj, 661, 95

\bibitem[{{Sun} {et~al.}(2009){Sun}, {Voit}, {Donahue}, {Jones}, {Forman}, \&
  {Vikhlinin}}]{2009ApJ...693.1142S}
{Sun}, M., {Voit}, G.~M., {Donahue}, M., {Jones}, C., {Forman}, W., \&
  {Vikhlinin}, A. 2009, \apj, 693, 1142

\bibitem[{{Tanaka} {et~al.}(2004){Tanaka}, {Goto}, {Okamura}, {Shimasaku}, \&
  {Brinkmann}}]{2004AJ....128.2677T}
{Tanaka}, M., {Goto}, T., {Okamura}, S., {Shimasaku}, K., \& {Brinkmann}, J.
  2004, \aj, 128, 2677

\bibitem[{{Teyssier} {et~al.}(2011){Teyssier}, {Moore}, {Martizzi}, {Dubois},
  \& {Mayer}}]{Teyssier+2011}
{Teyssier}, R., {Moore}, B., {Martizzi}, D., {Dubois}, Y., \& {Mayer}, L. 2011,
  \mnras, 618

\bibitem[{{Thompson} {et~al.}(2005){Thompson}, {Quataert}, \&
  {Murray}}]{2005ApJ...630..167T}
{Thompson}, T.~A., {Quataert}, E., \& {Murray}, N. 2005, \apj, 630, 167

\bibitem[{{Trac} {et~al.}(2011){Trac}, {Bode}, \&
  {Ostriker}}]{2011ApJ...727...94T}
{Trac}, H., {Bode}, P., \& {Ostriker}, J.~P. 2011, \apj, 727, 94

\bibitem[{{Vazza} {et~al.}(2012){Vazza}, {Br{\"u}ggen}, {Gheller}, \&
  {Brunetti}}]{2012MNRAS.421.3375V}
{Vazza}, F., {Br{\"u}ggen}, M., {Gheller}, C., \& {Brunetti}, G. 2012, \mnras,
  421, 3375

\bibitem[{{Vazza} {et~al.}(2011{\natexlab{a}}){Vazza}, {Brunetti}, {Gheller},
  {Brunino}, \& {Br{\"u}ggen}}]{2011A&A...529A..17V}
{Vazza}, F., {Brunetti}, G., {Gheller}, C., {Brunino}, R., \& {Br{\"u}ggen}, M.
  2011{\natexlab{a}}, \aap, 529, A17

\bibitem[{{Vazza} {et~al.}(2011{\natexlab{b}}){Vazza}, {Roncarelli}, {Ettori},
  \& {Dolag}}]{2011MNRAS.413.2305V}
{Vazza}, F., {Roncarelli}, M., {Ettori}, S., \& {Dolag}, K. 2011{\natexlab{b}},
  \mnras, 413, 2305

\bibitem[{{Vikhlinin} {et~al.}(2009{\natexlab{a}}){Vikhlinin}, {Burenin},
  {Ebeling}, {Forman}, {Hornstrup}, {Jones}, {Kravtsov}, {Murray}, {Nagai},
  {Quintana}, \& {Voevodkin}}]{2009ApJ...692.1033V}
{Vikhlinin}, A. {et~al.} 2009{\natexlab{a}}, \apj, 692, 1033

\bibitem[{{Vikhlinin} {et~al.}(2006){Vikhlinin}, {Kravtsov}, {Forman}, {Jones},
  {Markevitch}, {Murray}, \& {Van Speybroeck}}]{2006ApJ...640..691V}
{Vikhlinin}, A., {Kravtsov}, A., {Forman}, W., {Jones}, C., {Markevitch}, M.,
  {Murray}, S.~S., \& {Van Speybroeck}, L. 2006, \apj, 640, 691

\bibitem[{{Vikhlinin} {et~al.}(2009{\natexlab{b}}){Vikhlinin}, {Kravtsov},
  {Burenin}, {Ebeling}, {Forman}, {Hornstrup}, {Jones}, {Murray}, {Nagai},
  {Quintana}, \& {Voevodkin}}]{2009ApJ...692.1060V}
{Vikhlinin}, A. {et~al.} 2009{\natexlab{b}}, \apj, 692, 1060

\bibitem[{{Voit}(2005)}]{Voit2005}
{Voit}, G.~M. 2005, Reviews of Modern Physics, 77, 207

\bibitem[{{Vulcani} {et~al.}(2011){Vulcani}, {Poggianti},
  {Arag{\'o}n-Salamanca}, {Fasano}, {Rudnick}, {Valentinuzzi}, {Dressler},
  {Bettoni}, {Cava}, {D'Onofrio}, {Fritz}, {Moretti}, {Omizzolo}, \&
  {Varela}}]{2011MNRAS.412..246V}
{Vulcani}, B. {et~al.} 2011, \mnras, 412, 246

\bibitem[{{Walker} {et~al.}(2012){Walker}, {Fabian}, {Sanders}, {George}, \&
  {Tawara}}]{2012MNRAS.422.3503W}
{Walker}, S.~A., {Fabian}, A.~C., {Sanders}, J.~S., {George}, M.~R., \&
  {Tawara}, Y. 2012, \mnras, 422, 3503

\bibitem[{{Young} {et~al.}(2011){Young}, {Thomas}, {Short}, \&
  {Pearce}}]{2011MNRAS.413..691Y}
{Young}, O.~E., {Thomas}, P.~A., {Short}, C.~J., \& {Pearce}, F. 2011, \mnras,
  413, 691

\bibitem[{{Zibetti} {et~al.}(2005){Zibetti}, {White}, {Schneider}, \&
  {Brinkmann}}]{2005MNRAS.358..949Z}
{Zibetti}, S., {White}, S.~D.~M., {Schneider}, D.~P., \& {Brinkmann}, J. 2005,
  \mnras, 358, 949

\end{thebibliography}
\bibliographystyle{apj}

\begin{appendix}

\begin{table*}
  \caption{Summarizing the median and $1-\sigma$ percentiles of the enclosed gas, 
    stellar, and baryonic mass fractions, $f_\rmn{gas}$,  $f_\rmn{star}$, $f_\rmn{b}$ 
    for various simulated physics models, redshifts, and enclosing radii. We also 
    show the biased values for $f_\rmn{gas}$ measurements when assuming 
    hydrostatic equilibrium, $f_\rmn{gas,HSE}$, and additionally neglecting the 
    density clumping, $f_\rmn{gas,HSE+clump}$.}
\label{tab:Sum1}
\begin{center}
   \leavevmode
\small{
\begin{tabular}{l|ccc|ccc|ccc}
  \hline
  \hline
  $1.4 < M_{200}/10^{14} \rmn{M}_{\odot} < 2.6$ & & $z = 0$ &  & &$z = 0.5$ &  & & $z = 1$ & \\
  \hline
 Simulated physics$^a$  &  SH & CSF & AGN &  SH & CSF & AGN & SH & CSF & AGN\\
  \hline
  $r < R_{200}$ & & &  & & &  & & & \\
  \hline
  $f_\rmn{gas}$ & $  0.93_{-  0.03}^{+  0.03}$ & $  0.76_{-  0.03}^{+  0.03}$ & $  0.75_{-  0.04}^{+  0.04}$ & $  0.94_{-  0.03}^{+  0.03}$ & $  0.79_{-  0.02}^{+  0.03}$ & $  0.75_{-  0.05}^{+  0.04}$ & $  0.94_{-  0.01}^{+  0.03}$ & $  0.80_{-  0.01}^{+  0.02}$ & $  0.73_{-  0.03}^{+  0.04}$ \\
  $f_\rmn{gas,HSE}$ & $  1.12_{-  0.14}^{+  0.19}$ & $  0.96_{-  0.11}^{+  0.15}$ & $  0.89_{-  0.12}^{+  0.12}$ & $  1.15_{-  0.21}^{+  0.15}$ & $  1.04_{-  0.18}^{+  0.15}$ & $  0.90_{-  0.15}^{+  0.12}$ & $  1.15_{-  0.21}^{+  0.20}$ & $  1.05_{-  0.17}^{+  0.19}$ & $  0.87_{-  0.10}^{+  0.11}$ \\
  $f_\rmn{gas,HSE+clump}$ & $  1.23_{-  0.16}^{+  0.34}$ & $  1.06_{-  0.15}^{+  0.28}$ & $  0.95_{-  0.12}^{+  0.17}$ & $  1.30_{-  0.22}^{+  0.32}$ & $  1.17_{-  0.19}^{+  0.29}$ & $  1.01_{-  0.17}^{+  0.16}$ & $  1.36_{-  0.28}^{+  0.57}$ & $  1.26_{-  0.24}^{+  0.44}$ & $  1.00_{-  0.14}^{+  0.17}$ \\
  $f_\rmn{b}^b$ & $  0.93_{-  0.03}^{+  0.03}$ & $  0.95_{-  0.03}^{+  0.03}$ & $  0.86_{-  0.04}^{+  0.05}$ & $  0.94_{-  0.03}^{+  0.03}$ & $  0.96_{-  0.03}^{+  0.03}$ & $  0.86_{-  0.05}^{+  0.05}$ & $  0.94_{-  0.01}^{+  0.03}$ & $  0.96_{-  0.02}^{+  0.02}$ & $  0.83_{-  0.04}^{+  0.04}$ \\
  $f_\rmn{star}^b$ &   - & $  0.19_{-  0.02}^{+  0.02}$ & $  0.11_{-  0.02}^{+  0.02}$ & - & $  0.16_{-  0.02}^{+  0.02}$ & $  0.11_{-  0.02}^{+  0.02}$ & - & $  0.16_{-  0.01}^{+  0.02}$ & $  0.10_{-  0.02}^{+  0.02}$ \\
  \hline  
  $r < R_{500}$ & & &  & & &  & & & \\
  \hline
$f_\rmn{gas}$ & $  0.92_{-  0.04}^{+  0.04}$ & $  0.71_{-  0.03}^{+  0.03}$ & $  0.69_{-  0.05}^{+  0.04}$ & $  0.95_{-  0.03}^{+  0.03}$ & $  0.75_{-  0.02}^{+  0.03}$ & $  0.69_{-  0.04}^{+  0.05}$ & $  0.95_{-  0.04}^{+  0.03}$ & $  0.77_{-  0.02}^{+  0.02}$ & $  0.68_{-  0.04}^{+  0.04}$ \\
$f_\rmn{gas,HSE}$ & $  1.12_{-  0.11}^{+  0.20}$ & $  0.90_{-  0.09}^{+  0.15}$ & $  0.82_{-  0.09}^{+  0.12}$ & $  1.16_{-  0.15}^{+  0.18}$ & $  0.97_{-  0.11}^{+  0.13}$ & $  0.83_{-  0.10}^{+  0.14}$ & $  1.12_{-  0.15}^{+  0.23}$ & $  0.98_{-  0.13}^{+  0.18}$ & $  0.79_{-  0.10}^{+  0.12}$ \\
$f_\rmn{gas,HSE+clump}$ & $  1.21_{-  0.14}^{+  0.32}$ & $  0.97_{-  0.12}^{+  0.24}$ & $  0.87_{-  0.10}^{+  0.14}$ & $  1.31_{-  0.20}^{+  0.26}$ & $  1.09_{-  0.16}^{+  0.23}$ & $  0.93_{-  0.13}^{+  0.15}$ & $  1.28_{-  0.22}^{+  0.35}$ & $  1.12_{-  0.17}^{+  0.28}$ & $  0.89_{-  0.12}^{+  0.18}$ \\
$f_\rmn{b}^b$ & $  0.92_{-  0.04}^{+  0.04}$ & $  0.96_{-  0.04}^{+  0.04}$ & $  0.83_{-  0.06}^{+  0.05}$ & $  0.95_{-  0.03}^{+  0.03}$ & $  0.97_{-  0.03}^{+  0.04}$ & $  0.83_{-  0.05}^{+  0.06}$ & $  0.95_{-  0.04}^{+  0.03}$ & $  0.97_{-  0.03}^{+  0.04}$ & $  0.80_{-  0.05}^{+  0.05}$ \\
$f_\rmn{star}^b$ & - & $  0.25_{-  0.03}^{+  0.03}$ & $  0.14_{-  0.03}^{+  0.03}$ & - & $  0.22_{-  0.02}^{+  0.02}$ & $  0.14_{-  0.02}^{+  0.03}$ & - & $  0.20_{-  0.02}^{+  0.03}$ & $  0.13_{-  0.02}^{+  0.03}$ \\
  \hline
  $r < R_{1000}$ & & &  & & &  & & & \\
  \hline
$f_\rmn{gas}$ & $  0.90_{-  0.05}^{+  0.04}$ & $  0.64_{-  0.05}^{+  0.05}$ & $  0.60_{-  0.06}^{+  0.05}$ & $  0.93_{-  0.05}^{+  0.05}$ & $  0.69_{-  0.05}^{+  0.05}$ & $  0.61_{-  0.06}^{+  0.06}$ & $  0.95_{-  0.04}^{+  0.03}$ & $  0.71_{-  0.03}^{+  0.04}$ & $  0.61_{-  0.05}^{+  0.04}$ \\
$f_\rmn{gas,HSE}$ & $  1.10_{-  0.09}^{+  0.21}$ & $  0.81_{-  0.07}^{+  0.14}$ & $  0.74_{-  0.09}^{+  0.10}$ & $  1.19_{-  0.13}^{+  0.22}$ & $  0.91_{-  0.12}^{+  0.18}$ & $  0.78_{-  0.10}^{+  0.10}$ & $  1.24_{-  0.22}^{+  0.20}$ & $  0.97_{-  0.18}^{+  0.18}$ & $  0.75_{-  0.09}^{+  0.24}$ \\
$f_\rmn{gas,HSE+clump}$ & $  1.16_{-  0.10}^{+  0.25}$ & $  0.84_{-  0.08}^{+  0.17}$ & $  0.77_{-  0.10}^{+  0.13}$ & $  1.30_{-  0.19}^{+  0.33}$ & $  0.99_{-  0.14}^{+  0.26}$ & $  0.84_{-  0.12}^{+  0.17}$ & $  1.31_{-  0.18}^{+  0.35}$ & $  1.04_{-  0.20}^{+  0.35}$ & $  0.82_{-  0.10}^{+  0.25}$ \\
$f_\rmn{b}^b$ & $  0.90_{-  0.05}^{+  0.04}$ & $  0.99_{-  0.06}^{+  0.06}$ & $  0.79_{-  0.07}^{+  0.06}$ & $  0.93_{-  0.05}^{+  0.05}$ & $  1.00_{-  0.06}^{+  0.06}$ & $  0.80_{-  0.08}^{+  0.08}$ & $  0.95_{-  0.04}^{+  0.03}$ & $  0.99_{-  0.05}^{+  0.07}$ & $  0.79_{-  0.07}^{+  0.07}$ \\
$f_\rmn{star}^b$ & - & $  0.35_{-  0.04}^{+  0.04}$ & $  0.19_{-  0.05}^{+  0.04}$ & - & $  0.30_{-  0.04}^{+  0.04}$ & $  0.19_{-  0.04}^{+  0.05}$ & - & $  0.28_{-  0.04}^{+  0.06}$ & $  0.17_{-  0.05}^{+  0.05}$ \\
  \hline
\end{tabular}}
\end{center}
\begin{quote}
  $^a$ SH $\equiv$ {\it Shock heating}, CSF $\equiv$ {\it Radiative cooling}, AGN $\equiv$ {\it AGN feedback} \\
  $^b$ Note that $f_\rmn{star}$ and $f_\rmn{b}$ are similarly affected from the HSE bias as $f_\rmn{gas}$, which may increase the total baryon budget in galaxy clusters beyond the cosmic value. 
\end{quote}
\end{table*}

\begin{table*}
  \caption{Summarizing the median and $1-\sigma$ percentiles of the enclosed gas, 
    stellar, and baryonic mass fractions, $f_\rmn{gas}$,  $f_\rmn{star}$, $f_\rmn{b}$ 
    for various cluster masses, redshifts, and enclosing radii in our AGN feedback model. We also 
    show the biased values for $f_\rmn{gas}$ measurements when assuming 
    hydrostatic equilibrium, $f_\rmn{gas,HSE}$, and additionally neglecting the 
    density clumping, $f_\rmn{gas,HSE+clump}$.}  
\label{tab:Sum2}
\begin{center}
   \leavevmode
\small{
\begin{tabular}{l|ccc|ccc|ccc}
  \hline
  \hline
  AGN feedback & 
  \multicolumn{3}{c|}{$0.7 < M_{200}/10^{14} \rmn{M}_{\odot} < 1.4$} & 
  \multicolumn{3}{c|}{$1.4 < M_{200}/10^{14} \rmn{M}_{\odot} < 2.6$} & 
  \multicolumn{3}{c}{$2.6 < M_{200}/10^{14} \rmn{M}_{\odot} < 5.1$}\\
  \hline
  &$z = 0$  & $z = 0.5$& $z = 1.0$ &$z = 0$  & $z = 0.5$& $z = 1.0$ &$z = 0$  & $z = 0.5$& $z = 1.0$ \\
  \hline
  $r < R_{200}$ & & &  & & &  & & & \\
  \hline
$f_\rmn{gas}$ & $  0.72_{-  0.05}^{+  0.05}$ & $  0.72_{-  0.05}^{+  0.05}$ & $  0.69_{-  0.06}^{+  0.05}$ & $  0.75_{-  0.04}^{+  0.04}$ & $  0.75_{-  0.05}^{+  0.04}$ & $  0.73_{-  0.03}^{+  0.04}$ & $  0.77_{-  0.03}^{+  0.02}$ & $  0.78_{-  0.03}^{+  0.02}$ & $  0.78_{-  0.03}^{+  0.01}$ \\
$f_\rmn{gas,HSE}$ & $  0.86_{-  0.12}^{+  0.13}$ & $  0.88_{-  0.15}^{+  0.13}$ & $  0.86_{-  0.16}^{+  0.14}$ & $  0.89_{-  0.12}^{+  0.12}$ & $  0.90_{-  0.15}^{+  0.12}$ & $  0.87_{-  0.10}^{+  0.11}$ & $  0.92_{-  0.09}^{+  0.11}$ & $  0.92_{-  0.11}^{+  0.15}$ & $  0.90_{-  0.12}^{+  0.09}$ \\
$f_\rmn{gas,HSE+clump}$ & $  0.91_{-  0.13}^{+  0.17}$ & $  0.95_{-  0.14}^{+  0.21}$ & $  0.96_{-  0.18}^{+  0.24}$ & $  0.95_{-  0.12}^{+  0.17}$ & $  1.01_{-  0.17}^{+  0.16}$ & $  1.00_{-  0.14}^{+  0.17}$ & $  1.02_{-  0.13}^{+  0.19}$ & $  1.04_{-  0.14}^{+  0.23}$ & $  1.00_{-  0.03}^{+  0.22}$ \\
$f_\rmn{b}^a$ & $  0.79_{-  0.06}^{+  0.05}$ & $  0.80_{-  0.06}^{+  0.06}$ & $  0.78_{-  0.07}^{+  0.06}$ & $  0.83_{-  0.04}^{+  0.05}$ & $  0.83_{-  0.05}^{+  0.05}$ & $  0.80_{-  0.04}^{+  0.04}$ & $  0.86_{-  0.03}^{+  0.03}$ & $  0.86_{-  0.04}^{+  0.03}$ & $  0.90_{-  0.03}^{+  0.01}$ \\
$f_\rmn{star}^a$ & $  0.11_{-  0.03}^{+  0.02}$ & $  0.11_{-  0.02}^{+  0.03}$ & $  0.11_{-  0.03}^{+  0.02}$ & $  0.11_{-  0.02}^{+  0.02}$ & $  0.11_{-  0.02}^{+  0.02}$ & $  0.10_{-  0.02}^{+  0.02}$ & $  0.11_{-  0.02}^{+  0.01}$ & $  0.10_{-  0.01}^{+  0.02}$ & $  0.11_{-  0.01}^{+  0.01}$ \\
  \hline  
  $r < R_{500}$ & & &  & & &  & & & \\
  \hline
$f_\rmn{gas}$ & $  0.64_{-  0.05}^{+  0.06}$ & $  0.65_{-  0.06}^{+  0.05}$ & $  0.63_{-  0.07}^{+  0.07}$ & $  0.69_{-  0.05}^{+  0.04}$ & $  0.69_{-  0.04}^{+  0.05}$ & $  0.68_{-  0.04}^{+  0.04}$ & $  0.73_{-  0.04}^{+  0.03}$ & $  0.73_{-  0.04}^{+  0.03}$ & $  0.76_{-  0.02}^{+  0.01}$ \\
$f_\rmn{gas,HSE}$ & $  0.78_{-  0.09}^{+  0.11}$ & $  0.81_{-  0.12}^{+  0.13}$ & $  0.79_{-  0.16}^{+  0.13}$ & $  0.82_{-  0.09}^{+  0.12}$ & $  0.83_{-  0.10}^{+  0.14}$ & $  0.79_{-  0.10}^{+  0.12}$ & $  0.88_{-  0.11}^{+  0.17}$ & $  0.87_{-  0.14}^{+  0.11}$ & $  0.89_{-  0.09}^{+  0.13}$ \\
$f_\rmn{gas,HSE+clump}$ & $  0.82_{-  0.10}^{+  0.14}$ & $  0.86_{-  0.13}^{+  0.19}$ & $  0.86_{-  0.14}^{+  0.17}$ & $  0.87_{-  0.10}^{+  0.14}$ & $  0.93_{-  0.13}^{+  0.15}$ & $  0.89_{-  0.12}^{+  0.18}$ & $  0.92_{-  0.09}^{+  0.21}$ & $  0.98_{-  0.17}^{+  0.11}$ & $  1.07_{-  0.23}^{+  0.07}$ \\
$f_\rmn{b}^a$ & $  0.79_{-  0.07}^{+  0.07}$ & $  0.80_{-  0.07}^{+  0.07}$ & $  0.78_{-  0.08}^{+  0.07}$ & $  0.83_{-  0.06}^{+  0.05}$ & $  0.83_{-  0.05}^{+  0.06}$ & $  0.80_{-  0.05}^{+  0.05}$ & $  0.86_{-  0.04}^{+  0.04}$ & $  0.86_{-  0.04}^{+  0.04}$ & $  0.90_{-  0.03}^{+  0.02}$ \\
$f_\rmn{star}^a$ & $  0.15_{-  0.04}^{+  0.03}$ & $  0.14_{-  0.03}^{+  0.05}$ & $  0.14_{-  0.03}^{+  0.03}$ & $  0.14_{-  0.03}^{+  0.03}$ & $  0.14_{-  0.02}^{+  0.03}$ & $  0.13_{-  0.02}^{+  0.03}$ & $  0.13_{-  0.02}^{+  0.02}$ & $  0.12_{-  0.02}^{+  0.02}$ & $  0.14_{-  0.01}^{+  0.01}$ \\
  \hline
  $r < R_{1000}$ & & &  & & &  & & & \\
  \hline
$f_\rmn{gas}$ & $  0.55_{-  0.07}^{+  0.06}$ & $  0.56_{-  0.07}^{+  0.06}$ & $  0.55_{-  0.07}^{+  0.07}$ & $  0.60_{-  0.06}^{+  0.05}$ & $  0.61_{-  0.06}^{+  0.06}$ & $  0.61_{-  0.05}^{+  0.04}$ & $  0.66_{-  0.05}^{+  0.05}$ & $  0.67_{-  0.06}^{+  0.03}$ & $  0.67_{-  0.03}^{+  0.03}$ \\
$f_\rmn{gas,HSE}$ & $  0.69_{-  0.09}^{+  0.10}$ & $  0.72_{-  0.11}^{+  0.16}$ & $  0.71_{-  0.11}^{+  0.17}$ & $  0.74_{-  0.09}^{+  0.10}$ & $  0.78_{-  0.10}^{+  0.10}$ & $  0.75_{-  0.09}^{+  0.24}$ & $  0.80_{-  0.08}^{+  0.11}$ & $  0.84_{-  0.10}^{+  0.11}$ & $  0.97_{-  0.10}^{+  0.11}$ \\
$f_\rmn{gas,HSE+clump}$ & $  0.71_{-  0.09}^{+  0.11}$ & $  0.76_{-  0.12}^{+  0.19}$ & $  0.79_{-  0.15}^{+  0.20}$ & $  0.77_{-  0.10}^{+  0.13}$ & $  0.84_{-  0.12}^{+  0.17}$ & $  0.82_{-  0.10}^{+  0.25}$ & $  0.84_{-  0.08}^{+  0.14}$ & $  0.89_{-  0.09}^{+  0.15}$ & $  1.06_{-  0.15}^{+  0.11}$ \\
$f_\rmn{b}^a$ & $  0.75_{-  0.09}^{+  0.08}$ & $  0.76_{-  0.09}^{+  0.09}$ & $  0.75_{-  0.09}^{+  0.08}$ & $  0.79_{-  0.07}^{+  0.06}$ & $  0.80_{-  0.08}^{+  0.08}$ & $  0.79_{-  0.07}^{+  0.07}$ & $  0.84_{-  0.06}^{+  0.06}$ & $  0.83_{-  0.06}^{+  0.06}$ & $  0.86_{-  0.03}^{+  0.04}$ \\
$f_\rmn{star}^a$ & $  0.20_{-  0.06}^{+  0.05}$ & $  0.20_{-  0.06}^{+  0.06}$ & $  0.20_{-  0.05}^{+  0.05}$ & $  0.19_{-  0.05}^{+  0.04}$ & $  0.19_{-  0.04}^{+  0.05}$ & $  0.17_{-  0.05}^{+  0.05}$ & $  0.18_{-  0.04}^{+  0.03}$ & $  0.16_{-  0.02}^{+  0.05}$ & $  0.19_{-  0.01}^{+  0.03}$ \\
  \hline
\end{tabular}}
\end{center}
\begin{quote}
  $^a$ Note that $f_\rmn{star}$ and $f_\rmn{b}$ are similarly affected from the HSE bias as $f_\rmn{gas}$, which may increase the total baryon budget in galaxy clusters beyond the cosmic value. 
\end{quote}
\end{table*}

\section{Comparison between $\KU$ and centroid shift}
\label{sec:KU_w}

The kinetic-to-thermal energy ratio of a cluster, $\KU$, is the most direct
quantity to estimate the dynamical activity of a cluster and easily calculable
in simulations. However, it is not directly observationally accessible. As a
result, other proxies have been introduced to estimate a cluster's dynamical
state, such as the centroid shift $\bra w \ket$ \citep{1993ApJ...413..492M},
which measures the standard deviation between the projected separation of the
X-ray peak and the centroid of the X-ray emission that is calculated in
concentric circular apertures. In previous simulation work, $\bra w \ket$ was
found to be the most sensitive estimator of dynamical activity
\citep{2007MNRAS.380..437P}, thus motivating a comparison of the $\KU$ estimator
for the dynamical state of a cluster to $\bra w\ket$.

On projected X-ray surface brightness maps of our simulations, we follow
\citet{2009A&A...498..361P} and calculate
\begin{equation}
\bra w\ket = \left[\frac{1}{N-1}\sum_{i=1}^N
(\Delta_i - \bra\Delta\ket)^2\right]^{1/2} \times \frac{1}{R_{200}},
\end{equation}
where $\Delta_i$ is the projected distance between the surface brightness peak
and the centroid of the $i$th aperture. The centroids are calculated within
radial annuli ranging from $0.1 R_{200}$ to $0.05\, i\, R_{200}$ with
$i=3,4,\ldots,20$. The central region is traditionally avoided to minimize
biases associated with enhanced emission from cool cores (although inclusion
does not significantly change the results obtained).

\begin{figure}
\resizebox{\hsize}{!}{\includegraphics{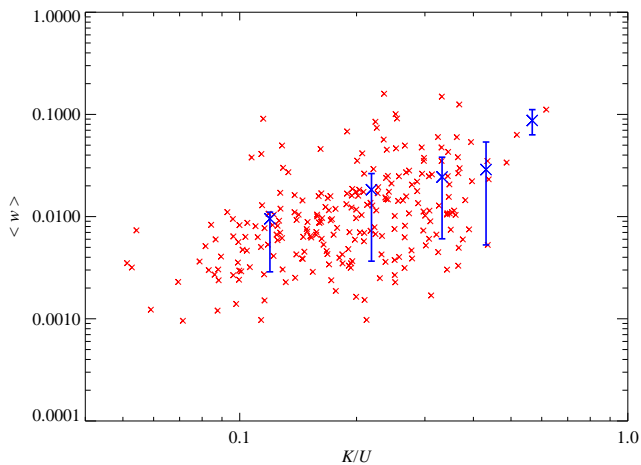}}
\caption{We compare the (theoretically accessible) kinetic-to-thermal energy
    ratio $\KU$ to the (observationally measurable) centroid shift $\bra w \ket$
    for the AGN feedback simulations at $z = 0$. Here the red crosses represent
    those values for each cluster with $M_{200} \ge 2\times 10^{14}\,
    \mathrm{M}_{\odot}$ in the simulations and the blue crosses are bin-averaged
    quantities. The linear correlation coefficient between $\KU$ and $\bra w
    \ket$ is $r\sim 0.4$ indicating a weak correlation between these two
    dynamical cluster measures.}
\label{fig:KU_w}
\end{figure}

Figure \ref{fig:KU_w} shows the correlation between $\KU$ and $\bra w \ket$.
Their correlation coefficient is $r\simeq 0.4$, which indicates that $\KU$ and
$\bra w \ket$ are only weakly correlated, making $\bra w\ket$ a biased proxy for
the dynamical state of a cluster. At any given $\KU$, the values for $\bra w
\ket$ scatter by up to two orders of magnitude.  This is in agreement with our
previous work \citepalias{2012ApJ...758...74B}, where we showed that the
correlation between $KU$ and $1 - \CA$, the ratio of 3D major to minor axis, is
$\simeq 0.58$. These two results suggest a reason for the weak correlation of
the centroid shift with $\KU$. This is partially caused by the substantial
scatter between the morphological irregularity and asphericity on the one side
and dynamical activity of a cluster on the other side, and partially because of
projection effects. More work is needed to combine different observationally
accessible proxies that correlate more tightly with theoretical measures of
dynamical cluster activity, such as the kinetic-to-thermal energy ratio, $\KU$.

\section{Comparing definitions for hydrostatic mass bias}
\label{sec:lau}

\begin{figure}
\resizebox{\hsize}{!}{\includegraphics{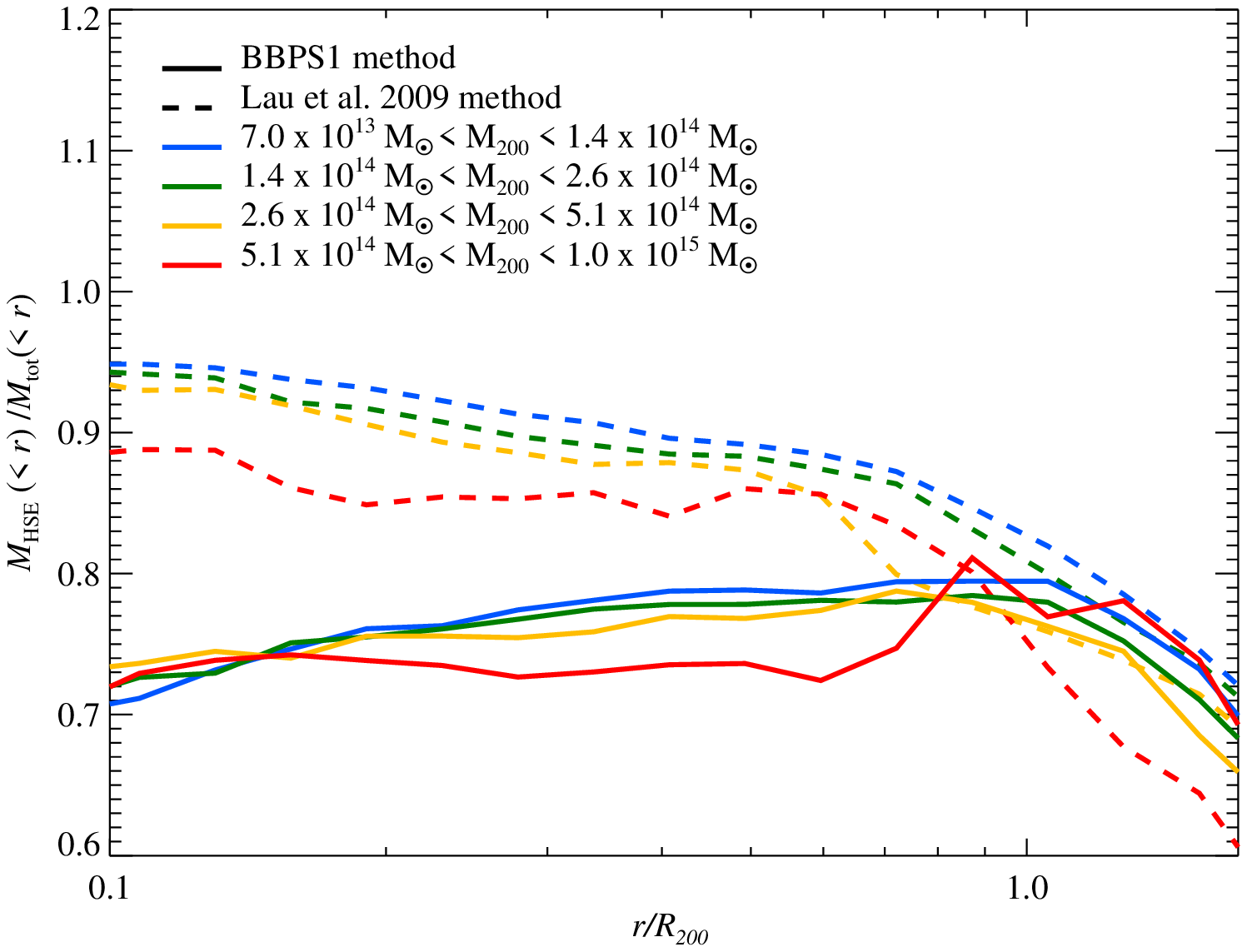}}
\caption{We show a comparison of two different methods for determining the HSE
  bias, $M_\rmn{HSE}/M_{\rmn{tot}}$, in various cluster mass bins for our AGN
  feedback simulations at $z=0$. The difference between
  $M_\rmn{HSE}/M_{\rmn{tot}}$ in this work
  \citepalias[and][]{2012ApJ...758...74B} and \citet{2009ApJ...705.1129L} is the
  definition of $M_{\rmn{tot}}$. Here, we adopt $M_{\rmn{tot}} \equiv M_\rmn{DM}
  + M_\rmn{gas} + M_\rmn{star}$ whereas \citet{2009ApJ...705.1129L} define
  $\tilde{M}_\rmn{tot} \propto \dd (P_\rmn{th}+P_\rmn{kin}) /\dd r$, which
  assumes spherical symmetry and a steady state. The radial trends differ the
  greatest at small radii, presumably due to violation of spherical symmetry and
  non-stationary, clumpy accretion.}
\label{fig:lau}
\end{figure}

Here we compare various definitions for the bias by hydrostatic equilibrium
assumption and review the differing methodologies and definitions used in the
(simulation) literature. Direct comparison between our work and others is
difficult since the methods used for calculating $M_\rmn{HSE}$ are
different. Previous simulation work used high resolution zoomed simulations of
individual clusters
\citep{2006MNRAS.369.2013R,2009ApJ...705.1129L,2012NJPh...14e5018R}, where the
authors were able to smooth and removed clumps from each cluster before
calculating $\dd P_\rmn{th}/\dd r$. Other simulations
\citep[e.g.,][]{2006MNRAS.369.2013R} produced mock X-ray images and performed
similar smoothing and clump removal techniques to observations. These {\it
  cleaning} procedures were not preformed in this work \citepalias[nor
in][]{2012ApJ...758...74B}, because of the different goals of this work. We
pursue a theoretically focused approach that should be equally applicable to
X-ray ({\em eROSITA}) and SZ cluster surveys (by {\em ACT} and {\em SPT}) and
aim at strengthening the understanding of the underlying connection of the hot
ICM rather than producing a specific mock observation tailored to a particular
observatory.

Another difference is in the definition of $M_\rmn{tot}$.
\citet{2009ApJ...705.1129L} define the total mass as $\tilde{M}_\rmn{tot}
\propto \dd (P_\rmn{th}+P_\rmn{kin}) /\dd r$, which assumes spherical symmetry
and a steady state. This is different from our approach \citepalias[and that
in][]{2012ApJ...758...74B} since we define $M_{\rmn{tot}} \equiv M_\rmn{DM} +
M_\rmn{gas} + M_\rmn{star}$. These different definitions for $M_{\rmn{tot}}$
produce significantly different HSE biases, $M_\rmn{HSE}/M_{\rmn{tot}}$, within
$R_{200}$ as illustrated in Figure \ref{fig:lau}. Note that the difference of
the solid and dashed curves in Figure \ref{fig:lau} (at each color and radius)
directly measures $\Delta M_\rmn{tot}=M_\rmn{tot}-\tilde{M}_\rmn{tot}$ (since we
use the identical $M_\rmn{HSE}$ in both cases).  In particular both approaches
yield significantly different radial trends at small radii, presumably due to
violation of spherical symmetry and non-stationary, clumpy accretion in the
definition of $\tilde{M}_\rmn{tot}$.  Adopting the definition of
\citet{2009ApJ...705.1129L} for the total mass, we recover the same radial
trends as these authors, which suggests that the main difference originates from
the different definition of $M_{\rmn{tot}}$ rather than $M_\rmn{HSE}$.

\section{Horizontally sliced SPH kernel integral}
\label{sec:SPH-fit}

To speed up the mass binning of a distribution of SPH particles into a spherical
shell, we integrate the mass portion of the SPH-kernel within that shell. We
make the approximation that the smoothing length $h$ is much smaller than the
radius of the shell in question, $r_{\rm{shell}}$, so the shell can be thought
of as two plane-parallel slices. We obtain a $9^{\rmn{th}}$ order polynomial fit
to the integrated SPH mass fraction below some normalized distance $x$, which is
constrained to be exact at $x=\{-1,0,1\}$:

\begin{equation}
\begin{aligned}
  M_{\rmn{gas}}(<x) &= 0.5 + 1.391850\, x - 2.543398\, x^3 \\
                  &+ 3.373049\, x^5 - 2.474134\, x^7 + 0.7526237\, x^9,
\end{aligned}
\label{eq:SPH-mass}
\end{equation}
where $x = (r - r_0)/h$ for an SPH particle at position $x_0$. The fit is
everywhere accurate to better than $2\times 10^{-3}$ of the total mass. Clearly,
this approximation breaks down near the center of the cluster (where the shell
radius could become comparable to the smoothing length). However, the constraint
at $\pm1$ guarantees conservation of mass even in the event the approximation
$r_{\rm{shell}}\gg h$ breaks down.

\section{Variance of projected $M_{\rmn{gas}}$ and $f_{\rmn{gas}}$ for differing number of sector elements}
\label{sec:sector-number}

\begin{figure*}
    \resizebox{0.5\hsize}{!}{\includegraphics{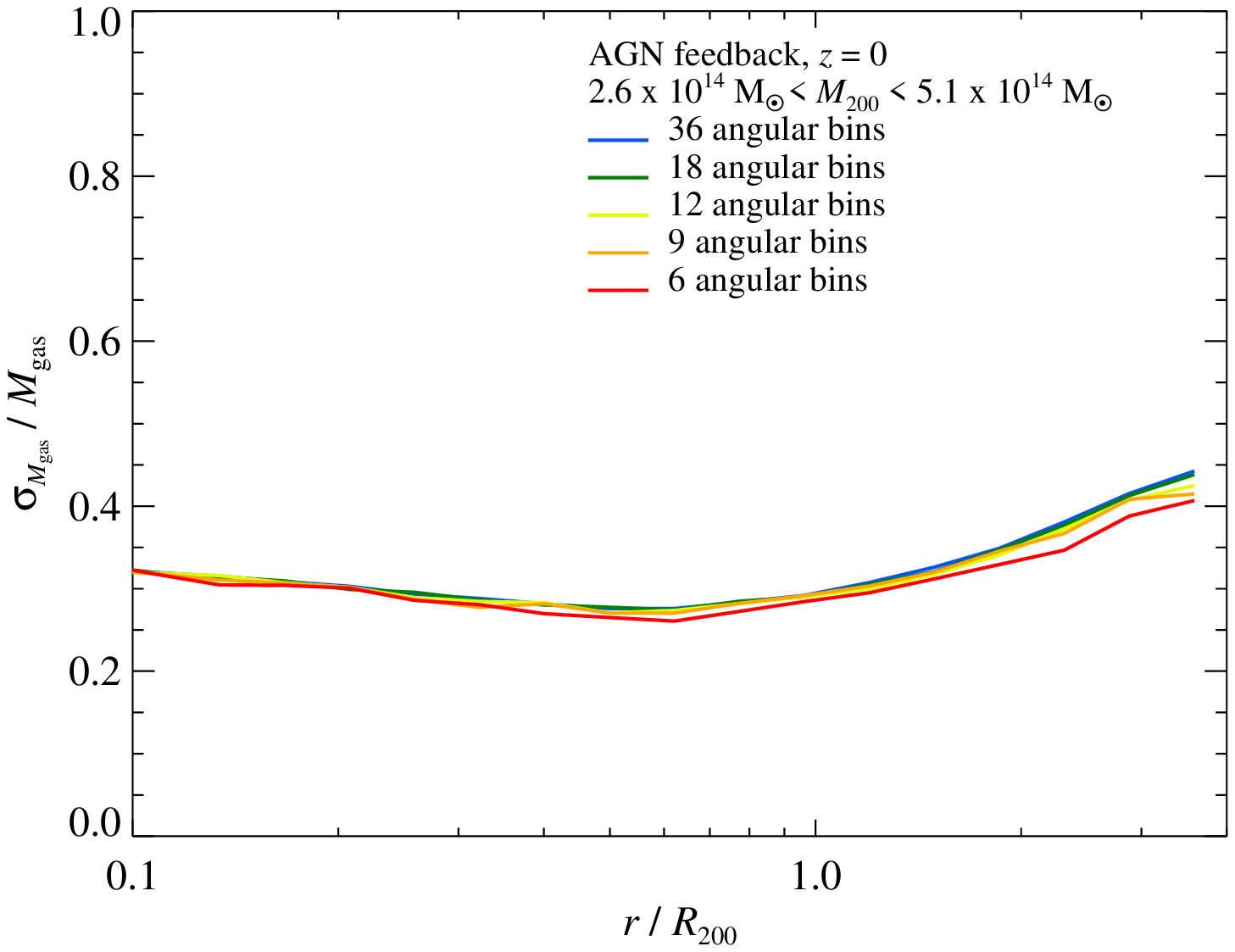}}%
    \resizebox{0.5\hsize}{!}{\includegraphics{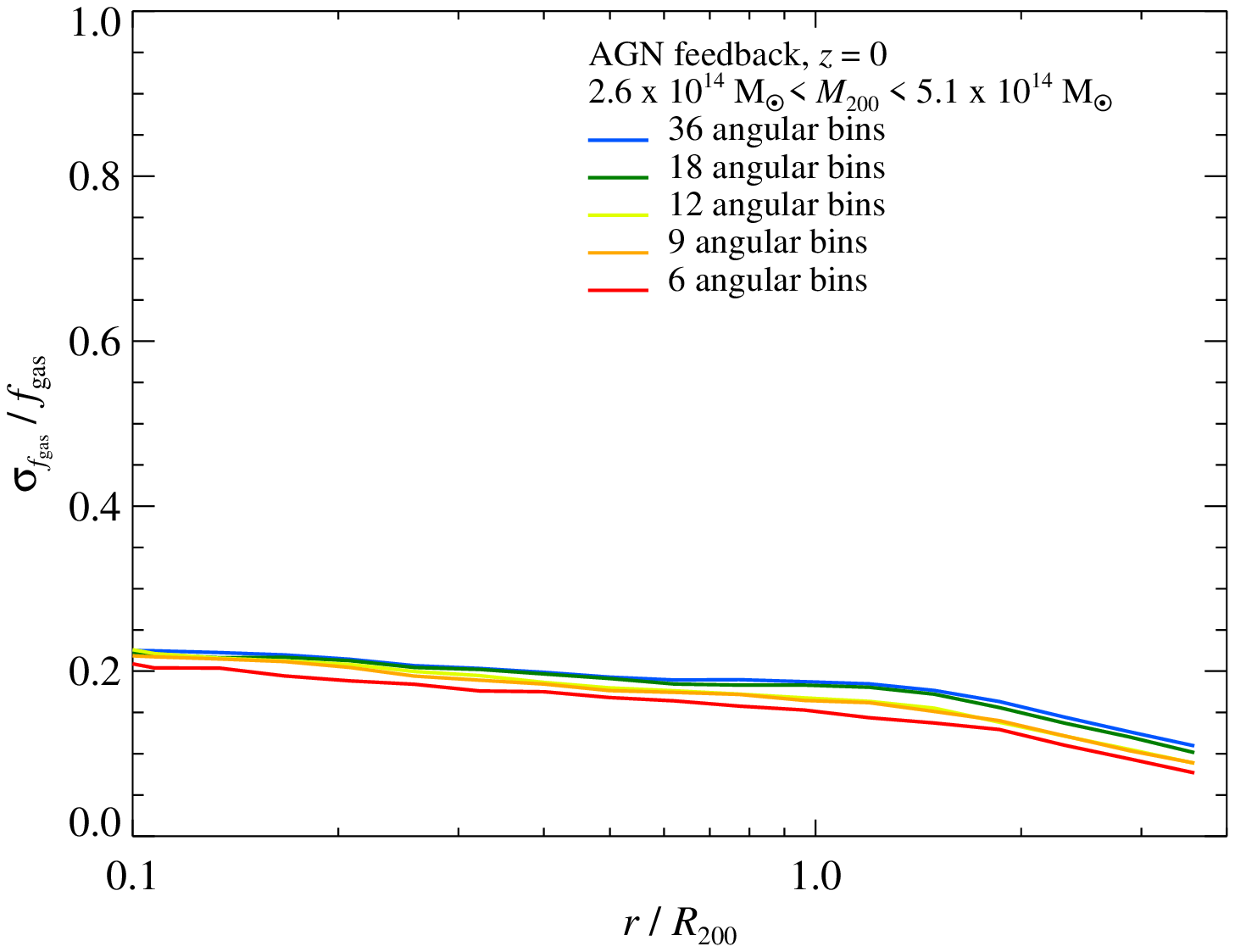}}\\
    \caption{Angular variance in projected (2D) maps of $M_{\rmn{gas}}$
        (left) and $f_{\rmn{gas}}$ (right) for differing number of sector
        elements. Here we show the measured variance (uncorrected for the
        sub-Poissonian noise term) and use all clusters in our sample with
        virial masses in the range $2.6\times 10^{14}\,\rmn{M}_\odot< M_{200}<
        5.1\times 10^{14}\,\rmn{M}_\odot$. There is only little difference in
      the 2D angular variances for varying number of sector elements.}
\label{fig:sectors}
\end{figure*}

In Figure~\ref{fig:sectors}, we address how the 2D angular variances of
$M_{\rmn{gas}}$ and $f_{\rmn{gas}}$ change if we vary the number of sector
elements. There is almost no difference in $\sigma_{M_{\rmn{gas}}}$ for varying
number of sector elements and the small systematic decrease in
$\sigma_{f_{\rmn{gas}}}$ is expected for the increase in bin volume (decrease in
sector numbers). Note that this is quite different for the angular variance of
the X-ray luminosity, which strongly depends on the number of sector elements,
especially in SPH simulations \citep[see Figure 4 of][]{2011MNRAS.413.2305V}. We
believe that this increased scatter solely derives from the increased clumping
factor with radius (see Figure~\ref{fig:fgas}).

\end{appendix}

\end{document}